\documentclass[10pt, fleqn, twocolumn, twoside]{article}

\usepackage{color}
\usepackage{epsfig}
\usepackage{amssymb}
\usepackage{graphicx}
\usepackage{special}

\begin{document}

          \onecolumn
          %\title{Galactic cosmic rays in a diffusion model}

	\thispagestyle{empty}
	
          \begin{center}
	{\sf \large To appear in Research Signposts\\
	"Recent Research Developments in Astrophysics"}
          \end{center}
          \vspace{5mm}
	
		            \begin{center}
	{\sf \huge Galactic Cosmic Ray Nuclei as a\\
	Tool for Astroparticle Physics}
          \end{center}

          \vspace{5mm}

          \begin{center}
	Maurin D.$^{1,2}$, Taillet R.$^{1,3}$, Donato F.$^4$, Salati 
	P.$^{1,3}$, Barrau A.$^{5,6}$, Boudoul G.$^{5,6}$
          \end{center}

          \vspace{5mm}
          \begin{center}
	$^1$ Laboratoire de Physique Th\'eorique ({\sc lapth}), 74941
Annecy-le-Vieux, France\\
	$^2$ Institut d'Astrophysique de Paris ({\sc iap}), 98 bis Bd
Arago, 75014 Paris,
	France\\
	$^3$ Universit\'e de Savoie, 73011 Chamb\'ery, France\\
	$^4$ Dipartimento di Fisica Teorica and {\sc infn}, via
	Giuria 1, Torino, Italy\\
	$^5$ Institut des Sciences Nucl\'eaires ({\sc isn}) de Grenoble, 53,
	Avenue des Martyrs, 38026 Grenoble, France   \\
	$^6$ Universit\'e Joseph Fourier, 38000 Grenoble, France
          \end{center}
		  
          \vspace{5mm}
		  
%           \begin{center}
% 	{\Large  Short running title : "Galactic cosmic rays"}
%           \end{center}

          \twocolumn
{\footnotesize {\bf Abstract: } Cosmic Ray nuclei in the energy
range 100 MeV/nuc - 100 GeV/nuc provide crucial information about the
physical properties of the Galaxy. They can also be used to answer
questions related to astroparticle physics. This paper reviews the
results obtained in this direction, with a strong bias towards the
work done by the authors at {\sc lapth}, {\sc isn} and {\sc iap}.
The propagation of these nuclei is studied quantitatively in the
framework of a semi-analytical  two-zone diffusion model taking into
account the effect of galactic wind, diffuse reacceleration and energy
losses.
The parameters of this model are severely constrained by an analysis
of the observed B/C ratio. These constraints are then used to study
other species such as radioactive species and light antinuclei.
Finally, we focus on the astroparticle subject and we study the flux
of antiprotons and antideuterons that might be due to neutralino
annihilations or primordial black hole evaporation.
The question of the spatial origin of all these species is also
addressed.
}

\tableofcontents

%%%%%%%%%%%%%%%%%%%%%%%%%%%%%%%%%%%%%%%%%%%%%%%%%%%%%%%%%%%%%%%%%%%%%%%%%%%%%%%%%%%%%
%%%%%%%%%%%%%%%%%%%%%%%%%%%%%%%%%%%%%%%%%%%%%%%%%%%%%%%%%%%%%%%%%%%%%%%%%%%%%%%%%%%%%

\section{Introduction}
\label{sec:intro}

Most of the available information about matter
in our Universe, and
in our Galaxy in
particular, comes indirectly from the collection of the electromagnetic
radiation
(from meter waves to $\gamma$ rays) that was emitted or
absorbed by this matter.
A completely different information is provided by the cosmic ray
nuclei, which constitute a genuine sample of galactic matter.
Many different nuclei species are observed, in a wide range of energy,
and with different origins. Some of them come unaltered from the sources
(they are called {\em primaries}\/), others ({\em secondaries}\/) come from
nuclear reactions between the primaries and the interstellar medium,
or from the  disintegration of unstable species.
Moreover, the trajectories of these nuclei from creation to detection
are rather erratic, due to the influence of the galactic magnetic
field on all the charged particles, and it is generally not possible
to follow the direction of an incoming nucleus back to the source.
If we were able to understand clearly the processes by which all these
nuclei are produced, accelerated and propagated in the Galaxy,
the wealth of data available now or in the near future would yield
most valuable
information about the matter content and magneto-hydrodynamical
properties of our Galaxy.
In principle, it would even be possible to discover some
evidence for new physics (e.g. supersymmetry) or new objects
(e.g. primordial
black holes or stars made of antimatter) as they can give rise to the
emission of charged  antinuclei and make an extra
contribution to the
observed cosmic ray fluxes.
This review presents a summary of the work made in
this direction by the authors from 1999 to 2002, in a {\sc lapth-isn-iap}
collaboration.
As a first step, we tried to reach a {\em quantitative}\/ understanding
of the propagation
of cosmic ray nuclei in the energy range 100~MeV/nuc-100~GeV/nuc.
More precisely, we described propagation with a diffusion
model, in which the free parameters are adjusted to account for the
available data on cosmic rays. This provides the regions of the
parameter space allowed by the data.
As a second step, we took advantage of this model to investigate several
points concerning astrophysics and astroparticle physics.

During this study, various aspects  related to the ``standard"
or to more speculative processes were examined in detail.
These may be summarized as follows:
\begin{itemize}
          \item Standard cosmic rays
          \begin{itemize}
	\item Diffusion parameters from secondary-to-primary ratio
	\item The flux of standard secondary antiprotons and antideuterons
	\item Spatial origin
	\item Radioactive species and the local bubble
	\item Evolution of composition with energy
          \end{itemize}
          \item Exotic cosmic rays
          \begin{itemize}
	\item Baryonic Dark Matter
	\item Antimatter
	\item Supersymmetric particles
	\item Primordial Black Holes
          \end{itemize}
\end{itemize}
The extraction of the diffusion parameters is the central goal since
all conclusions follow from their values.

%%%%%%%%%%%%%%%%%%%%%%%%%%%%%%%%%%%%%%%%%%%%%%%%%%%%%%%%%%%%%%%%%%%%%%%%%%%%%%%%%%%%%
%%%%%%%%%%%%%%%%%%%%%%%%%%%%%%%%%%%%%%%%%%%%%%%%%%%%%%%%%%%%%%%%%%%%%%%%%%%%%%%%%%%%%

\section{Propagation models}
\label{sec:diff.propag.models}

		%---------------------------------%
		%---------------------------------%

\subsection{The context}
\label{subsec:context}

The study of cosmic ray radiation raises a great number of questions.
The most obvious ones are how the charged nuclei are
accelerated and how do they propagate in chaotic magnetic fields.
These questions are actually still actively studied and debated,
but the nature of the sources and propagation media on which cosmic ray
physicist focus has slightly changed in the last twenty years.
The pioneering work of Parker~\cite{Parker01},
who first used the diffusion-convection equation, focused on Solar
modulation. The same equation was used in the late
seventies~\cite{Krimsky_et_cie} to investigate the
acceleration of Galactic Cosmic Rays (hereafter GCR, in opposition to
Solar Cosmic Rays, SCR)
in sources that were strongly suspected to be supernov\ae\ (SN)
remnants. Whereas
SCR deal with the Sun magnetic field and plasma in the Solar cavity,
GCR have to face the question of transport in the largely unknown
galactic magnetic field. These studies reached an even larger scale,
since the discovery of ultra high energy cosmic rays (UHECR) that
are certainly extragalactic in origin. The exact
nature of this radiation is not clearly established but if these are
charged nuclei, the relevant medium in which these nuclei propagate is the
extra-galactic magnetic field. From the UHECR source point of view,
one has to imagine some powerful astrophysical sources
where ``standard" acceleration occurs, but an {\em astroparticle}\/ solution
(i.e. heavy meta-stable decaying new particles) could furnish
the energetic particles as well.
New particles could thus be discovered by the study of cosmic
radiation, which is an interesting coming back to the very first
concern of this field of research, when the muon and the positron
were first observed in cosmic rays.

Notwithstanding the fact that UHECR is one of the driving subjects
in the development of the so-called astroparticle field,
the study of much less energetic particles, with energies ranging
from GeV to PeV, may also put constraints on the existence of new
particles.
As discussed at length in the following, these low energy particles have
a galactic origin, and by their study,
many questions concerning the Galaxy may be addressed,
such as LiBeB primordial abundances, dark matter content of our
Galaxy, or the nature of the sources.
It might be possible to reach a consistent
picture of either conventional (SNs, Wolf-Rayet stars) or possible more exotic
(micro-quasars~\cite{Sunyaev}, anti-globular clusters~\cite{Belotsky})
galactic sources.
As a consequence, we underline that GCR nuclei in the GeV-PeV energy
range  provide a quite interesting laboratory
for both astrophysicists and particle physicists.
  From now on, we will only consider cosmic rays in this energy range.

The great amount of data in various energy ranges
has led to a quite good understanding of charged nuclei
propagation (see e.g.~\cite{Webber97}) along with the induced $\gamma$-ray
production at low and high latitude in the galactic plane.
As the gyration radius of charged particles in the galactic magnetic field
is small, the propagation is intimately related to the
detailed structure of this turbulent magnetic field.
The latter is not observed directly and one
would like to use the cosmic rays to infer its properties.
It turns out to be a difficult task as, despite the success outlined
above, several unknowns and inconsistencies
remain at the quantitative level.
Unlike the Solar case for which we have {\em in situ}\/ observations
of this turbulence -- as early as in the mid-sixties (see e.g.~\cite{Jokipii})
-- favoring a Kolmogorov spectrum, some recent MHD
simulations~\cite{Ptuskin97}
along with our secondary/primary studies point towards greater spectral
index of turbulence. Even though a satisfying global picture emerges,
some nuclei resist to a simple interpretation.
If the first enduring problem --  the depletion of the
grammage distribution at low values for sub-Fe/Fe~\cite{Modif_PLD,DuVernois96} --
seems now to be solved thanks to new cross section
measurements~\cite{Webber_PLD}, another lasting discussion is related to
acceleration and selection mechanisms that lead to the observed abundances.
Is it a chemical selection, i.e. First Ionization Potential
bias (see e.g.~\cite{Meyer}), or a volatility
bias related to grain destruction in SN explosions~\cite{Meyer_et_al}?
The accuracy of current data is not sufficient to conclude and the
question remains in suspense, albeit the importance to elucidate
the injection mechanism in acceleration models. It is at least known to a
great certainty from radioactive primaries ($^{57}$Co, $^{59}$Ni)
synthesized in SN that $\sim 10^5$ years have past
between synthesis and acceleration~\cite{Wiedenbeck}.
The energy spectrum produced by the sources has also some indeterminacy.
The acceleration models agree about a power-law dependence in rigidity
${\cal R}^{-\alpha}$ (where the {\em rigidity}\/ is defined as momentum per
unit charge ${\cal R} \equiv p/Z$)
but numerical estimates of the spectral index $\alpha$ can be rather different.
Values $\alpha\lesssim2.0$ are preferred
by acceleration theory (see~\cite{pente_source} and in
particular~\cite{Jones00}
for a short and readable
introduction on diffusive shock acceleration), but if Kolmogorov
spectrum for diffusion is
retained, one is left with $\alpha\approx 2.5$ from observed spectra
that seems more problematic.
Spectra may also differ from pure power laws. This depends also on the kind
of sources involved (SN, explosion in wind bubbles, superbubbles)  along
with their plasma and magnetic
states. There could be less explosive sources such a Wolf-Rayet stars
which eject species through a powerful wind (this is
the $^{22}$Ne abundance anomaly, see e.g.~\cite{Soutoul01}).

This list is far from being exhaustive, even if we restrict ourselves
to the case of $\sim$GeV/nuc charged nuclei.
To conclude on what we call ``standard" CR
studies, we emphasize that the limiting factor is related to our
poor knowledge of most production cross sections.

		%---------------------------------%
		%---------------------------------%

\subsection{Physical motivations for diffusive propagation}
\label{subsec:motivation}

\shortversion{%% version longue
The history of cosmic ray propagation theory traces back to
the pioneering work of Fermi, which provides a statistical
description of the way random magnetic
irregularities scatter the particles and accelerate them.
This leads to diffusion in energy.
Earlier in the same decade,
Chandrasekhar~\cite{Chandra} had shown rigorously that diffusion could be
equivalently described in terms of random walk, i.e. sequence of small
erratic steps.
By then, it was possible to imagine that the scattering mentioned
above could lead to spatial diffusion.
Actually, this hypothesis was first proposed on a  phenomenological
basis, as emphasized in Berezinskii {\em et al.}\/~\cite{Berezinskii},
in the context of Solar particles transport.
It was then applied to cosmic ray transport in the Galaxy.
The notion of diffusive motion was confirmed and refined by more fundamental
approaches, based on relativistic Boltzmann equations~\cite{Jones}.
It now appears as a most valid description, that accounts for a wide range
of observations (both in acceleration and propagation mechanisms).
The reader is referred to~\cite{Jokipii} for a historical background
about transport equations (see also~\cite{Gleeson_Axford} for a
derivation from the kinetic equation level and~\cite{Jokipii_Parker} for
the diffusion/convection equation).
}
{ %% version courte
}

The linearized kinetic theory approach provides grounds for a
consistent derivation
of the transport equation, which reads, neglecting spallations and
energy losses for the sake of clarity~\cite{Berezinskii}
\begin{eqnarray}
    \lefteqn{\frac{\partial f}{\partial t}-\vec{\nabla} (K\vec{\nabla} f
    -\vec{V_c} f)-
    \frac{\vec{\nabla}.\vec{V_c}}{3}\frac{1}{p^2}
    \frac{\partial}{\partial p}
    (p^3 f)} \nonumber \\
    && =
    \frac{1}{p^2}\frac{\partial}{\partial p}p^2 K_{pp}
    \frac{\partial}{\partial p}f +\frac{dQ}{dp}\;\;,
    \label{linearize}
\end{eqnarray}
where $f\equiv f(t,\vec{r},\vec{p})$ is the phase space
distribution.
This equation contains most of the effects described below,
like spatial diffusion, galactic wind, with the adiabatic energy loss
associated, and diffusion in momentum space.
In this equation, spatial diffusion has been assumed to be isotropic.
Actually, the diffusion coefficient $K$ should be replaced by a tensor,
with parallel and transverse components.
As regards the first one, there is a strong consensus
about a form
\begin{equation}
          K_{\parallel}({\cal R})=K_0 \, \beta  \, {\cal R}^{2-\kappa}\;,
\end{equation}
where $\kappa$ is the spectral index of the turbulence spectrum.
The transverse component is still debated, and the two main
propositions (given by quasi-linear or the Bohm conjecture)
are probably wrong~\cite{Casse}.
All the results presented here are based on the usual assumption
that diffusion is isotropic, with a diffusion coefficient
\begin{equation}
          K({\cal R})=K_0 \, \beta \, {\cal R}^{\delta}\;,
\end{equation}
where the normalization $K_0$ and the spectral index $\delta$ should
ideally be related to the astrophysical properties of the interstellar medium.
Unfortunately, our knowledge in this field is still demanding, and
the value of the two parameters $K_0$ and  $\delta$ can only
be determined indirectly by the analysis of cosmic ray observations.

The different diffusion schemes mentioned above also lead
to different forms for the energy dependence of the reacceleration
term $K_{pp}$~\cite{Schlicheiser01}. The reader is referred to
\cite{Maurin02}, and we will not discuss this point further.

		%---------------------------------%
		%---------------------------------%

\subsection{Overview of the effects affecting propagation}
\label{subsec:effect_propag}

This section is devoted to a brief overview of the physical effects
that play a role in the propagation of cosmic rays.
The next section (see \ref{subsec:geometry})
will enter into more details and focus on the
modelling of these effects.

			%###############%
\subsubsection{Geometry and content of the Galaxy}

The propagation of cosmic rays ceases to be of diffusive nature beyond
some surface where they can freely stream out of the diffusive volume.
The density then drops to nearly zero, so that this surface may be
considered as an absorbing boundary.
The exact shape and dimensions of this boundary are not known, but
direct observations of the radio halo of external galaxies suggest
that it might radially follow the galactic disc, with a greater thickness.
In this work, the diffusive halo will be modelled as a
cylinder of radius $R= 20$ kpc and half-height $L$ whose numerical
value, still to be determined, is probably greater than a few kpc.
The boundary thus imposes that the density satisfies
$N(r=R,z)=N(r,z=\pm L)=0$.
Embedded in this diffusive halo lies the galactic disc (hereafter ``the
disc", in short) containing the stars and the gas .
The gas is mostly made of hydrogen (90\%),
neutral and ionized, and helium (10\%)
(the heavier nuclei that may be present are of negligible importance
for our concerns).
The different components (stars and gas) have different half
heights $h_i$, of the order of $h \sim 100\unit{pc}$;
they all satisfy $h\ll L$, so that the disc will
be considered as infinitely thin for all practical purposes.
The density of interstellar matter is observed to be about
$n_{\rm ISM} \sim 1 \unit{part} \unit{cm}^{-3}$ for all radii, so
that we take $n(r,z) =
2h\delta(z) n_{\rm ISM}$.
Sources and interactions with matter are confined to the thin disc and
diffusion which occurs throughout disc and halo with the same strength,
is independent of space coordinates. The Solar System is located in
the galactic disc ($z=0$) and at a Galactocentric distance
$R_\odot=8 \unit{kpc}$~\cite{centrogal}.
A schematic view of the galactic model is shown in Fig.~\ref{schema_galaxie}.

			%###############%,
\subsubsection{Spallations: the importance of cross sections}
\label{subsub:spall_overview}

When a cosmic ray crosses the disc, it may interact with an
interstellar hydrogen or helium nucleus and initiate a nuclear reaction
({\em  spallation}\/).
The importance of this effect is governed by the corresponding
cross sections.
Actually, it is important to know not only the {\em reaction} or {\ total}
cross section, which
determines the rate of destruction of a given CR species, but also
the branching ratio into a particular channel, which gives the
formation rate of new nuclei. The cross section for a given channel
is often referred to as the {\em spallation}\/ or {\em fragmentation}\/
cross section.
The determination of all these cross sections is a nuclear physics
problem that shall be addressed in details in Sec.~\ref{subsubsec:losses2a}.

\subsubsection{Energy losses from interaction with the ISM}
\label{subsubsec:losses}

There are two types of energy losses which are relevant for nuclei:
ionization losses in the ISM neutral matter ($90$\% H and $10$\% He),
and Coulomb energy losses in a completely ionized plasma,
dominated by scattering off the thermal electrons.
The other effects like bremsstrahlung, synchrotron radiation and
inverse Compton are negligible in the conditions 
considered here.

\shortversion{ %% version longue
The Coulomb energy loss rate is given in~\cite{losses,Strong_Moskalenko}
\begin{displaymath}
          \left(\frac{dE}{dt}\right)_{\rm Coul}\approx -4\pi r_e^2 c m_e c^2
Z^2 n_e \ln
          \Lambda\frac{\beta^2}{x_m^3+\beta^3}\;,
\end{displaymath}
where
\begin{displaymath}
          x_m\equiv (3\sqrt\pi/4)^{1/3}\sqrt{2kT_e/m_ec^2}\;\;;
\end{displaymath}
\begin{displaymath}
          \ln \Lambda\approx \frac{1}{2}\ln\left(\frac{m_e^2c^4}{\pi r_e \hbar^2
          c^2n_e}.\frac{M\gamma^2\beta^4}{M+2\gamma m_e}\right)\;\;.
\end{displaymath}
In these expressions $r_e$ and $m_e$ denote the classical radius and
rest mass of the electron (Particle Data
Group~\footnote{http://pdg.lbl.gov/}),
$\langle n_e \rangle \sim 0.033$~cm$^{-3}$ and $T_e\sim10^4$~K
denote the density and temperature of the interstellar
electrons~\cite{Nordgren},
$Z$ and $M$ are the charge and mass numbers of the incoming nucleus
and $\ln\Lambda \sim 40-50$ is the Coulomb logarithm.

The relativistic expression giving the ionization losses is
\begin{displaymath}
          \left(\frac{dE}{dt}\right)_{\rm Ion} \!\!\! (\beta\geq\beta_0) \approx
          - \frac{2\pi r_e^2 m_e c^3 Z^2}{\beta}\sum_{\rm s=H,He}n_sB_s
\end{displaymath}
where
\begin{displaymath}
          B_s\equiv  \ln \left( \frac{2m_ec^2\beta^2\gamma^2Q_{\rm max}}
          {I_s^2}\right)-2\beta^2\;\;;
\end{displaymath}
\begin{displaymath}
          Q_{\rm
max}\equiv\frac{2m_ec^2\beta^2\gamma^2}{1+[2\,\gamma \,m_e/M]}\;\;;
\end{displaymath}
and $\beta_0 c\sim 0.01\,c$ is the typical velocity of bound electrons in
the hydrogen atom,
$I_s$ is the geometrical mean of all the ionization
and excitation potentials of the considered atom,
($I_{\rm H}=19\unit{eV}$ and $I_{\rm He}=44\unit{eV}$),
$M\gg m_e$ is the incident nucleon mass,
and  $n_s$ is the density of the target atom in the ISM.
}
{ %% version courte
The expressions for the corresponding energy loss rates can be found
in~\cite{losses}.
}

			%###############%
\subsubsection{Adiabatic losses from convective wind}

Among the phenomena affecting the propagation of cosmic rays, the
magnitude of those presented in this section and the
following are more subject to debate because
of the uncertainty associated to parameters $V_c$ and $V_a$.

It is very likely that the medium responsible for diffusion
is moving away from the disc, with a velocity $V_c$.
This is referred to as {\em convective}\/ or {\em galactic wind}\/,
in analogy with the {\em Solar wind}\/.
One of the effects of this galactic wind is to dilute the energy of the
particle located in the disc in a larger volume~\cite{Parker02}. This adiabatic
expansion results in a third type of energy loss, depending on
$\nabla.V_c$.
Throughout our works,
a very simple and tractable form for $V_c$ is adopted,
following~\cite{Webber_Lee_Gupta}.
It is considered to
be perpendicular to the disc plane and to have a constant magnitude throughout
the diffusive volume, so that $dV_c/dz=0$ except at $z=0$ where a
discontinuity occurs, due to the opposite sign of the wind velocity
above and below the galactic plane.
In this case, the dilution effect on the energy is given by a term that can be
expressed in the same form as ionization and coulomb losses
\begin{equation}
\left( \frac{dE}{dt}\right)_{\rm Adiab}=-E_k\left(\frac{2m+E_k}{m+E_k}\right)
\frac{V_c}{3h}\;.
\end{equation}
$E_k$ stands for the total kinetic energy and it should not be confused
with the kinetic energy per nucleus frequently used in this paper.
We emphasize that this term corresponds to a process occurring only in the disc
but not in the halo.

\subsubsection{Reacceleration}
\label{subsubsec:reac}
Along with a the spatial diffusion coefficient $K$,
Eq.~(\ref{linearize}) also contains the momentum diffusion 
coefficient $K_{pp}$;
both are related to the diffusive nature of the process.
The latter coefficient $K_{pp}$ is related to the velocity of
disturbances in the hydrodynamical plasma, called {\em Alfv\'en
velocity}\/.
 From the quasi-linear theory, it is given by (see e.g.~\cite{Seo_Ptuskin})
\begin{equation}
K_{pp}=
\frac{h_{\rm
reac}}{h}\times\frac{4}{3\delta(4-\delta^2)(4-\delta)}{V_a}^2p^2/K(E).
\label{kpp}
\end{equation}
In this expression,
$h_{\rm reac}$ stands for the half-height of the cylinder in which
reacceleration occurs.
In our model, $h_{\rm reac}\equiv h$, but as $K_{pp}$ depends only on the
combination $V_a^2 h_{\rm reac}/h$, the same diffusion in momentum
space is obtained for $h_{\rm reac}\neq h$, provided that the true
Alfv\'en velocity is given as a function of the parameter $V_a$ by
$V_a \times h/h_{\rm reac}$.

\subsubsection{Full propagation equation}
\label{subsubsec:full}

\begin{figure}[hbt!]
\begin{center}
     \includegraphics[width=\columnwidth]{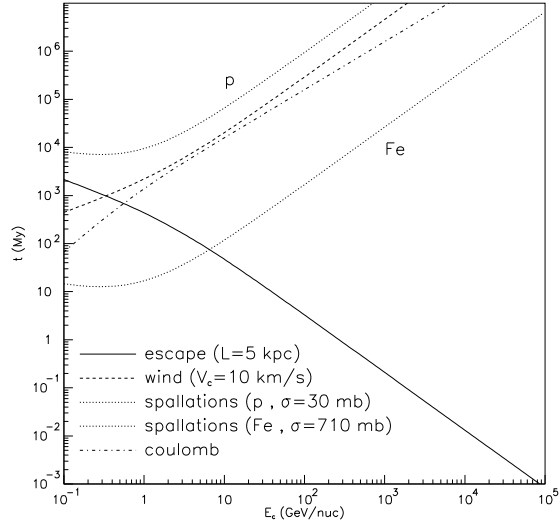}
     \caption{Characteristic times of several processes affecting the
     propagation of cosmic rays are displayed in the 100 MeV/nuc-100 
GeV/nuc energy range.
     Typical values $K_0=0.03 \unit{kpc}^2 \unit{Myr}^{-1}$, $\delta =
     0.6$ and $V_c = 10 \unit{km} \unit{s}^{-1}$ were considered.
     The dominant process at energies higher than a few GeV is the
     escape through the boundaries of the diffusive volume.
     The effect of spallations is seen to be small for the
     propagation of protons, whereas it is crucial for heavy nuclei
     such as Fe.}
     \label{fig:temps_car}
\end{center}
\end{figure}

\begin{figure}[hbt!]
\begin{center}
\includegraphics[width=\columnwidth]{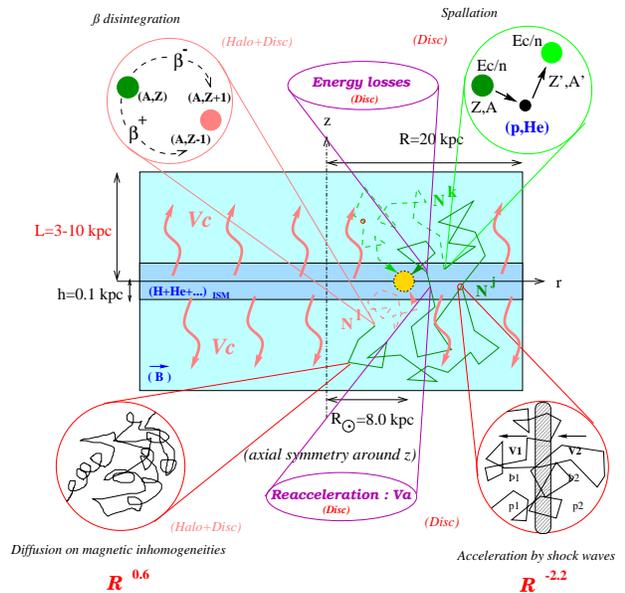}
\caption{Schematic view of our Galaxy as well as all propagation steps
included in our model.}
\label{schema_galaxie}
\end{center}
\end{figure}

The transport equation~(\ref{linearize}) can be rewritten for each
species $j$ using the cosmic ray differential density $N^j(E) \equiv dn^j/dE$.
As the momentum distribution function is normalized
to the total cosmic ray number density ($n=4\pi \int dp \; p^2f$),
we have $N^j(E)=(4\pi/\beta) p^2 f^j$ to finally obtain, assuming
steady-state (see Sec.~\ref{subsubsec:sec_sources}),
\begin{eqnarray}
          \lefteqn{
          -\vec{\nabla} \left[ K\vec{\nabla} N^j(E) -\vec{V_c} N^j(E)\right]
          - \Gamma^j N^j }
          \nonumber\\
          &&
          -\frac{(\vec{\nabla}.\vec{V_c})}{3} \frac{\partial}{\partial E}
          \left[\frac{p^2}{E}N^j(E) \right] = {\cal Q}^j(E) + \nonumber \\
          && \frac{\partial}{\partial E}\left[
          -b_{\rm tot}(E) N^j(E)+ \beta^2 K_{pp}\frac{\partial
N^j(E)}{\partial E}
          \right]\;;
          \label{transport_CR}
\end{eqnarray}
where the following notation has been used for the total energy loss term
$b_{\rm tot}=b_{\rm loss}+b_{\rm reac}$, with
\begin{displaymath}
          b_{\rm loss}(E) = \left( \frac{dE}{dt}\right)_{\rm Ion}
          \!\!\! +\left( \frac{dE}{dt}\right)_{\rm Coul}
          \!\!\! +\left( \frac{dE}{dt}\right)_{\rm Adiab}
\end{displaymath}
and the reacceleration drift term defined as
\begin{displaymath}
          b_{\rm reac}(E) = \frac{(1+\beta^2)}{E}K_{pp}\;.
\end{displaymath}
We also use a compact notation to describe the most general form for a
source term
\begin{equation}
\label{SOURCES}
{\cal Q}^j(E)=q_0Q^j(E)+\sum_k^{m_k>m_j}
                  \Gamma^{kj}N^{k}(0)\;,
\end{equation}
which includes primary sources -- normalized abundance $q_0$,
spectrum $Q^j(E)$ --, but also secondary sources,
coming from spallations (see Sec.~\ref{subsubsec:sec_sources})
or radioactive decay of a heavier species (see Sec.~\ref{subsubsec:beta_EC}).
The relative magnitude of all the effects affecting propagation
can be estimated from the typical timescale associated with these
effects, as displayed in Fig.~\ref{fig:temps_car}.

Taking advantage of cylindrical symmetry and adding
radioactive contributions localized in the disc and the halo, the
previous equation may be rewritten as (making implicit the energy dependence)
\begin{eqnarray}
          \lefteqn{0=\left( {\cal L}_{\rm diff} -\Gamma^j_{\rm rad}\right)
          N^{j}(r,z)+
          \sum_k^{m_k>m_j}\Gamma_{\rm rad}^{kj}(E)N^{k}}\nonumber\\
          &&+2h\delta(z) \left( q^j_0Q(E)q(r) \nonumber -\Gamma^j(E) 
          N^{j}(r,0) \right)\\
          &&+ 2h\delta(z) \sum_k^{m_k>m_j}\Gamma^{kj}(E)N^{k}(r,0) 
          \label{eq_a_resoudre1}
\end{eqnarray}
with
\begin{displaymath}
          \label{eq: operateur diffusion}
          {\cal L}_{\rm diff} =-V_{c} \frac{\partial}{\partial z}
          +K(E)\left(\frac{\partial^{2}}{\partial z^{2}}+
          \frac{1}{r}\frac{\partial}{\partial r}
          \left(r\frac{\partial}{\partial r} \right) \right)\;.
\end{displaymath}

One needs to solve a complete  triangular-like set of coupled equations since
a given nucleus can only be obtained by spallation of a heavier one and
never by a lighter.
Quantities in this equation are functions of spatial coordinates
(not time, steady-state being assumed) and of
kinetic energy per nucleon (energy for short)
since this is the appropriate parameter to be used, as it is conserved in
spallation reactions.

			%###############%

\subsubsection{Solar modulation}
\label{subsubsec:modulation}

The cosmic rays that we detect on Earth had to penetrate the Solar
cavity, a process by which they lose energy. This phenomenon is called
Solar modulation (see~\cite{Duldig} for a review), and may be
pictured as follows.
The Sun emits low energy particles in the form of a fully
ionized plasma having $v\sim 400$~km~s$^{-1}$.
This so-called {\em Solar wind}\/ shields the Solar cavity from penetration
of low energy GCR.
It was first studied by Parker~\cite{Parker02} who established the evolution of
the flux in the Solar cavity.
Analog to the propagation in the Galaxy, it is
a diffusion equation in a quite different geometry (spherical).
For practical purposes, this equation can be solved numerically,
or one can use the force-field approximation.

\paragraph{Force-field} Perko~\cite{Perko} provided a useful and compact
approximation to the full modulation equation.
The final result for a nucleus with charge $Z$ and atomic number $A$
is a mere shift in the total energy
\begin{equation}
          E^{\rm TOA} / A \; = \;
          E^{\rm IS} / A \, - \, \left| Z \right|  \phi / A\;.
\end{equation}
Here $E^{\rm TOA}$ and $E^{\rm IS}$ correspond to the top-of-atmosphere
(modulated) and interstellar total energy, respectively.
The Solar modulation parameter $\phi$ has the  dimension of a rigidity
(or an electric potential), and its value varies according to the
11-years Solar cycle, being greater for a period of maximal Solar
activity.
Another equivalent quantity often used is
$\Phi=\left|Z\right|\phi/A\simeq \frac{1}{2}\phi$.
Once the momenta at the Earth $p^{\rm TOA}$ and at the boundaries
of the heliosphere $p^{\rm IS}$ are determined, the interstellar flux of
the considered nucleus is related to the TOA flux according to
the simple rule
\begin{equation}
          \frac{\Phi^{\rm TOA} \left( E^{\rm TOA} \right)}
          {\Phi^{\rm IS}  \left( E^{\rm IS} \right)}
          \; = \;
          \left\{ \frac{p^{\rm TOA}}{p^{\rm IS}} \right\}^{2}\;.
\end{equation}

The determination of the modulation parameter $\Phi$ is totally
phenomenological and suffers from some uncertainties.
As explained in the following, we will deal with data taken around periods
of minimal Solar activity, for which we fixed $\Phi=250\unit{MV}$
(or equivalently $\phi = 500 \unit{MV}$).

The effect of Solar modulation may be decoupled from the problem of
interstellar
propagation.
Would a more careful treatment of Solar modulation be needed,
e.g.~\cite{Bieber}, an interstellar flux can easily be
obtained from force-field modulated fluxes by demodulation
(the force-field approximation we used is reversible).
This interstellar flux thus obtained could then be
used as an input for any other preferred treatment of Solar
modulation. Of course, Solar modulation induces some uncertainty, but this
question is still debated, and a rigorous treatment of this effect is
beyond the scope of this paper and moreover has not the
same impact on antiprotons/protons and on other nuclei.
For the first ones, the effect is greater but as one is mostly interested in
setting upper limits and drawing exclusion plots on exotic sources,
it is not worth using complex modulation schemes.
The other nuclei are less sensitive to the exact modulation
parameter and to the scheme used.
Thus, our results should not be too sensitive to them.

		%---------------------------------%
		%---------------------------------%

\subsection{Modelling the effects affecting propagation}
\label{subsec:geometry}

We gave above the basic ingredients that enter in propagation models.
We now come back to several aspects for a more detailed discussion, the purpose
being twofold: first, some ingredients support enduring developments
and for these, it is not always an easy task to choose among the variety
of parameterizations and approaches available. Thus, taking a sort of
step by step walk -- leading sometimes to kind of zoological approach --,
we will see how problems were tackled in the literature in the past,
up to the point they reached today. Secondly, as modelling the galactic
environment is a never ending story, because the description is richer
and richer as one dives into fine structure subtleties,
we find it advantageous to stress the level of detail we wish to
include in our model.
Finally, some choice will be made for the sake of tractability of the
propagation
problem.

\subsubsection{Matter content, ISM and LISM}
\label{subsub:ILM_LISM}
The propagation of Galactic Cosmic Rays is influenced by the distribution
of matter in the disc and the magnetic structure of the Galaxy.
These two ingredients require a careful modelling.
As regards the first point, the matter content is quite well known
(see~\cite{Ferriere} for a review) from {\em in situ}\/ observations of
gas thanks to multi-wavelength surveys.
As regards the second point, it is more difficult to determine because of our
particular position in the galactic disc.
Studies of other spirals are particularly useful in that respect.
For instance, the presence of a diffusive halo or a galactic wind
is more easily established in these other galaxies by radio
observations.

\begin{figure}[hbt!]
\begin{center}
\includegraphics[bb=-4 160 618 626,clip,width=0.85\columnwidth]{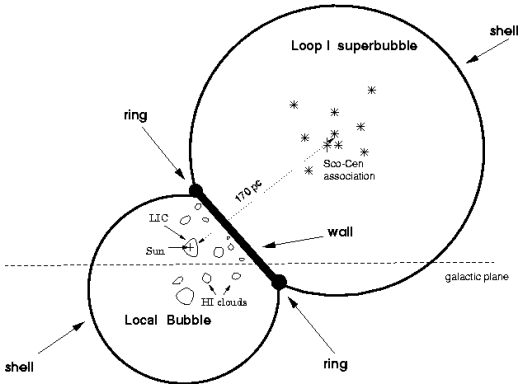}
\caption{Schematic representation (not drawn to scale) between
the local bubble and the neighboring Loop~I superbubble
(from~\cite{Breitschwerdt}).}
\label{fig:LISM}
\end{center}
\end{figure}

\paragraph{Local interstellar medium (LISM)}
As will be discussed in Sec.~\ref{subsec:rad}
some radioactive species of cosmic rays are sensitive to the very
fine structure of the interstellar medium (ISM),
in a region of about $\sim$ 100 pc around the Solar neighborhood,
usually referred to as the local interstellar medium (LISM).
A general review of our local environment can be found in~\cite{Cox}.
The LISM is defined as a region of radius $\lesssim$ 65-250 pc
(the {\em bubble}\/) containing extremely hot ($\sim 10^5-10^6$ K)
and low density ($n\lesssim$ 0.005 cm$^{-3}$) gas,
surrounded by a dense neutral gas boundary
({\em hydrogen wall}\/)~\cite{Sfeir_Linsky}.
A finer description of this bubble tells that the Sun
is located in a local fluff with  $N_{\rm HI}\sim$ 0.1 cm$^{-3}$,
$T\sim 10^4$ K and a typical extension of $\sim 50$ pc.
To end with, it is of great importance for further modelling to realize
that the local bubble is highly asymmetric~\cite{Cox,Fruscione_Welsh}, and
that several cloudlets are present in the bubble (for a schematic
representation, see Fig.~\ref{fig:LISM}).
Various models have attempted to explain the formation of this local
bubble~\cite{Smith_Cox,Breitschwerdt}, but this subject is far beyond our
concern.

Several approximation levels may be considered to model this bubble.
The first level is that for most species of cosmic rays  it may be ignored, 
as the presence of the bubble has almost no effect on 
their propagation.
A next level is necessary for some radioactive species which are very sensitive
to the local environment.
In the thin disc approximation, this bubble may then be represented by a
lower density in some disc area surrounding the Sun.
In our model, it is a circular hole of radius $r_{\rm hole}$ (to be determined)
with a null density. It will be called ``the hole" for short. A third
level of approximation is the one proposed in~\cite{Ptuskin_LISM}, who
use a three-layer model.

\paragraph{Interstellar medium (ISM)}
Apart from the radioactive species, cosmic rays
travel for hundreds of kpc before reaching us, so that they are not
sensitive to the  kind of structures mentioned above but only to
the smoothed out properties of the ISM.
      From this coarse point of view, the ISM is a rather homogeneous
mixture of neutral hydrogen, helium, molecular
hydrogen, ionized gas and dust (see e.g.~\cite{Strong_Moskalenko})
located in a narrow disc.
The total density of matter varies with the distance $r$, but
for the sake of analytical tractability, we consider a constant
density of $\sim 1 \unit{particle}  \unit{cm}^{-3}$.
This is not a crucial assumption, as the cosmic rays detected on
Earth have only probed a region of a few kpc of extension, so that
this effective number can be considered as the mean density in this region.

To end with the ISM question, a last point deserves attention. Gas forms
clouds, and stars form with gas, especially where dense clouds are.
In~\cite{Ptuskin90}, authors have studied the consequence of having most
of these sources in dense clouds. Notwithstanding the
question of a larger production
of secondaries in these clouds, they could have different
magnetic fields and then be more or less transparent to cosmic rays.
As a matter of fact, we would be left with two more
adjustable parameters (filling factor, transparency), and in a certain
way, other usual propagation parameters fit to observations
would have the meaning of averaged effective quantities,
analogously  to a uniform distribution of source and gas.
Anyway, with our model, we do not intend precisely to provide better than
effective parameters, useful and applicable for all species that
seem to have the
same propagation history. This notion of effectiveness and domain of validity
of our model will often come back all the paper long.

\subsubsection{Diffusive halo, regular and erratic magnetic field}
\label{subsubsec:halo_et_la}
Many aspects of cosmic rays propagation call for the existence
of an extended magnetic halo (see e.g.~\cite{Ginzburg}).
First, clear evidences were obtained
from non--thermal radio emission in NGC 4631~\cite{Ekers}.
Two components of the galactic magnetic field coexist: a regular
one (average value about a few $\mu$G, parallel to the galactic plane,
responsible for confinement) and a stochastic one which is responsible for
charged nuclei diffusion (as well as diffusive reacceleration), that has
about the same strength. Tab.~\ref{tab:halo}, taken from~\cite{Dogiel},
summarizes several observational evidences.
\begin{table*}[htb!]
\begin{center}
\begin{tabular}{|ll|l|}   \hline
          1- Observations  &  Evidence for a halo  & half-height $L$ \\
\hline \hline
          2- Diffuse radio emission & yes & $>5\unit{kpc}$ \\
          3- Existence of galactic frontiers& yes (for some of them)&
$1-5\unit{kpc}$ \\
          4- Chemical composition & model-dependent & -\\
          &	(yes if combined the the radio data)& \\
          5- Anisotropy of UHECR& yes (?) & $\gtrsim 3\unit{kpc}$ \\
          6- Gradient in CR density & &\\
          ~~~~~~~a- data from {\sc sas}-2& yes & $\sim 3\unit{kpc}$ \\
          ~~~~~~~b- data from {\sc cos-b}& yes & $\sim 15\unit{kpc}$ \\
          7- High latitude $\gamma$ excess& &\\
          ~~~~~~~a- $p+p$& yes& $\gtrsim 1\unit{kpc}$\\
          ~~~~~~~b- Inverse Compton& yes& $\sim10\unit{kpc}$\\  \hline \hline
\end{tabular}
\caption{Estimations of the cosmic ray halo extension  (adapted
from~\cite{Dogiel}).}
\label{tab:halo}
\end{center}
\end{table*}
In the last ten years, several studies tried to constrain the halo size
using a combination of stable plus radioactive nuclei ratio measurements.
They found $L\sim 4$~kpc~\cite{Webber_Lee_Gupta},
$L\lesssim 4$~kpc~\cite{Bloemen93}, $L\sim4-12$~kpc~\cite{Strong_Moskalenko}
and $L\sim 2-4$~kpc~\cite{Webber_Soutoul}.
As all the authors used quite different diffusion parameters,
with or without wind, reacceleration, and as they took quite different
diffusion slopes $\delta$, the trend seems to favor small halos.
However, the radial $\gamma$-ray distribution in our Galaxy
is rather flat, which points towards a large halo (see e.g.~\cite{Bloemen93}).
To face the problem, \cite{Breitschwerdt02} have
recently proposed a more complex diffusion model, where some
quantities (galactic wind) have a spatial dependence.
We feel that such an approach becomes unavoidable with the present development
of cosmic rays studies.
However, we wish to mention two arguments against the claim
of inconsistency advanced by these authors to motivate their model:
first, we showed in~\cite{Maurin01,Maurin02} that considering B/C ratio
in diffusion/convection/reacceleration models, all halo sizes are
possible.
Second, focusing on radioactive species, we showed~\cite{Donato01}
that if we agreed with usual results for a homogeneous
LISM, the existence of a local underdensity leaves the choice of $L$
almost free (too large halo sizes, i.e. $L>12$~kpc are excluded).
Thus, from our point of view, the discussion about the halo size is far
from being closed.

Finally, it must be kept in mind that halos are dynamical objects.
In particular, there seems to be a transition zone
(thick disc) between the more quiet halo and the active gaseous disc.
This is deduced from the clouds, hot ionized gas and
stars kinematics.
The halo could be substructured in
chimneys, galactic fountains and galactic holes~\cite{Bloemen_Ed}.
The importance of this substructure for cosmic ray studies is an open
question.

			%###############%

\subsubsection{Standard sources in the disc}
\label{subsubsec:sources}

One easily imagines that the location of sources plays an important
role for the density of all species, primaries, secondaries and
radioactive nuclei.
Moreover, it is intimately related to the question of validity of the
stationary hypothesis in cosmic ray propagation models.
Nearby sources are for example invoked in~\cite{Erlykin} to explain
the observed behavior at PeV energy, i.e. the knee.
If one goes one step further
into this line of thought, nothing would guarantee that even at low energy
the steady state approximation is justified. It could be that
propagation parameters provided by stationary models are just effective
parameters, that would need to be reinterpreted in terms of time dependent
propagation models. Such a view could seemingly solve some present problems
of cosmic rays (Maurin {\em et al.}\/, in preparation). Anyway, this paper
focuses on stationary models.

\begin{figure}[hbt!]
          \includegraphics*[width=\columnwidth]{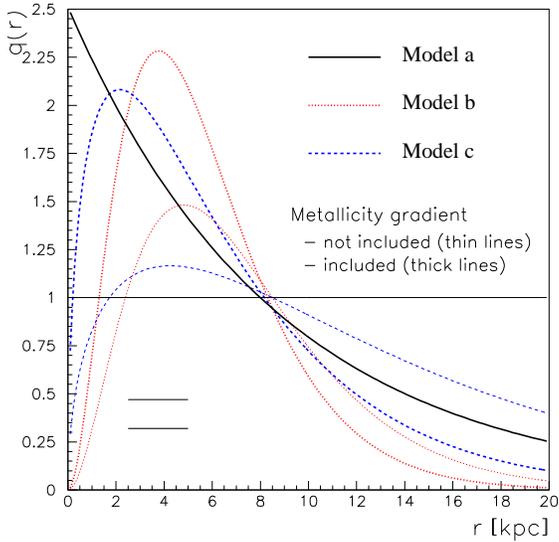}
          \caption{Radial distribution of the sources.
          The thin lines correspond to the p and He component, with the
          three types (a), (b) and (c) discussed above.
          The thick lines correspond to the other species, for which the
          distribution id modified by the metallicity gradient, according
          to Eq.~(\ref{metallicity_gradient2}).}
          \label{fig:dist_rad}
\end{figure}

\paragraph{Radial distribution of sources $q(r)$}
Measurements of galactic $\gamma$ rays in the seventies
have raised the question of the radial distribution
of cosmic rays. This distribution is needed in order to evaluate
the resulting $\gamma$ emissivity at different galactocentric locations.
The first distribution used was that of Kodaira~\cite{Kodaira}
following the radial distribution of supernov\ae\ which is also close
to that of pulsars. This is consistent with the present picture of
cosmic rays where supernov\ae\ provide the energetic budget and mechanism
to accelerate nuclei. The description
of the galactocentric distribution has been steadily improved~\cite{Lyne}
thanks to new observations of pulsars and supernovae.
We take here the distribution of Case \&
Bhattacharya~\cite{Case-Bhattacharya2}
which is an improvement of their earlier
analysis~\cite{Case-Bhattacharya1} and happens to be closer
to the distribution adopted by Strong \& Moskalenko~\cite{Strong_Moskalenko}.
All these distributions could be compared to flat distribution, i.e. $q(r)=1$,
that is also widely used in one-dimensional propagation
models~\cite{Jones_Lukasiak_Ptuskin_Webber}.
These three radial distributions are
\begin{itemize}
          \item Model a: flat distribution
          \begin{displaymath}
              q(r) = 1\;.
          \end{displaymath}
          \item Model b: Case \& Bhattacharya~\cite{Case-Bhattacharya2}
          \begin{displaymath}
	q(r)=\left(\frac{r}{8.5}\right)^{2.0}
	\exp\left(-3.53\times\frac{(r-8.5)}{8.5}\right)\;.
          \end{displaymath}
          \item Model c: Strong \& Moskalenko~\cite{Strong_Moskalenko}
          \begin{displaymath}
	q(r)=\left(\frac{r}{8.5}\right)^{0.5}
	\exp\left(-1.\times\frac{(r-8.5)}{8.5}\right)\;.
          \end{displaymath}
\end{itemize}
They are displayed in Fig.~\ref{fig:dist_rad} along with the metallicity
gradient discussed below.

We emphasized in~\cite{Maurin02} that the difference
between a flat and a more realistic $q(r)$ was a mere rescaling of the
propagation parameters.
However, this difference is not so important if one considers these
parameters as {\em effective} and uses them in a consistent framework.
To be more specific, one should make the same hypothesis on $q(r)$ when
i) determining the propagation parameters from the observation of
some cosmic species and
ii) predicting the abundance of some species from these parameters.
On the other hand, if one is interested in a physical interpretation
of these parameters and wants to deduce some galactic property or 
to predict some quantity which does not depend only on
the diffusion properties (such as
proton-induced $\gamma$-ray production), one should be careful to
consider the right radial distribution.

\paragraph{Metallicity gradient} It is also expected that the
composition of SN should depend on its position in the Galaxy.
Indeed, the inner Galaxy is richer in heavy stellar
material than the outer Galaxy;
this is the case for most spiral galaxies.
Early studies showed a gradient for O/H from observations of
ionized nebulae in galaxies like M33, M51, and M101, but later works
observed this trend in our Galaxy for many other abundances
(see~\cite{Maciel1} for a review and Sec.~2 of~\cite{Gummersbach}).
Several recent observations (see Tab.~1 of~\cite{Chiappini} for
a compilation and~\cite{Andrievsky} for latest results) lead to very
similar conclusions for the metallicity gradient. It can roughly be
parameterized, for all ion species
X but He, as
\begin{equation}
          \frac{d[{\rm X}/{\rm H}]}{dr}=-0.05 \mbox{~dex~kpc}^{-1},
          \label{metallicity_gradient}
\end{equation}
where $[{\rm X}/{\rm H}]$ is defined as 
$log_{10}({\rm X}/{\rm H})-log_{10}({\rm X}/{\rm H})_\odot$.
Associated with the radial distribution, this gives an additional factor
\begin{equation}
          q(r) \rightarrow 10^{-0.05(r-8.)}\times q(r)\;,
          \label{metallicity_gradient2}
\end{equation}
except for H and He.

       \paragraph{Spiral structure}
Because of the presence of spiral arms, there is also an
angular dependence of the source distribution.
As~\cite{Li_Bartunov}
showed, SNe~II and Ib are likely to follow the gas distribution.
As also confirmed in~\cite{Dragicevich}, SNe are
seemingly correlated with the spiral arms position. The structure can be
found in a recent meta-analysis of the Milky Way~\cite{Vallee} where
the structure is likely to resemble a 4 logarithmic spiral arms (see its
Fig.~2).
Notice that, despite their possible usefulness, such models already fall to
describe the arms outside the range $3-13$~kpc.
Anyway, it is shown in~\cite{Taillet01} that this structure actually
has very little effect on the
origin of sources that contribute to the flux detected at Earth.
However, we are left once more with the question of the meaning of
the propagation parameters we will derive.

Nevertheless, if a spiral structure is retained, the other ingredients should
follow as well (magnetic field structure, gas distribution,\dots);
such a modelling
is beyond the scope of this study, though this is probably an important point.
Joint progress in several research fields
(diffusion mechanisms, better knowledge of galactic structure, etc) associated
with numerical developments will certainly allow to cope with these
questions in a couple of years. Anyway, we will not discuss further this
point in this paper.

\paragraph{Source spectrum}
\label{page:source_spectrum}
The energy spectrum of the particles emitted by the sources is
determined by the acceleration process at work.
There is a strong belief, based on theoretical work, that this energy
spectrum $dQ/dp$ is a power-law in rigidity~\cite{accel_power_law}.
For a species $j$, the differential spectrum in energy $dQ^j/dE\equiv
Q^j({\cal R})$
is then given by
\begin{equation}
          Q^j({\cal R}) = \frac{q^j_0}{\beta} \left(\frac{{\cal R}}{1
          \unit{GV}}\right)^{-\alpha}
          \label{pure_power_law}
\end{equation}
in our Eq.~(\ref{transport_CR}) (see details in~\cite{Maurin02}), where
the value of $\alpha$ is still debated (grossly from about 1.5 to
2.5~\cite{pente_source}).
We will refer to this case as the {\em pure power law spectrum}\/.
Among the various acceleration models, some of them
are able to produce various spectra for different
nuclei~\cite{biermanie_Ellison}, but also a break in slopes
around PeV energies (see references in~\cite{Maurin03}).

Because of collective effects, one can also
consider a possible deviation from this power-law at low
energy, and we choose
\begin{equation}
          Q^j({\cal R}) = q^j_0 \left(\frac{{\cal R}}{1
\unit{GV}}\right)^{-\alpha}\;.
\end{equation}
We refer to this as the {\em modified}\/ spectrum.

			%###############%
			%###############%

\subsubsection{Exotic sources in the halo}
\label{subsubsec:exotic_sources}

Cosmic rays may also be created by the pair annihilation of exotic
particles -- such as the neutralino predicted by supersymmetry
(SUSY) --,  as discussed in
Sec.~\ref{subsec:exotic_susy}, or by the evaporation of primordial black
holes  -- hereafter PBH -- as discussed in Sec.~\ref{subsec:pbh}.
Both populations are supposed either to fill (SUSY) or to follow (PBH)
the dark matter profile, which can be assumed to be a spherical
isothermal profile, i.e. for each species $\rho_i(r,z)=\rho_i^{\odot}\times f(r,z)$ with
\begin{equation}
          f(r,z) =
          {\displaystyle \frac{R_{c}^{2} + R_{\odot}^{2}}{R_{c}^{2} +
          r^{2} + z^{2}}} \;.
          \label{distribution_isotherme}
\end{equation}
For definiteness, the dark matter halo has been assumed to have a core
radius $R_c = 3.5$ kpc. The uncertainties on $R_c$ and the consequences
of a possible flatness have been shown to be irrelevant at least
in the PBH case \cite{barrau02}.

The source density in the Solar neighborhood is about
$\rho_{\chi}^{\odot} = 0.4$ GeV cm$^{-3}$. For PBH the normalization
$\rho_{\rm PBH}^{\odot}$ is very different (far less) and can be constrained
through CR signatures (see Sec.~\ref{subsec:pbh}). This is at variance with
the SUSY case where the SUSY space parameter is constrained
instead of source density (see Sec.~\ref{subsec:exotic_susy}).

\paragraph{Effective source term}
Finally, the production rate, i.e. the effective source term is
different in the two cases: it is proportional to $\rho_{PBH}^{\odot}\times
f(r,z)$
for cosmic rays evaporating from  PBH, but to
$(\rho_{\chi}^{\odot}\times f(r,z))^2$ for those coming from neutralino
{\em annihilations}\/.
\begin{figure}[hbt!]
          \includegraphics[width=\columnwidth]{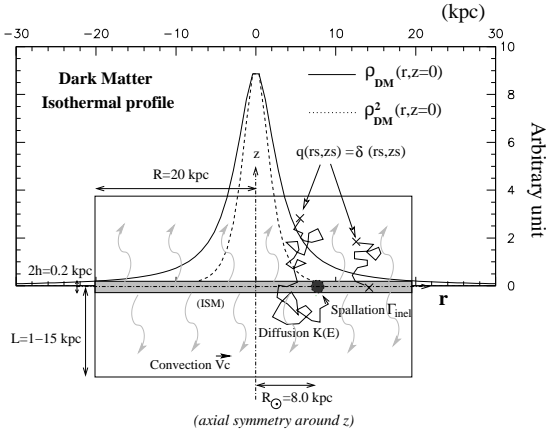}
          \caption{Schematic view of our Galaxy: exotic primary CR
          production follows either the dark matter profile (PBH, solid line) 
	  or the square of the dark matter profile (SUSY, dashed line). 
	  They have been arbitrary rescaled.
          }
          \label{fig:rep_sources}
\end{figure}
The resulting source profiles
are shown in Fig.~\ref{fig:rep_sources}, compared to the typical radial
extension of the Galaxy. One easily
imagines that sources closer to the galactic center will contribute more
to the detected flux, even more in the SUSY case than in the PBH case
(see also discussion in Sec.~\ref{subsec:origine_primaires_halo}).

			%###############%
			%###############%

\subsubsection{Catastrophic losses: destruction on interstellar matter}
\label{subsubsec:losses2a}

The overview in Sec.~\ref{subsub:spall_overview} emphasized the role of nuclear
reactions and decay in the propagation of cosmic rays.
The importance of these events is not the same for all species, and
one should compare the typical propagation
time $\sim 20$~Myr to the effective time of the process involved.
The most important loss is from spallations, i.e. inelastic interaction
of the nucleus with the matter content of the thin gaseous disc
(as a crude estimation, H~:~He~:~CNO~:~Fe~$\approx10:1:0.1:0.01$, about
the same proportions as for cosmic rays). This statement follows from
the mere presence of secondary species:
LiBeB and sub-Fe nuclei (Sc, Ti, V) are absent in sources but they are
observed with the proportion B/C~$\approx$~sub-Fe/Fe~$\approx 0.1$ in
cosmic ray
radiations; they were formed by spallations.
Of course, the heavier nuclei have larger cross sections and break
more easily during their wandering through the Galaxy. This preferential
destruction in flight is, as a matter of fact, responsible for part of
the nuclear enrichment in heavies observed till PeV energies (\cite{Maurin03},
see also Sec.~\ref{subsec:iap01}).

When a cosmic ray nucleus A of energy $E_A$ interacts with a nucleus
H or He (for simplicity, we write the formulae for H only)
at rest in the interstellar medium, a nuclear reactions may occur,
producing some daughter species B, C, \ldots of different
energies $E_B$, $E_C$, \ldots Actually, it turns out that the kinetic
energy per nucleus $E$ is approximately conserved during the reactions, so that
all the information we need is contained in the set of differential
cross sections
$d\sigma_{A+H \rightarrow B+C+\ldots}/dE$.
      From these, the total inelastic cross section for the species $A$ may
be obtained by summing over the final states $B$, $C$, \ldots
The production cross section for the species $B$ can also be found by
summing over the initial states $A$.

On a purely theoretical point of view, the computation of  the
$d\sigma_{A+H \rightarrow B+C+\ldots}/dE$
is a well-defined but hard to solve nuclear physics problem
(see~\cite{Bondorf} for a review).
A nucleus-nucleus collision is generally an out of
equilibrium process that need to be dynamically treated.
At a few hundreds of MeV/nuc, the system expands and the equation of state
of nuclear matter enters in the spinodal region (region of great
mechanical instability):
small perturbations are amplified until they reach the size of the system.
At GeV energies, one is left with several nuclei (multi-fragmentation); a
statistical description can be used (the number of fragments does not depend
on energy but only on the nucleus involved). More specific processes, such as
single and double nucleon removal are indeed the dominant processes for many
nuclei at these energies. Purely hadronic processes arise at a few hundreds of
GeV; more and more resonances are excited until quark fragmentation becomes
dominant. Among microscopical
approaches~\cite{Bondorf,Micro_approach}, one could
quote probabilist models (statistical multi-fragmentation), dynamical models
(quantum molecular dynamics) and kinetic models
(Boltzmann/Vlasov-Uehling-Uhlenbeck
equations). However, none of these models enables a quick and precise
evaluation of
the desired cross section. At most, they can reproduce data with only
a reasonable
agreement. This approach is certainly promising as empirical formulae have now
probably reached their highest point of refinement and development.

      From the experimental point of view, the accurate determination of
the cross sections $d\sigma_{A+H \rightarrow B+C+\ldots}/dE$
is difficult, as it requires the
identification of all the nuclei produced in the reaction.
The total reaction cross section for a species $A$ is much easier to
measure, as one has only to know that the reaction actually occurred.
This is the reason why the question of total reaction cross section is
usually separated from the problem of production cross section of
secondaries.
They will accordingly be considered separately in the next two parts.

\paragraph{Semi-empirical formulae~\cite{semi_empirical}}
The first ingredient is the {\em reaction}\/ (or {\em inelastic}\/)
cross section
$\sigma_{\rm inel}(E)$, giving the probability that a
cosmic ray nucleus of energy $E$ undergoes a nuclear reaction with an
interstellar nucleus.
It should ideally be given by a fundamental approach, but as discussed
above, this program has not been fulfilled yet.
Simpler considerations lead to semi-empirical formulae which
turn to be quite powerful in reproducing all existing data on nuclear
reactions.
In the simplest approach, the cross section is proportional to the
geometrical area of the nucleus ($\pi R_{ef\!f}^2$) that scales as
$\sim A^{2/3}$.
In 1950, Bradt \& Peters have proposed a first correction to take into account
the overlap of the two nuclei wavefunctions,
\begin{displaymath}
          \sigma_{\rm inel}=\pi r_0(A^{1/3}_{\rm target}+ A^{1/3}_{\rm
proj}-b_0)^2\;,
\end{displaymath}
where $r_0$ is the nucleon radius and $b_0$ the overlap
  parameter
(or transparency).

\shortversion{ %% version longue
In the eighties, Letaw and coworkers provided a refinement which is still
widely used, in particular for $p+A$ reactions.
At high energy, the cross section tends to the asymptotic energy
independent form
\begin{eqnarray*}
          \sigma^{\rm HE}_{\rm
inel}\mbox{(mb)}=45A^{0.7}[1+0.016\sin(5.3-2.63\ln A)]\;.
\end{eqnarray*}
At lower energy, the behavior becomes energy dependent: one finds a slight
hollow around 200~MeV/nuc, followed by
a sharp enhancement and a sharp decrease till $\sim20$MeV/nuc;
\begin{eqnarray*}
          \lefteqn{\sigma^{\rm HE}_{\rm inel }(E_k)=\sigma^{\rm HE}_{\rm
inel} \mbox{(mb)}
          }\\
          &&
          \times [1-0.62\exp(-E_k/200)\sin(10.9E_k^{-0.28})]\;.
\end{eqnarray*}
In the above formula, $E_k$ denotes the kinetic energy per nucleon
in MeV/nuc. More generally, there are some necessary corrections
for light nuclei (due to their particular nuclear structure) and
the reactions He+A are evaluated through an energy dependent rescaling
factor $[\sigma_{\rm He}/\sigma_{\rm p}](E_k)$. Later on, in 1993,
Sihver and coworkers,
gave a formulation closer to the geometrical picture given above, to take
also into account A+A reactions, thanks to new experimental data. Finally,
in 1996, Wellish and Axen introduced refined functions in Sihver {\em et al.}\/
formul\ae\ in order to be able to describe the whole data set with a
2\% accuracy
from 6.8 MeV/nuc to 9 GeV/nuc (more than 1400 data points, but they only retain
points whose accuracy is better than 4\%).
}
{ %% version courte
}

\shortversion{ %% version longue
\paragraph{Universal parameterization~\cite{universal_param}}
As was shown in~\cite{Silberberg98}, the parameterization that yields
the better fit to most measurements of nucleus-nucleus inelastic cross
sections is currently the universal parameterization proposed by
\cite{universal_param}. It is given by
\begin{eqnarray}
\label{tripathi_general}
          \sigma_{\rm inel}=\pi
r_0^2\left(A_{\rm proj}^{1/3}+A_{\rm target}^{1/3}+\delta_E\right)^2\\
          \times\left(1-R_c\frac{B}{E_{cm}}\right)\nonumber\;.
\end{eqnarray}
Here, $r_0=1\unit{fm}$ and $E_{cm}$ denotes the kinetic energy per nucleus in
the rest mass frame. $R_c$ is a parameter introduced for
light systems in order to keep the same formalism for all reactions. The term
\begin{displaymath}
          \delta_E=1.85S+\frac{0.16S}{E_{cm}^{1/3}}-C_E+
          \frac{0.91(A_t-2Z_t)Z_p}{A_tA_p}
\end{displaymath}
describes several effects:
\begin{itemize}
          \item $S=A_p^{1/3}A_t^{1/3}/(A_p^{1/3}+A_p^{1/3})$ is the asymmetrical
          mass term that is related to the overlapping volume of the
colliding system;
          \item $C_E=D\left[1-\exp\left(
          -E/T_1\right)\right] 
-0.292\exp\left(-E/792\right)\cos(0.229E^{0.453})$
          models the energy dependence at intermediate and large energies,
mainly through
          transparency and Pauli blocking ($E$ is the colliding kinetic energy
          per nucleus in MeV/nuc). The number $D$ is related to the system
density and as
          $T_1$, it is adjusted once for all for almost all reactions
with a more
          careful treatment and specific adjustment for light systems
($A\leq4$).
          \item Finally, the last term in $\delta_E$ takes into account isotopic
          dependence of cross sections.
\end{itemize}
The remaining terms in Eq.~(\ref{tripathi_general}), respectively
\begin{displaymath}
          B=\frac{1.44Z_pZ_t}{R}
\end{displaymath}
and
\begin{displaymath}
          R=r_p+r_t+
          \frac{1.2\left(A_p^{1/3}+A_t^{1/3}\right)}{E_{cm}^{1/3}}\;,
\end{displaymath}
correspond to the coulomb barrier (that depends on the energy) and
the radius to evaluate the barrier height. In the latter term, $r_i$ is the
equivalent radius of the hard sphere and is related to $r_{rms,i}$ thanks to
$r_i=1.29\;r_{rms,i}$ (this last quantity is obtained directly through
experiment~\cite{Devries}).
}
{%%version courte
\paragraph{Universal parameterization~\cite{universal_param}}
As was shown in~\cite{Silberberg98}, the parameterization that yields
the better fit to most measurements of nucleus-nucleus inelastic cross
sections is currently the universal parameterization proposed by Tripathi
{\em et al\/.} The reader is referred to \cite{universal_param} for the
details.}

\begin{figure*}[hbt!]
          \begin{center}
	\includegraphics[width=\textwidth]{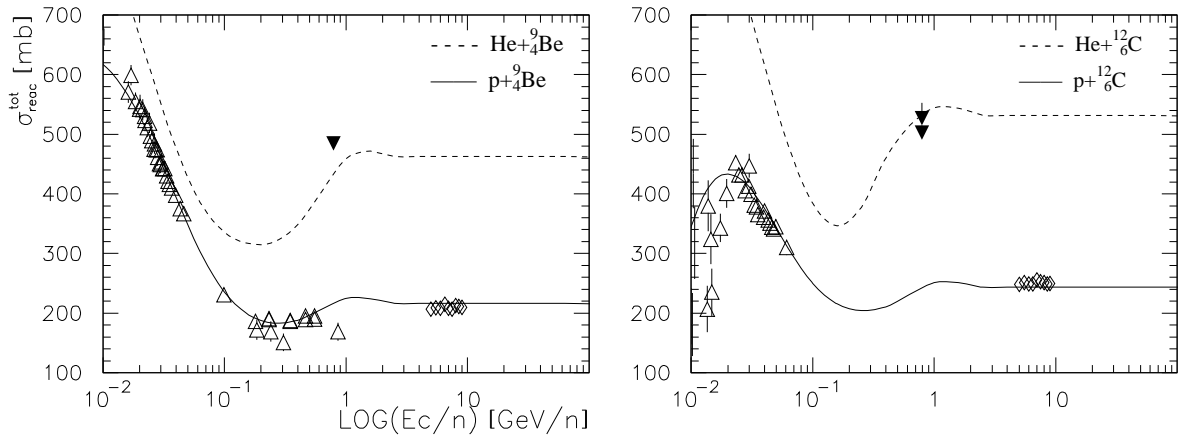}
	\caption{Comparison of total inelastic cross sections
	from universal parameterization~\cite{universal_param}
	to data~\cite{data_inel}: empty triangles
	are from Bauhoff (1986) and Carlson (1996), full ones
	from Tanihata {\em et al.}\/ (1985), and diamond from
	Bobchenko (1979).}
	\label{fig:reaction}
          \end{center}
\end{figure*}

\paragraph{Experimental data and accuracy~\cite{data_inel}}
As an illustration, Fig.~\ref{fig:reaction} shows these inelastic
cross sections
for the reactions (p,He)+$^9$Be, (p,He)+$^{12}$C along with the
experimental data points.
We see that this parameterization reproduces well the non-trivial
features of the cross section.
However, it must be stressed that no new measurement has been made since
1985 for the reaction of protons on light nuclei,
and that many isotopes have even never been measured to date.
Moreover, the helium induced reactions are even more
difficult to probe experimentally.
Some of the results concerning the abundance of some species of
cosmic rays rely on the faith that the parameterization is valid for
these nuclei.
Considering that when data exist, the parameterization gives an
accuracy of a few \%, it is reasonable to consider that the above
parameterization is $2-5$\% accurate for proton induced reactions and $10-20$\%
for He induced reactions.
			%###############%
\subsubsection{Secondary production from spallations on the
interstellar matter}
\label{subsubsec:sec_sources}

When a nucleus undergoes a spallation, it may generate a large variety
of lighter nuclei, which are not all present in standard sources.
These nuclei may typically be classified as $^2$H, $^3$He (p and He
induced), LiBeB (CNO induced) and
sub-Fe (Fe induced).
They give important clues to understand the characteristics of
galactic propagation.
As another example of the importance to have a good knowledge of
these production
cross sections, one can also mention cosmogenic nuclei (due to the penetration
of GeV CR protons in extraterrestrial bodies)  which allow an estimation
of the stability of cosmic ray fluxes in the past billion years~\cite{Reedy}.
To study the nuclei population $N_2$ and to draw an inference for
propagation models,
we need to now, as accurately as possible, processes such as
\begin{equation}
          N_1+(p,He)\longrightarrow N_2\;.
          \label{spal_sec}
\end{equation}
In a more rigorous way, one should also take into account
contributions from heavier species of the ISM (e.g. $N_1$+CNO).
However, these cross
sections are only poorly estimated, not to say not known; moreover,
these have been
correctly included for antiproton production but were shown to be negligible
contributors.

\paragraph{Semi-empirical and empirical formul{\ae}~\cite{Codes_Web}}
In the literature, there exists basically two fast procedures to evaluate the
spallation cross sections corresponding to reactions (\ref{spal_sec}).
They were provided by cosmic ray physicists because the energy range
in which these reactions are needed is generally not of great
interest for nuclear
physicists. Silberberg \& Tsao~\cite{Tsao_Silberberg} approach has
well-motivated
theoretical grounds and takes advantage of some observed regularities
       in i) the mass difference
between target and daughter nuclei ($\Delta A\equiv A_{\rm
target}-A_{\rm produced}$);
ii) the ratio between the number of neutrons and protons in the
daughter nucleus.

In the nineties, many new data issued from various targets
were obtained, and this semi-empirical parameterization gave only a
$\pm 35$\% accuracy for these new data.
It led Webber and coworkers~\cite{Webber_spal} to develop a new
approach fully based upon
experimental regularities.
\shortversion{%% version longue
In this parameterization, having remarked that production of different
isotopes has similar $E$ dependence, parameters were fit to one kinetic energy
per nucleon, and then extended to other energies,
\begin{eqnarray*}
          \lefteqn{\sigma^{p+N_i}(Z_i,A_i\;;Z_f,A_f;E)=\sigma_0
          (Z_f,Z_i)} \\
          && \times f_1(Z_f,A_f,Z_i,A_i)f_2(E,Z_f,Z_i)\;\;.
\end{eqnarray*}
We do not enter into the details (the reader is referred to quoted papers), but
only briefly comment the three terms that enters in this equation:
the first one
describes the dependence on the fragment
charge (the production is exponentially suppressed far from the
stability valley)
and contains in particular two parameters ($\sigma_{Z_f}$ and $\Delta_{Z_f}$)
that are chosen to fit the data.
Notice  that these parameters, thanks to new measurements,
were updated in a later version of their code; we implemented in our
propagation
code parameters as given in Tab.~V of~\cite{Webber_spal98}. The second term
describes isotopic distribution of fragments for a given species. This is
independent of energy as long as we are not
too far from the stability valley. The third term is the energy
dependence and is
only related to charges involved. Added to these three terms, the particular
single and double nucleon-removal reactions are also taken into account, and
He-induced spallations are tabulated from~\cite{Ferrando} formul{\ae}.
}
{%% version courte
}

\shortversion{%% version longue
\paragraph{Some remarks about these parameterizations}
Actually, for our purpose, the most suitable set of formulae is given by Webber
and coworkers' empirical formulae. It seems that the more general approach
of Silberberg and coworkers is more useful to propagate $Z>30$ heavy nuclei
(see also~\cite{Heavies}) as the corresponding cross section have not
been measured yet. It was found in~\cite{Silberberg98} that for the existing
data points,
the dispersion obtained with Webber's code is better than the
dispersion using theirs.
However, the two approaches have probably reached their latest development.
As remarked recently in~\cite{Webber_spal98}, any refinement of these
formulae asks for
complicate modifications of the present formulae. If one takes into account
isospin effect, odd-even effect, neutron and proton stripping due to coulomb
barrier, it would require an adjustment nucleus by nucleus. Webber's code
is only valid for $Z>3$: light systems require further modelling
because they behave
very differently from other nucleus (see e.g.~\cite{Light_systems}).
To improve these codes, some authors have
focused on peculiar single and double nucleon removal (coulomb
fragmentation)
that are dominant for most secondary
productions~\cite{nucleon_removal}. A further
point of discussion is the reliability of the straight-ahead
approximation, i.e.
\begin{eqnarray*}
         \int_0^{\infty}\beta 'N^k(T')\sigma^{kj}(T,T')dT'=\beta
         N^k(T)\sigma^{kj}(T)\;.
\end{eqnarray*}
In practice, this corresponds to the fact that each fragment has the
same kinetic energy per nucleon than has the parent nucleus.
If this approximation seems quite justified at high energy (above a
few GeV/nuc),
its relaxation may affect, for example, the B/C ratio around 1
GeV/nuc~\cite{Straight_ahead}.
In our propagation code, we use Webber {\em et al.}\/'s code
that is not only very well suited, but also widely used by almost all cosmic
ray physicists (for the nuclei we deal with).
}
{%% version courte
In our propagation code, we use Webber et al's code
that is not only very well suited, but also widely used by almost all cosmic
ray physicists (for the nuclei we deal with).
}

\paragraph{Experimental data and accuracy~\cite{data_spall}}
At the present time, it is estimated that roughly $\sim 95$\% of the
reactions involving cosmic ray nuclei have been measured at least at
one energy.
As an illustration, we give in Tab.~\ref{tab:sigma_prod} the
cross sections for the reactions involving $^{12}$C.
It illustrates the importance of single nucleon removal ($^{12}$C
$\rightarrow$ $^{11}$C, $^{11}$B) as mentioned above. Double
nucleon removal ($^{12}$C $\rightarrow$ $^{10}$B) is less marked but visible.
Let us notice also a maximum around
$^6$Li that corresponds to half of the initial nucleus. This table
also emphasizes
the multi-fragmentation character of spallations.
%%%%%%%%%%% TABLE DES NOYAUX EQUIVALENTS POUR LES RAD %%%%%%%
\begin{center}
          \begin{table}[hbt!]
	\vbox{\vspace{0.35cm}}
	\hspace{0.75cm}
	\begin{tabular}{|c|c|}  \hline
	    Reaction  &  $\sigma_{\rm spal} \pm \Delta \sigma_{\rm spal}$  \\
	    $^{12}$C+p$\rightarrow\dots$ & (mb) \\  \hline \hline
	    $^{11}$C        &       $29.2\pm2.5$\\
	    $^{10}$C        &       $3.6\pm0.5$\\
	    $^{9}$C         &       $0.24\pm0.0.5$\\ \hline
	    $^{12}$B        &       $0.12\pm0.05$\\
	    $^{11}$B        &       $27.7\pm0.7$\\
	    $^{10}$B        &       $12.3\pm3.0$\\
	    $^{8}$B         &       $0.44\pm0.04$\\ \hline
	    $^{10}$Be       &       $4.2\pm0.6$\\
	    $^{9}$Be        &       $6.7\pm0.9$\\
	    $^{7}$Be        &       $10.1\pm1.3$\\ \hline
	    $^{9}$Li        &       $0.25\pm0.06$\\
	    $^{8}$Li        &       $1.47\pm0.23$\\
	    $^{7}$Li        &       $12.5\pm1.8$\\
	    $^{6}$Li        &       $19.8\pm2.7$\\ \hline
	    $^{6}$He        &       $0.87\pm0.31$\\
	    $^{4}$He        &       $159.\pm21.$\\
	    $^{3}$He        &       $24.8\pm3.2$\\ \hline
	    $^{3}$H         &       $88.\pm31.$\\
	    $^{2}$H         &       $138.\pm41.$\\
	    $^{1}$H         &       $143.\pm42.$\\ \hline
	\end{tabular}
          \caption{Cross sections of the
	fragmentation of $^{12}$C on a proton target at $3.66\unit{Gev/nuc}$
	(adapted from Korejwo {\em et al.\/}, 1999)~\cite{data_spall}.}
          \label{tab:sigma_prod}
          \end{table}
\end{center}
\begin{center}
          \begin{table}[hbt!]
	\begin{tabular}{|c||c|c|c|c|c|} \hline
	    {\scriptsize $_|$} $\;\!\!\!\!\!\rightarrow $    & & &  &  & \\
	    (p,He)&$^{11}$B  & $^{10}$Be  & $^{10}$B& $^9$Be&     $^7$Be\\
	    $ + $ &  &  &  & & \\\hline \hline
	    $^{28}$Si&     -     	& 1.7   & -   & 1.9  &        1.9\\
	    $^{24}$Mg&     2.3     & 2.7   & 2.5   & 3.0  &        3.1\\
	    $^{20}$Ne&       2.1     & 2.6   & 2.2   & 2.8  &        3.0\\
	    $^{16}$O&       20.3    & 20.0   & 23.0  & 21.7 &        22.7  \\
	    $^{15}$N&       3.2       & 5.5    & 1.1   & 4.2  &        2.9 \\
	    $^{14}$N&       5.4     & 5.0   & 5.6   & 5.5  &        5.7 \\
	    $^{13}$C&       4.2       & 1.2   & 1.2   & 2.9  &        1.9  \\
	    $^{12}$C&       56.9    & 22.8  & 44.8  & 26.5 &        27.9 \\
	    $^{11}$B&       -       &  30.6   & 16.1  & 15.6 &        9.5  \\
	    $^{10}$Be&       -       &  -   & -  & 1.6 &        1.2 \\
	    $^{10}$B&       -       &   -   &  -    & 7.3  &       7.0\\
	    $^{9}$Be&       -       &   -   &  -    & -  &       6.9
	    \\ \hline\hline
	    Total      &       94.4\%  &  92.1\% & 96.5\% &  93.0\%&
93.7\% \\\hline
	\end{tabular}
	\caption{Fraction in \%
	of the isotopes contributing to the formation of secondary B and Be,
	(when this fraction is greater than 1 \%), at $1.8\unit{GeV/nuc}$.
	The contribution of each channel is weighted by the propagated
	abundance of the father nucleus.}
	\label{tab:sigma_prod_moyenne}
          \end{table}
\end{center}
%%%%%%%%%%%%%%%%%%%%%%%%%%%%%%%%%%%%%%%%%%%%%%%%%%%%%%%%%%%%%%%%%%%%
Tab.~\ref{tab:sigma_prod_moyenne} illustrates another aspect of spallations in
cosmic ray propagation: the output is from Webber's spallation code
and has been weighted by
the relative abundances of sources to isolate the most important
contributions. This table underlines the fact that, if most secondaries are
created through spallation of the most abundant species, all the others
come from small
contributions of a great numbers of nuclei.
Finally, the present data accuracy ranges from $\sim 2-3$\% to $\sim
20$\% depending
on the produced fragment.
The old activation measurements seem to be unusable,
and despite numerous experiments since the beginning of the eighties,
not all the
nucleus have been at least measured twice.
Recently, Flesh {\em et al.}\/ (1999)~\cite{data_spall} observed
systematic deviations from Webber measurements.
It is thus not an easy task to estimate the accuracy of the above
parameterizations.
For the case of B/C ratio that is particularly important in our studies,
one can estimate  the accuracy to lie in the range of about $5-10$\%
for p induced reactions and $10-20$\% for He induced reactions.
It is worth noting that production cross sections will probably be
the limiting factor
of cosmic ray studies in the near future,
when high statistic experiments ({\sc pamela, ams}) will obtain new very
accurate data on CR fluxes.
This is already the case for antiproton predictions (see~\cite{Donato02}
and Sec.~\ref{sub:sec_anti_p_d}).
For secondary nuclei such as B, sub-Fe, consequences
of these errors are difficult to estimated, because of the complex
reaction chain involved (see~\cite{Webber_cross_section} for a discussion).

\paragraph{Ghost nuclei}
There remains a last point which is only rarely addressed.
As mentioned above, a complete grid of nuclear reactions is
needed to implement the cross sections, including short-lived
species which are not seen directly but that contribute to some other
species through their decay products.
As an example, let us consider the production of $^9$Be.
One has to add all the contributions from all heavier nuclei, including
the specific contribution from intermediate reactions such as
\begin{displaymath}
          N^k+\mbox{(p,He)}\rightarrow
\end{displaymath}
\begin{displaymath}
          \left\{
          \begin{array}{ccc}
	^9\mbox{Li} &\stackrel{\beta^-\, {\scriptstyle
	(t_{1/2}=178\;ms)}}{\longrightarrow}&
	\;^{9}\mbox{Be} \; ({\cal B}r=0.49)\;.\\
	^{11}\mbox{Li}
	&\stackrel{\beta^-\!+\:2n \,{\scriptstyle
	(t_{1/2}=8.6\;ms)}}{\longrightarrow}&
	\;^{9}\mbox{Be} \; ({\cal B}r=0.041)\;.
          \end{array}
          \right.
\end{displaymath}
The nuclei $^9$Li and $^{11}$Li are called {\em ghost nuclei}\/: they
contribute
indirectly to the flux, but do not have to be propagated as their half-lives are
negligible in the face of propagation time scale.
This will be the case for all the nuclei having $t_{1/2}\lesssim20$~Myr.
It turns out that there is a gap in the lifetimes of
nuclei between 10 kyr and a few Myr, so that all nuclei having
$t_{1/2}\lesssim 1$~kyr could be considered as ghosts.
Among them, those which have too short a lifetime are not seen in the
experiments and are already accounted for in the measured cross
sections.
We emphasize that, would we be interested
in supernovae explosions, the nuclei relevant would not be the same because
of a much shorter typical evolution time.

In conclusion, even if it will be transparent in the rest of the
paper -- and also
in almost all propagation papers --, one has to keep in mind that each time we
evaluate production cross section, in the true computation intervenes the
quantity (${\cal B}r(X \rightarrow j)$ is the branching ratio of $X$ into $j$)
\begin{displaymath}
            \sigma^{\rm effective}_{i\rightarrow j}=\sigma_{ij}^{\rm direct}+
            \sum_{
            \hbox{\scriptsize \it ghosts}\;\; X}
            \sigma_{i\rightarrow X}\;{\cal B}r(X \rightarrow j)\;.
\end{displaymath}
Nuclear grids for all species (including very heavy nuclei) can be found
in~\cite{Letaw_grid}. Thanks to recent nuclear compilations~\cite{recent_grid},
we reevaluated
this data grid (and the associated ghost nuclei) for $Z<30$; we found many new
ghost nuclei or modified branching ratio. However, these new nuclei are
generally far from the stability valley, such that their production cross
section is clearly very small. We checked that for most species,
this completion have a negligible effect, so that the~\cite{Letaw_grid}'s table
of ghost nuclei is sufficient and does not need to be further
revised~\footnote{A complete
table of ghost nuclei for $Z<30$ can be found in Maurin 2001 (PhD
dissertation).}.

\paragraph{Production of antiprotons by spallations}
Among the nuclei that are created by spallations, the
antiprotons require a specific treatment, as their creation relies on
the conversion of collision energy into mass and not on the mere breaking
of a heavier nucleus.
In order to evaluate the contribution from the ${\rm p-H_{\rm ISM}}$
interactions -- the dominant process at first approximation -- the
Tan \& Ng parameterization based on experimental data is
used~\cite{TanNg82}.
Collisions that involve heavier nuclei are more difficult to obtain.
The contribution  ${\rm p-He_{\rm ISM}}$
has been considered firstly by means of a simple geometrical
approach~\cite{gaisser}. However, using a more sophisticated nuclear
Monte Carlo treatment, it was noticed in~\cite{Simon_Molnar_Roesler} that this
reaction
does not only enhance the antiproton flux as a whole but also change
its low energy spectrum, mostly for kinematical reasons. By the way,
these authors also considered ${\rm He-p_{\rm ISM}}$, ${\rm He-He_{\rm ISM}}$
as well as CNO contributions, the latter being negligible (see their
Fig.~7). Unfortunately, very few experimental data are available on
antiproton production cross sections in nuclear collisions and a
model-based evaluation -- such as the {\sc dtunuc}
program~\cite{Simon_Molnar_Roesler} -- seems
to be unavoidable. This is what we will use in the following.

\subsubsection{Antideuteron production}
\label{subsub:dbar}
When an antiproton and an antineutron are formed by a reaction, they
may merge into an antideuteron nucleus.
The calculation of the probability for the formation of an antideuteron
proceeds in two steps. We first need to estimate the probability for
the creation of an antiproton-antineutron pair. Then, there is some
probability that those antinucleons merge together to yield an antinucleus
of deuterium.

As regards the first step, the differential probability
for the production
of a single antiproton or antineutron (resp. ${\cal P}_{\bar{p}}$  or
${\cal P}_{\bar{n}}$)
is known for each initial process -- spallation, neutralino
annihilation or black hole evaporation --
and a first guess would be that
the production of two antinucleons is proportional to their product.
This so-called factorization hypothesis is fairly well established at 
high energies.
For spallation reactions, however, the bulk
of the antiproton production takes place for an energy $\sqrt{s} \sim 10$ GeV
which turns out to be of the same order of magnitude as the antideuteron mass.
Pure factorization should break in that case as a result of energy
conservation and needs to be slightly adjusted.
We have therefore assumed that the center
of mass energy available for the production of the second antinucleon is
reduced by twice the energy carried away by the first antinucleon.
This yields the following  factorization of the
probability to form a $\bar{p}$-$\bar{n}$ pair (see~\cite{Orloff} for
a more detailed discussion)
\begin{eqnarray*}
          \lefteqn{{\cal P}_{\bar{p} , \bar{n}} (\sqrt{s} ,
\vec{k_{\bar{p}}} , \vec{k_{\bar{n}}}) = }\\
          && \frac{1}{2} {\cal P}_{\bar{p}} (\sqrt{s} , \vec{k_{\bar{p}}}) \,
          {\cal P}_{\bar{n}} (\sqrt{s} - 2 E_{\bar{p}} , \vec{k_{\bar{n}}}) \\
          && + \left( \vec{k_{\bar{p}}} \leftrightarrow \vec{k_{\bar{n}}}
\right)\;.
\end{eqnarray*}

Once the antiproton and the antineutron are formed, there is a finite
probability that they combine together to give an antideuteron.
The coalescence function ${\cal C}(\vec{k_{\bar{p}}} -
\vec{k_{\bar{n}}})$ describes the probability
for a $\bar{p} - \bar{n}$ pair to yield by fusion an antideuteron, as
a function of the difference of the initial momenta.
An energy of $\sim 3.7$ GeV is required to form by spallation an
antideuteron whereas the binding energy of the latter is
$B \sim 2.2$ MeV. The coalescence function is therefore strongly
peaked around $\vec{k_{\bar{p}}} -\vec{k_{\bar{n}}} = \vec{0}$.

\begin{figure}
\centerline{\includegraphics*[width=\columnwidth]{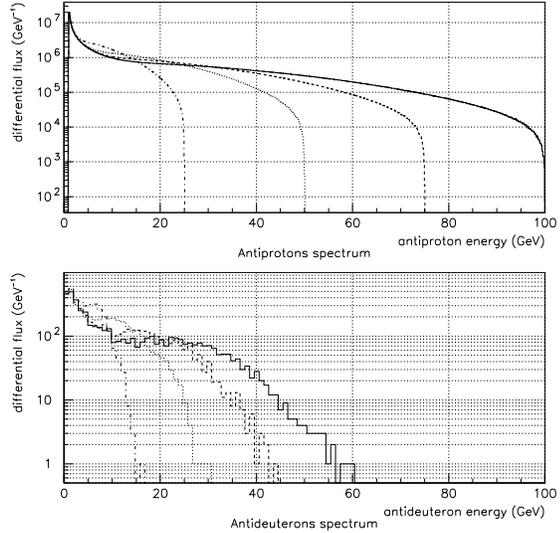}}
\caption{Upper plot: antiproton differential spectrum obtained with
$1.9\times 10^8~\bar{u}$ quark jets generated by {\sc pythia} 
at 100, 75, 50, 25 GeV.
Lower plot: antideuteron spectrum obtained in the same conditions with a
coalescence momentum $p_0=160$ MeV.}
\label{fig:coalescence}
\end{figure}

We considered that the antinucleons merge together if the momentum of the
corresponding two-body reduced system is less than some critical value
$P_{\rm coal}$. That coalescence momentum is the only free parameter of this
factorization and coalescence scheme. As shown in~\cite{Orloff},
the resulting antideuteron production cross section in proton-proton
collisions is well fitted by this simple one-parameter model.
The result, for a typical value~$p_0=160$ MeV~\cite{barrau02}, 
as well as the 
comparison with the antiproton yield is displayed in 
Fig.~\ref{fig:coalescence}.

New calculations for such astrophysical applications of the antideuteron
production are under progress in the diagrammatic approach to the coalescence
model~\cite{Protasov}.

			%###############%

\subsubsection{$\beta$ and EC decay}
\label{subsubsec:beta_EC}

A second catastrophic loss to add to inelastic reactions
is the decay of unstable nuclei.
One can distinguish two decay processes:
$\beta$ decay which is spontaneous and electronic capture decay
(hereafter denoted EC for short)
which can happen only if an electron of the interstellar medium has been
first attached in the K shell (so that the presence probability of the
electron is finite in the region of the nucleus).

\paragraph{$\beta$ decay} The ubiquitous example of this case is
$^{10}$Be$\rightarrow$$^{10}$B whose rest half-life is $t_{1/2}=1.51$~Myr.
The ratio $^{10}$Be/$^{9}$Be observed in cosmic rays is smaller than
the ratio of two normal secondaries coming from the same primary progenitors,
which is merely the ratio of the production cross sections.
This has often been interpreted as an evidence for an extended
diffusive halo where the cosmic rays would spend most of their time.
This interpretation may be misleading and could lead to a wrong
intuition~\cite{Donato01}. One of the main interest of these nuclei is
that the short-lived unstable species propagate only very locally,
and thus are very sensitive to the local environment.
In particular, they may be used to give an independent
evidence for the existence of a local  underdense bubble, as
shown in~\cite{Donato01} (see Sec.~\ref{subsec:rad} for more details).

We compile in Tab.~\ref{tab:AU_FINAL_BETA-}
the nuclei that can be propagated as pure $\beta$ unstable.
Those with half-time shorter than $^{14}$C are ignored
(see Sec.~\ref{subsubsec:sec_sources}). 
%%%%%%%%%%%%%%%%% Table DES RADIOACTIFS BETA- %%%%%%%%%%%%%%%%
\begin{table}[hbt!]
          \begin{center}
	\begin{tabular}{|c|c|c|c|c|}   \hline
	    $Z$  & Nucleus  &  Daughter    & $t^{\rm unit.}_{1/2}$ (error)\\
	    \hline \hline
	    \vbox{\vspace{0.4cm}}
	    $4$  &  $^{10}_{4}$Be  & $^{10}_{5}$B & $1.51^{\rm Myr}\;(0.06)$ \\
	    \vbox{\vspace{0.4cm}}
	    $6$ & $^{14}_{6}$C  & $^{14}_{7}$N & $5.73^{\rm kyr}\;(0.04)$  \\
	    \vbox{\vspace{0.4cm}}
	    $13$  &  $^{26}_{13}$Al & $^{26}_{12}$Mg & $0.91^{\rm
Myr}\;(0.04)$\\
	    \vbox{\vspace{0.4cm}}
	    $17$ & $^{36}_{17}$Cl  & $^{36}_{18}$Ar & $0.307^{\rm
Myr}\;(0.002)$ \\
	    \vbox{\vspace{0.4cm}}
	    $26$ & $^{60}_{26}$Fe & $^{60}_{28}$Ni & $1.5^{\rm Myr}\;(0.3)$ \\
	    \hline \hline
	\end{tabular}
          \end{center}
          \caption{Pure $\beta$ unstable isotopes
($1\unit{kyr}<t_{1/2}<100\unit{Myr}$)
          from a propagation point of view (see~\cite{Donato01} for details).}
          \label{tab:AU_FINAL_BETA-}
\end{table}
%%%%%%%%%%%%%%%%%%%%%%%%%%%%%%%%%%%%%%%%%%%%%%%%%%%%%%%%%%%%%%%%
In particular, we have checked that the EC mode
can be neglected for
$^{27}$Al and $^{36}$Cl but not for $^{54}$Mn and $^{56}$Ni
(EC half-life determination for these two last species is particularly
difficult, see resp.~\cite{half-life_Mn} and~\cite{half-life_Ni}).

\paragraph{Electronic capture~\cite{EC}} A second kind of unstable
species is given by
nuclei decaying under EC process.
A nucleus such as $^{59}$Ni is formed during last stages of stellar
nucleosynthesis; it decays as $^{59}$Ni$\rightarrow$$^{59}$Co in
$t_{1/2}=80$~kyr once it attaches an electron in its K-shell.
Studies of this species allows for example to show that there was a
delay of at least $\sim 10^5$~yr between
nucleosynthesis  and acceleration.
Attachment and the converse process, electron stripping, are crucial
processes that determine the effective lifetime of the EC unstable
nucleus. Notice that because of these processes, the energy dependence
of lifetimes is more complex in the
EC mode than in the $\beta$ mode.

There are two ways to attach an electron, radiative and non-radiative process:
after attachment of a free electron, the momentum and energy
balance are restored by the emission of a photon or the recoil of the
nucleus respectively. Radiative capture is dominant around a few
hundreds of MeV/nuc in hydrogen. Non radiative capture scales as
$Z_t^5$ and dominates for the heavier species of the ISM.
Only a few measurements are available for the latter process, and
they suggest that this
contribution is anyway less than a few \%; it is neglected in this work.
It must be said that these processes suffer from the
same limitations that the production cross sections, i.e. they are not
studied by atomic physicist, but by cosmic ray astrophysicists.
The results widely used nowadays have been derived in the seventies and to our
knowledge no updates have been made at GeV energies.

\shortversion{%% version longue
Fig.~\ref{fig:tau_s_et_tau_a} displays the rate of attachment
and stripping for several charges.
Some general conclusions can be drawn from this single result:
i) comparing $\tau_{\rm attach}$ to the general propagation time
$\sim 20$~Myr, only low energy and
large $Z$ nuclei are likely to attach an electron and subsequently decay;
ii) comparing $\tau_{\rm attach}$ to the stripping time, once an
electron is captured,
decay is much more likely than stripping;
iii) consequence of the previous point is that for
$Z<30$ species, nuclei are always at most singly attached.
\begin{figure}
\centerline{\includegraphics*[width=\columnwidth]{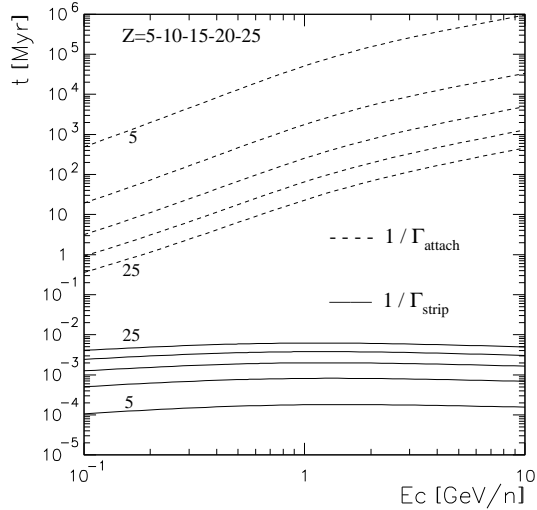}}
\caption{Energy dependence of attachment and stripping rates
$\tau_{\rm attach}=1/(n_{\rm ISM}.v.\sigma_{\rm attach})$ et
$\tau_{\rm strip}=1/(n_{\rm ISM}.v.\sigma_{\rm strip})$ for several charges.}
\label{fig:tau_s_et_tau_a}
\end{figure}

Actually, as attachment arises only when there are available electrons,
one should take into account the local properties of the interstellar medium
(LISM) in all studies involving EC-species (as for radioactive $\beta$-decay,
for different reasons). These attachment and stripping cross sections
are included in the propagation code used for the work presented here.
We should mention that these effects may be neglected in most cases,
e.g. for B/C studies.
}
{%% version courte
We should mention that these effects may be neglected in most cases,
e.g. for B/C studies.
}
			%###############%
\subsubsection{Convection and reacceleration}
\label{subsubsec:conv.reac.}
The ingredients discussed here were introduced with the first theoretical
development of cosmic ray physics, but they were only taken into account quite
recently in the propagation models.

\paragraph{Convective wind}

It has been recognized for a long time that a
thin disc configuration would be disrupted by cosmic ray
pressure~\cite{V_disruption}.
It can be stabilized by the presence of a halo, but
further considerations imply that this halo would not be
static either.
The stellar activity and the energetic phenomena associated to the late stage
of stellar evolution may push the interstellar plasma and the
magnetic field associated with it out of the galactic plane.
The net result would be the presence of a convective current directed
outwards from the galactic plane and called {\em galactic wind}\/,
which adds a convective term to the diffusion equation.
The properties of this wind may be studied by hydrodynamical~\cite{V_HD} and
magneto-hydrodynamical (MHD)~\cite{V_MHD} methods.

Consequences of a wind have been first investigated by
Ipavitch~\cite{Ipavitch}.
Since then, it has been observed in other
galaxies through a flattening of the electron induced radio
spectrum around 1 GHz in the halo~\cite{V_observation} that could be neither
predicted by a diffusion model, nor by a pure convection model. In our
own Galaxy, this effect is more difficult
to be clearly established~\cite{V_our_gal}, and its effects
have been investigated in particular on nuclei in various
models~\cite{V_nuclei}.

Tab.~\ref{tab:Vgal} compiles values obtained in several studies for very simple
forms of galactic wind.
As one can see, almost all the authors set only upper limits on this
possible wind.
\begin{table*}[hbt!]
          \begin{center}
	\begin{tabular}{|c|c|c|c|c|c|}   \hline
	    \label{tab:Vgal}
	    $V_c$[km s$^{-1}$] & $z$-dependence & model
	    & resolution & data & ref. \\ \hline\hline
	    $\lesssim 60$ & constant ($V_c$)& thin disc+halo (1D) &
Monte-Carlo &
	    B/C, $^{10}$Be/$^9$Be & (Owens, 1977) \\
	    $\sim 8$ & linear ($2V_c z$)& - & analytical & - &
(Jones, 1979) \\
	    $\lesssim 20$ & - & - & numerical & -  & (K\'ota \& Owens, 1980)\\
	    $\lesssim 16$ & constant ($V_c$) & - & analytical & - &
	    (Freedman {\em et al.}\/, 1980)\\
	    $\lesssim 20$ & - & thin disc +halo (2D) & - & - &
	    (\cite{Webber_Lee_Gupta}, 1992)\\
	    $\lesssim 15$ & linear ($3V_c z$) & - &- & - & (Bloemen {\em et
	    al.}\/, 1993) \\
	    $\lesssim 20$ & constant ($V_c$) & - & - & - & (Lukasiak {\em et
	    al.}\/, 1994)\\
	    $\lesssim 20$ & - & - & - & $^{10}$Be & (\cite{Webber_Soutoul},
	    1998)\\
	    $\lesssim 7$ & linear ($V_c z$)& - & numerical & B/C,
	    $^{10}$Be/$^9$Be &  (\cite{Strong_Moskalenko}, 1998)\\
	    \hline\hline
	\end{tabular}
	\caption{Constraints on galactic convective wind obtained
	from nuclei adjustments in various models listed in~\cite{V_nuclei}.}
          \end{center}
\end{table*}
Actually, the exact form of galactic winds is not known.
      From a self-consistent analytical description
including MHD calculations of the galactic
wind flow, cosmic ray pressure and the thermal gas in a rotating Galaxy,
a wind increasing linearly with $z$ up to $z\sim 15$~kpc, with a $z=0$
value of about 22.5 km~s$^{-1}$ was found in~\cite{Ptuskin97} (see also
references therein). Following a completely different approach,
\cite{Soutoul_Ptuskin_Vent} extract the velocity form able to reproduce
data from a one-dimensional diffusion/convection model. They obtain a decrease
from 35~km~s$^{-1}$ to 12~km~s$^{-1}$ for $z$ ranging from 40~pc to~1~kpc
followed by an increase to~20~km~s$^{-1}$ at about 3~kpc.
These two recent works raised a positive conclusion about presence of a wind.
For reference, the values for our best B/C fits correspond to about
15~km~s$^{-1}$. The difficulty to compare constant wind values to other
$z$-dependences is related to the fact that cosmic rays do not spend the
same amount of time at all $z$, so that there cannot be a simple correspondence
from one model to another.
As a result, all the above-mentioned models are formally different,
with, moreover, many different inputs (spectral index, diffusion slope) that
complicate a possible comparison.
Nevertheless, their values are roughly compatible,
the grounds for a physical motivation for this wind being provided
in~\cite{Ptuskin97}.
One can notice that other models (e.g.~\cite{Strong_Moskalenko}) dislike
convective
wind in their
fit, but we will come back to this point in Sec.~\ref{subsec:res.from_B/C}.
To conclude about galactic winds, a radial dependence is not
excluded~\cite{Breitschwerdt02}.

\paragraph{Reacceleration}
The term {\em acceleration}\/ is used for the process by which
low energy nuclei gain a huge amount of energy in a short time,
promoting them to the rank of Cosmic Ray Nuclei.
Once the cosmic rays are accelerated,
there are other processes that can lead to less sudden
energy gains, referred to as {\em reacceleration}\/.
There are two classes of such processes.
For the first, called  {\em sporadic reacceleration}\/~\cite{reac_spor},
it is assumed
that there are some well localized reaccelerating centers (e.g.
supernov\ae), on which
the cosmic rays can scatter and gain a small amount of energy at each
of these scatterings.
For the second, referred to as {\em continuous}\/~\cite{reac_cont},
reacceleration
occurs during the wandering of the charged nuclei through the Galaxy, as
an unavoidable consequence of spatial diffusion.
The quasi-linear theory of Boltzmann equation, where all effects are averaged
over statistical properties of the plasma (see e.g. \cite{Kulsrud}),
indicates that the net effect is a diffusion in energy
space~\cite{reac_theorie}.
The typical times for diffusion in space and energy are related through
the relation $\tau_{\rm spatial}.\tau_{\rm reac}\propto L^2/V_a^2$ (see
e.g.~\cite{Schlick86}), where $V_a$ is the Alfv\'en speed of the medium.
Some basic features of these two modes of reacceleration
can be found in Giler {\em et al.}\/ (1989)~\cite{reac_spor}.
It is likely that both these modes are present
(see as a good example, the treatment of~\cite{Medina_tanco}).

The most direct observational evidence for reacceleration comes from
cosmic ray species which are unstable via electronic capture.
Their abundance relative to stable isotopes shows that they had some
time to decay and therefore to attach an electron, which is more
difficult at high energy.
This means that these EC unstable species had a lower energy at the beginning
of their propagation stage: the nuclei have gained about
100-200~MeV/nuc for a kinetic
energy of a few hundreds of MeV/nuc~\cite{reac_traceur}.
A less direct evidence is related to the question of isotropy of cosmic rays
measurements, which favors small diffusion slope (e.g. Kolmogorov
$\delta=1/3$)
whereas the classical propagation models (no reacceleration)
prefer larger values $\delta \sim 0.6$ (see~\cite{Strong_Moskalenko}
and~\cite{reac_cont}). It turns out that reacceleration allows to redeem
small $\delta$.

In our studies~\cite{Maurin02,Maurin01}, we further
investigated the allowed values of the reacceleration parameter, i.e.
the Alfv\'en speed $V_a$.
We tried to compare these values with those used by other
authors~\cite{Seo_Ptuskin,Strong_Moskalenko,Jones_Lukasiak_Ptuskin_Webber}.
The difficulty is that the secondary to primary ratios are determined by
an effective value:
\begin{equation}
          V_a^{\rm eff}=V_a^{\rm true} \times \sqrt{\frac{h}{h_{\rm 
reac}}}\times
          \sqrt{\omega}\;,
          \label{trueVa}
\end{equation}
where $h$ and $h_{\rm reac}$ are respectively the height of the diffusive and
reacceleration zone.
The parameter $\omega$, which may depend on $z$, characterizes the level
of turbulence and is often set to 1~\cite{Seo_Ptuskin}.
Our model, as others, uses
\begin{eqnarray}
          \omega(z)= \left\{
          \begin{tabular}{cl}
	$1$ & if $z<h_{\rm reac}$ ,\\
	$0$ & otherwise ;
          \end{tabular}
          \right.
\end{eqnarray}
as a crude approximation of the more complex reality.
The total reacceleration rate, at least in a first approximation,
is given by the convolution of the time spent in the
reacceleration zone with the corresponding {\em true}\/ Alfv\'en speed
in this zone.
There are no direct observational clues about the size of the
reacceleration zone, or  about $\omega(z)$.
As a result, even if the analysis of cosmic rays can give values
of $V_a^{\rm eff}$, this does only give the value $V_a^{\rm true}$
up to an unknown factor $h/h_{\rm reac}$.
The situation is even more complex if no assumption is made about
$\omega(z)$. If $\omega(z)$ strongly depends on $z$ in a large reacceleration
zone, then the expression above may still be used, with $h_{\rm
reac}$ being an effective size depending on the vertical distribution
of cosmic rays.
However, this situation seems to be unfavored by MHD
simulations~\cite{Ptuskin97}.

To conclude with an order of magnitude, our best models indicate that
$\sqrt{h/h_{\rm reac}}$ must be $\sim 1/4$ in order to give 
realistic values~\footnote{There is a misprint in~\cite{Maurin02}
where 1/4 should be read instead of 4.}
for $V_a$ (with $\omega \sim 1$)~\cite{Maurin02}.

		%---------------------------------%
		%---------------------------------%

\subsection{Different approaches of propagation}
\label{subsec:history}

The equations describing diffusion have been written
down in Sec.~\ref{subsubsec:full}, as Eq.~(\ref{eq_a_resoudre1}).
There are many approaches to solve the same diffusion equation, i.e.
direct
resolution (if possible), extraction of the Green function of the problem,
or separation of astrophysical and nuclear aspects through the weighted
slab technique.
Most of the theoretical work on cosmic rays makes an extensive use of
the weighted slab model and of the so-called Leaky Box model.
We now present  a synthetic overview of these models and give some clues
about their successes and failures.

\subsubsection{Leaky Box}
One of the simplest (though very simple) model
is the so-called Leaky Box, in which the Galaxy is described as a finite
propagation volume, delimited by a surface.
Inside this volume, the densities of sources, interstellar matter and
cosmic rays are homogeneous.
Moreover, each nucleus has a probability per unit time $1/\tau_{\rm esc}$
to escape from the box.
In the stationary regime, the densities are given by
\begin{equation}
          \frac{N^j}{\tau_{\rm esc}}+nv\sigma^j N^j=q^j+\sum_{{\rm heavier}~k}
          nv\sigma^{kj}N^k\;.
\label{LB_eq}
\end{equation}
This model has been successful to explain most
of observed cosmic ray stable fluxes at different energies
by a single function $\tau_{\rm esc}(E)$.
This function can be either adjusted to the data,
its physical interpretation being found afterwards, or
extracted  directly from more complete propagation
equations~\cite{Jones_Lukasiak_Ptuskin_Webber}.
The second approach is a good way to understand why Leaky Boxes work
so well, and is
more easily seen in the framework of the weighted slab formulation.
Thus we now turn to this point, introducing first the slab model.

\subsubsection{Slab model}
\label{subsubsec:slab}

Independently of the exact processes which are responsible for
propagation,
the first certainty we have about cosmic ray nuclei is that
they have crossed some interstellar matter between their
creation in the sources  and their detection.
This leads to spallation reactions that alter the initial (source)
composition by
destroying primary species and producing secondaries.
The ratio of secondary to primary species fluxes gives information
about the quantity of matter crossed. It is convenient to introduce
the column density of matter crossed by a particle, also called
the {\em grammage}\/, expressed in g~cm$^{-2}$.
All the nuclei of a given species, with a given energy, do not have
the same propagation history, in particular they have not crossed the
same amount of matter, so that a distribution of grammages is
associated with each species.

In a first step, though, one can assume that all nuclei of a given
species have crossed the same grammage.
This is called the slab model.
Formally, the number $\tilde{N}^j(x)$ of nuclei $j$ that have crossed
the grammage $x$ is
related to the destruction rate
(the inelastic cross section is denoted $\sigma^j$) and the creation
rate (i.e. the spallation rate of all heavier nuclei $k$ giving $j$,
the cross section being denoted $\sigma^{kj}$) by
\begin{equation}
           \frac{d\tilde{N}^j(x)}{dx}+\frac{\sigma^j}{\bar{m}}\tilde{N}^j(x)=
           \sum_{{\rm heavier}~k}
           \frac{\sigma^{kj}}{\bar{m}}\tilde{N}^k(x)\;,
           \label{slab1}
\end{equation}
with the initial condition $\tilde{N}^j(x=0)=q^j$.
The resolution of this equation yields the secondary to primary ratios
as functions of $x$. Comparison with observations give the value of $x$.
This leads to a contradiction, as
weighting correctly the production rate by all the parent nuclei,
the observed LiBeB/CNO ratio $\sim 0.28$ gives $x\sim 4.8$~g~cm$^{-2}$,
whereas the ratio sub-Fe/Fe$\sim 1.5$ gives $x\sim 0.8$~g~cm$^{-2}$
(see e.g.~\cite{Longair}).
			%###############%

\subsubsection{Weighted slab and other propagation models}
\label{subsubsec:weighted_slab}

The previous model is too simple and one should take into account
that the nuclei of the same species and same energy
may cross different amounts of matter.
Introducing the probability $G(x)$ that a nucleus $j$ has crossed
the grammage $x$, the density of a nucleus $j$ is given by
\begin{equation}
           N^j=\int_0^\infty \tilde{N}^j(x)G(x)dx\;.
           \label{separation_x}
\end{equation}
The function $G(x)$ is called the {\em Path Length Distribution}\/ (hereafter
referred to as PLD).
It is remarkable that the use of the same function $G(x)$ for all
nuclei can solve the apparent contradiction between the
grammage seen by CNO and Fe:
the Fe nuclei have a larger destruction cross section, so that they
are more easily destroyed.
The associated function $\tilde{N}^j(x)$ decreases more rapidly and
the sub-Fe/Fe
ratio is more sensitive to the low $x$ part of the grammage distribution
function $G(x)$.

\paragraph{Weighted slab as models or as a general technique}
The probability distribution $G(x)$ can be determined by the choice of
a propagation model. For example, the slab model is obtained by
considering
$G(x)=\delta(x-x_0)$.
Actually, the weighted slab approach has been widely used in literature,
under two slightly different  forms.
In the weighted  slab {\em model}\/, one tries empirically to
modify the Path Length
Distribution $G(x)$ to account for the data.
The physical meaning of $G(x)$ may be explored in a second step.
The  weighted slab {\em technique}\/ is more general, and the name
refers to the possibility to introduce Eq.~(\ref{separation_x})
containing the additional and meaningful
variable $x$ (the grammage) in any propagation equation (such as
the LB model, diffusion model, diffusion/convection model\dots).
Example of direct extraction
of PLD can be found in~\cite{Lez_Web_Owens76a,grammage_moyen}.
Up to the end of this section, we will omit this subtility.
The weighted slab approach allows to link Leaky Box models with more
realistic diffusion models, explaining why these Leaky Boxes work so well.

\paragraph{Path Length Distribution of the Leaky Box model}
As an alternative to the direct analytical resolution of
Eq.~(\ref{LB_eq}), one can
insert Eq.~(\ref{separation_x}) in Eq.~(\ref{LB_eq})
which separates the nuclear part from the other effects.
This gives
\begin{eqnarray}
         \label{eq:PLD_LB}
         \left\{
         \begin{array}{ll}
	\displaystyle
	G^{\rm LB}(x)=\frac{1}{\lambda_{\rm
esc}}\exp\left(-\frac{x}{\lambda_{\rm esc}}\right)
	\;\;;\\
	\displaystyle \frac{d{\tilde N}^j (x)}{dx} +
	\frac{\sigma^j}{\bar{m}} \tilde{N}^j(x)=
	\!\!\sum_{k>j}\!\frac{\sigma^{kj}}{\bar{m}}\tilde{N}^k(x)\;\;.
         \end{array}
         \right.
\end{eqnarray}
In this expression, the PLD $G(x)$ is expressed as
a function of $\lambda_{\rm esc}(E)=\bar{m}nv\tau_{\rm esc}(E)$.
An interesting feature of this model is that the average 
grammage,
given by $\langle x\rangle=\int xG(x)dx$ is exactly the escape length, i.e
$\langle x\rangle= \lambda_{\rm esc}(E)$. Actually, this quantity solely
depends
on energy if $G^{\rm LB}(x)$ is the same for all species.
This PLD may be modified to give new weighted slab models.
For example, some discrepancies have led authors in the past to
introduce the double
PLD~\cite{double_PLD} which is a mixture of two $G^{\rm LB}(x)$ with
two different
characteristic functions $\lambda_{\rm esc}$ and $\lambda'_{\rm esc}$.
This can be interpreted as the presence of two different 
propagation zones
(Simon {\em et al.}\/, 1977, 1979~\cite{double_PLD}), which supports
the more realistic disc-halo model.
Let us notice that the terms {\em escape length}\/ and
{\em Path Length Distribution}\/ are sometimes mixed in the literature,
which can generate confusion.

Finally, it is worth noting a specific point quite independent of the
rest of the discussion. It has been noticed in~\cite{Protheroe_Heinbach}
that a different but equivalent formulation could be obtained
from the above set~(\ref{eq:PLD_LB}) of LB Weighted Slab representation,
by the substitution
\begin{displaymath}
         G^{\rm LB}(x)=(1/\lambda_{\rm esc}) \exp(-x/\lambda_{\rm esc})
\longleftrightarrow
         G(x)=1\;,
\end{displaymath}
and the addition of a term $N^j/\lambda_{\rm esc}$ in the nuclear part
of Eq.~(\ref{eq:PLD_LB}), i.e.
\begin{displaymath}
         \frac{d{\tilde N}^j (x)}{dx} + \frac{N^j}{\lambda_{\rm esc}}+
         \frac{\sigma^j}{\bar{m}} \tilde{N}^j(x)=
         \!\!\sum_{k>j}\!\frac{\sigma^{kj}}{\bar{m}}\tilde{N}^k(x)\;.
\end{displaymath}
The meaning of the new set of equations is clear:
either one solves the {\em stationary}\/ Eq.~(\ref{LB_eq}), or
one solves the same but {\em non-stationary}\/ equation integrating
over all times, i.e. $\int \tilde{N}^k(x) G(x)dx\equiv\int \tilde{N}^k(x)dx$
(here, $x$ is a dummy variable, so that it can be called time).
Such a procedure is very general. It is for example used
in~\cite{Stephens_Streitmatter}.

\paragraph{Weighted slab technique applied to diffusion models}
The link between the diffusion equation and the escape time
introduced in Leaky Box models was clarified by Jones in \cite{Jones70}.
He showed that diffusion models, energy losses included,
can be reinterpreted in terms of a Leaky Box model
(with some exceptions, such as $e^-$ for which the synchrotron or inverse
Compton losses are too important).
This reinterpretation, called the {\em leakage lifetime approximation}\/
can be understood by writing the diffusion equation, neglecting
energy gains and
losses and assuming steady state,
\begin{equation}
           \label{big_eq}
           -\nabla . (K^j\nabla N^j)
           + nv\sigma^j N^j=
           q^j+\!\!\sum_{k>j}\!nv\sigma^{kj}N^k\;.
\end{equation}
When applying the weighted slab technique (Eq.~\ref{separation_x}),
the previous equation can be split into a purely diffusive
and a purely  spallative part
\begin{eqnarray}
           \label{eq: PLD}
           \left\{
           \begin{array}{ll}
	\displaystyle -K \triangle
	G({\bf r},x)=Q({\bf r},x)\;\;;\\
	\label{def: WSM}
	\displaystyle \frac{d{\tilde N}^j (x)}{dx} +
	\frac{\sigma^j}{\bar{m}} \tilde{N}^j(x)=
	\!\!\sum_{k>j}\!\frac{\sigma^{kj}}{\bar{m}}\tilde{N}^k(x)\;\;.
           \end{array}
           \right.
\end{eqnarray}
The Path Length Distribution $G({\bf r},x)$ encodes all the
propagation properties.
This function has been extracted for various geometries (thin disc,
spherical halo)
and various forms of source spatial distribution
$Q(r)$~\cite{Lez_Web_Owens76a}.
A general result is that as the operator $\triangle$ is hermitic, the
function $G$ can always be set under the form
\begin{displaymath}
           G({\bf r},x)\propto \sum_{j=1}^{\infty}
           c_n({\bf r}) \exp \left(-\frac{x}{\lambda_n}\right)  \;.
\end{displaymath}
Comparing to $G^{\rm LB}(x)$ derived above in Eq.~(\ref{eq:PLD_LB}),
this shows that diffusion models
can be equivalently expressed in terms of a sum of Leaky Boxes.
Jones showed \cite{Jones70} that in most situations, only the
first terms contribute in the sum above. This explains why the
simplest Leaky Box models are so successful to describe diffusion,
even in a complex geometry.

The specific case of primaries is noticeable.
In this case, the nuclear part of Eq.~(\ref{eq: PLD})
reduces to a very simple form to finally
give
\begin{eqnarray}
           N^j({\bf r})=\int_0^{\infty} G({\bf r},x) \exp (-\sigma^j x)  dx\;.
\end{eqnarray}
Mathematically, the solution is the Laplace transform
of parameter $\sigma^j$ of the function $G({\bf r},x)$.
Several solutions corresponding to several diffusive geometries
can be tabulated with this Laplace transform~\cite{Margolis}.
This approach gives an alternative way to evaluate
the average grammage $\langle x\rangle$, i.e. the equivalent
Leaky Box description, from $N^j$
\begin{displaymath}
           \langle x\rangle = \frac{\int_0^{\infty}x\:G(x)dx}{\int_0^{\infty}
G(x)dx}
           = -\left.\!\! \left( \frac{d}{d\sigma^j}\ln N^j \right)
           \right|_{\sigma^j=0}\;\;.
\end{displaymath}

To end with these relations between treatment of the same equations,
the latter formulation is close to what one obtains using the random
walk approach~\cite{Taillet01,Jones71}.
This should be not too surprising since~\cite{Chandra} showed a long time
ago the equivalence of random walk description and diffusion
processes.
			%###############%

\subsubsection{Limitation and usefulness of all these models}
\label{subsubsec:links_between_approaches}
Actually, the fluxes derived with the weighted slab technique
as depicted above differ slightly from the direct evaluation. Whereas
Lezniak~\cite{Green}
tried to extend the simple scheme presented in Eq.~(\ref{separation_x})
to include energy losses, Jones~\cite{Jones91} demonstrated that
the separation between the nuclear side and the propagation side is only
approximately valid even in the high energy regime.
The reason is that in the rewriting $N^j({\bf r})=\int_0^{\infty}{\tilde
N}^j (x)G({\bf r},x)dx$, the separation between
${\tilde N}^j(x)$ and $G({\bf r},x)$ can have only one parameter in
common,
otherwise, equations obtained from the initial propagation equation are
generally inconsistent. Hence, one can think that energy
appears only as a mute parameter, but this is not 
exactly the case: the astrophysical part,
i.e. $G({\bf r},x)$ is the same for all species as long as rigidity is
considered. Conversely, for the nuclear part, the equation obtained
above is only valid for a fixed kinetic energy per nucleus.

However, redefining some parameters and under several conditions,
the technique can be made exact~\cite{Ptuskin_PLD_exact},
and this idea has been numerically and quantitatively validated
in~\cite{Stephens_Streitmatter}. Despite that, the leakage lifetime
approximation is not valid in some cases a  mentioned above.
For some nuclei, a description in terms of Leaky Box may lead to
wrong results. This is the case for radioactive species in a
realistic Galaxy, as was shown by \cite{Prishchep}.

\paragraph{Concluding remarks}
The Leaky Box models, due to their simplicity, are very well suited
to the extraction of source abundances (elemental as well
as isotopic). They can also be used to compute the secondary
antiproton production, since the same processes as for
secondary stable nuclei are at work. However, as emphasized
in~\cite{Donato01} (see also Sec.~\ref{subsubsec:sec_sources}),
they are not able to predict any primary contribution in the antiproton
signal, since it requires the knowledge of the spatial distribution
of primary progenitors. It is also well known that the Leaky Box
parameters are just phenomenological with only a distant connection
to physical quantities. However, even if apparently many existing effects
cannot be correctly included, it is shown in~\cite{Jones_vent}
that the galactic wind
can be accounted for in a Leaky Box description and emphasized that
the phenomenological behavior of the escape length at low energy could
be due to the presence of  this galactic wind.
This idea was investigated further in~\cite{Jones_Lukasiak_Ptuskin_Webber}
with a generation of several equivalent phenomenological escape lengths from
several possible physical configurations of a one-dimensional diffusion
model.
The relation between one-dimensional models and Leaky Box models is thus
firmly established and very well understood. This relation also
elucidates some of the physical contents of Leaky Box models.

Our model furnished the following step towards realistic
description, because it is equivalent, up to several minor
modifications, to the one-dimensional model
of~\cite{Jones_Lukasiak_Ptuskin_Webber}. 
Finally, in the Strong {\em et al.}\/
model~\cite{Strong_Moskalenko}, all subtle effects can be studied
and modelled, with the counterpart that the numerical approach makes
the physical intuition of the results less straightforward and the
computation slower. However, the physical input is almost the same as
in our model, which explains why the results are very close.

Even if all these models are equivalent to describe
the {\em local}\/ observations of charged cosmic rays, they lead to
very different conclusions and interpretations when the spatial variation
of the cosmic ray density is considered.
As an illustration of the poor current understanding of this global
aspect, we mention the ever-lasting problem of
the gamma ray excess about 1~GeV towards the galactic
center~\cite{Excess_gamma}
or the too flat radial $\gamma$-ray distribution observed in the
disc~\cite{dist_rad_plate}.
		%---------------------------------%
		%---------------------------------%

\subsection{Numerical implementation}
\label{subsec:numerical_simu}

      From the above discussions, it should be clear that the evaluation of
the fluxes involves the computation of a nuclear reactions grid and
the resolution of differential equations describing diffusion.

\paragraph{Nuclear part}
The flux of a given nucleus is the sum of the primary (source)
contribution and all the secondary contributions, from the spallations
of heavier nuclei.
It may be convenient to write a matrix relation
(easily diagonalized, see coefficients in~\cite{book_Syrovatskii})
between the fluxes $\{ N^j\}^{j=1\dots n}$ and the sources $\{
q^j\}^{j=1\dots n}$
\begin{displaymath}
          [N]=[\alpha][q] + [\beta][N]\;\;,
\end{displaymath}
where the matrices $[\alpha]$ and $[\beta]$ contain the information
about destructive spallations and secondary creation.
This is particularly useful in the Leaky Box model, as this relation
is straightforwardly
inverted and can provide source abundances as well as associated errors
without too much efforts~\cite{Letaw_grid,Margolis}.
However, as soon as energy losses are considered, this becomes more
complicated.

\begin{figure}[hbt!]
          \includegraphics*[width=\columnwidth]{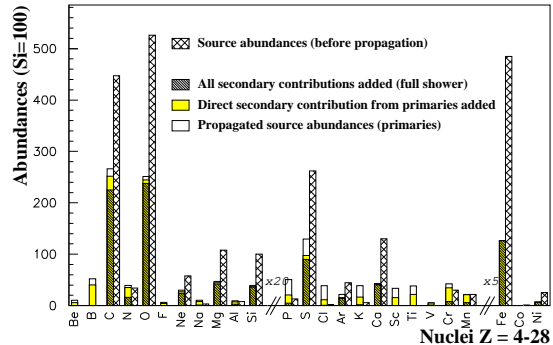}
          \caption{This figure shows the source abundances,
          along with propagated abundances. For the secondary species, we
          indicate the abundance that would be obtained by neglecting the
          secondary contribution due to the spallation of heavier secondaries.
          This second order effect is discussed in the text.
          For the sake of clarity, the scale has been expanded
          for species heavier than P, by a factor $\times 20$ for species
          lighter than Mn and $\times 5$ for heavier species.}
          \label{fig:abondance_moi}
\end{figure}

We chose a more direct method, i.e. the {\em cascade}\/ method.
The nuclei are classified according to their masses.
The heaviest nucleus $N_1$ has no secondary contribution and is given
directly by its source contribution $q_1$.
Eq.~(\ref{eq_a_resoudre1}) is solved for this first nucleus.
The flux of the second heavier nucleus $N_2$ is a mere
combination of the secondary contribution coming from $N_1$
and its own source contribution $q_2$.
Then, Eq.~(\ref{eq_a_resoudre1}) is solved for this second nucleus,
using the result obtained for $N_1$.
The flux of the third $N_3$ is a combination of its own source
contribution $q_3$, plus spallative
contributions from the two previous ones $N_1$ and $N_2$.
This procedure is repeated for all the nuclei.
These implicit contributions are displayed
in Fig.~\ref{fig:abondance_moi}: source abundances have been arbitrarily
normalized to 100 (dashed right bars, Solar abundances~\cite{SSabund}
times first ionization potential, see e.g.~\cite{FIP}).

Left bars are for propagated fluxes
at about 1 GeV/nuc in a simple Leaky Box model without energy losses.
A first point is to notice that heavy nuclei are more subject to
destruction during propagation than light nuclei.
Secondary species also appear clearly on the plot.
Coming back to the cascade procedure, three quantities are displayed:
one has (left bars) the sole propagation of sources (dashed bars),
i.e. primaries;
faint shaded bars show the sole secondary contribution (direct contribution of
heavier sources); the higher bars correspond to the summation of all
intermediary
steps. The effect of these higher order secondaries is particularly
important for heavier species.
An iterative method can also be used~\cite{Strong_Moskalenko}, implementing
the previous step propagated abundances as sources for the next step,
until convergence is reached.

\paragraph{Spatial part}
The spatial part of the diffusion equation involves second order
differential equations.
In our semi-analytical approach, one deals with a second order
equation in energy but no
spatial derivative. The resolution is much simplified, compared for example
to \cite{Strong_Moskalenko} where the full second order partial
derivative transport equation is solved.
As regards the energy part, the Runge-Kutta method is not suited and
one must turn to
more refined methods, such as  the Crank-Nicholson scheme.
The general method can be found in many numerical books~\cite{NumRec}.
An alternative approach is random-walk Monte Carlo simulations.
A diffusive process can be mimicked by a random walk,
with step related to the diffusion coefficient~\cite{Chandra}.
This approach is very time consuming but a few authors has used this
technique since mid-seventies (e.g. Owens, 1976b~\cite{Green}) to
propagate self-consistently proton
and electrons~\cite{Webber_Rockstroh}.
Such an approach is useful when one wants to model very complex and
inhomogeneous environments.

		%---------------------------------%
		%---------------------------------%

\subsection{Quick survey of charged nuclei behaviors and their interest}
\label{subsec:CR_overview}

\begin{figure}[hbt!]
          \begin{center}
	\includegraphics[width=\columnwidth]{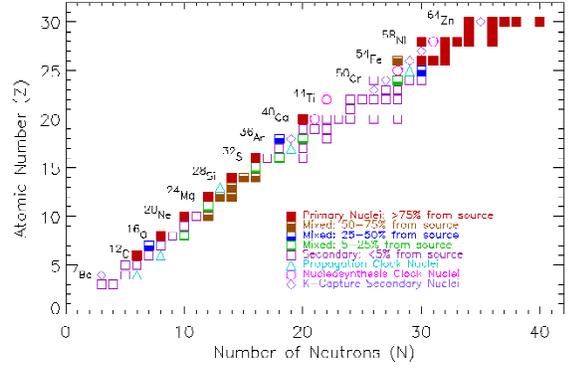}
	\caption{Stable and unstable cosmic rays
	in propagation model and their primary or
	secondary character (taken from~\cite{site_ouebe_CRIS}).}
	\label{fig:isotopes_scan_CRIS}
          \end{center}
\end{figure}

We now have all the elements needed to compute the fluxes of all the
cosmic ray nuclei. They all have different propagation histories, and
they carry information of different nature (see~\cite{revue_decay}
for a review).
They can be classified as follows:
\begin{itemize}
          \item {\bf Stable primaries and mixed species} (i.e. {\em secondary}\/
          plus {\em primary}\/ contributions): most elemental abundances
          point towards a ``standard" origin (from a particle physicist
point of view).
          Some nuclei present anomalous contributions that could be due to
wind-induced
          enrichment from Wolf-Rayet stars surfaces. The correct abundance
          determination of nuclei belonging to this class also allows a better
          characterization of the acceleration processes.
          \item {\bf Pure stable secondary species}: the three groups of
secondary species
          ($Z=1-2$, $Z=3-5$ and $Z=21-23$) give information on the
	propagation history
          for quite different charges. Their importance for the
          determination of the propagation parameters
          is extensively demonstrated in this review.
          \item {\bf $\boldmath \beta$ unstable secondary species}:
          though radioactive species
          were first used to show the existence of a large diffusive
          halo~\cite{Garcia77}, we underlined in~\cite{Donato01} the interest
          of $^{10}$Be, $^{26}$Al and $^{36}$Cl ($^{54}$Mn is more
difficult to tackle
          since it has a mixed decay mode, $\beta$ and EC) for the study
          the LISM, or as a possibility to derive the
          diffusion coefficient once LISM properties are
fixed~\cite{Ptuskin_LISM}.
          \item {\bf \boldmath Primary EC clocks}: through Co/Ni,
Co/Fe or the more
          precise $^{59}$Ni/$^{60}$Ni ratios, these EC clocks indicate
	the time elapsed between
          synthesis and acceleration and are shown as circles in
	Fig.~\ref{fig:isotopes_scan_CRIS}
          (the latest {\sc ace} results point towards $t\gtrsim 5\times 10^5$
yr~\cite{Wiedenbeck}).
          \item {\bf Secondary reacceleration indicators}: other 
K-capture nuclei
          (open diamonds in
	Fig.~\ref{fig:isotopes_scan_CRIS}) indicate that
reacceleration during propagation
          could be as large as a few hundreds of MeV/nuc at low energy.
          The latest results from the {\sc ace} experiment seem to confirm this
reacceleration,
          but the propagation of EC unstable nuclei in a refined
propagation model
          and a thorough analysis including peculiarities of the LISM has not yet
          been performed (Maurin {\em et al.}\/, in preparation).
\end{itemize}
This classification is summarized in Fig.~\ref{fig:isotopes_scan_CRIS}.
%%%%%%%%%%%%%%%%%%%%%%%%%%%%%%%%%%%%%%%%%%%%%%%%%%%%%%%%%%%%%%%%%%%%%%%%%%%%%%%%%%%%%
%%%%%%%%%%%%%%%%%%%%%%%%%%%%%%%%%%%%%%%%%%%%%%%%%%%%%%%%%%%%%%%%%%%%%%%%%%%%%%%%%%%%%

\section{Solutions of the diffusion/convection equation for two-zone
cylindrical models}

\subsection{General remarks}
The density of cosmic rays of energy $E$ at the position $(r,z)$
is obtained by solving the energy-dependent diffusion
equation~(see Sec.~\ref{subsubsec:full}).
It turns out that it is possible to first focus on the spatial
diffusion and, then to take the energy changes into account, as
long as these energy changes take place in the disc.
The density is then obtained by solving a Laplace equation in a
cylindrical geometry.
The standard method is to develop all the quantities over a suitable
complete set
of orthogonal functions involving the first Bessel function $J_0$.
Introducing $\rho = r/R$, one can write
\begin{displaymath}
          N(r,z) = \sum_{i=1}^\infty N_i(z) J_0(\zeta_i \rho)\;,
\end{displaymath}
with
\begin{displaymath}
          N_i(z) = \frac{2}{J_1^2(\zeta_i)} \int_0^1 \rho N(\rho R,z)
J_0(\zeta_i \rho)
          d \rho\;;
\end{displaymath}
the $\zeta_i$ are the successive zeros of the function $J_0$.
This development is inserted into the diffusion equation, which can
then be rewritten, using the properties of Bessel functions, as
\begin{eqnarray}
          \lefteqn{N_i''(z)  -
          \left(\frac{2h\Gamma_{\rm inel}}{K}\delta(z)
          + \frac{\Gamma_{\rm rad}}{K} +\frac{\zeta_i^2}{R^2} \right) N_i(z)}
          \nonumber \\
          && - \frac{V_c}{K} N_i'(z) = -\frac{q_i^{\rm source}(z)}{K}\;,
          \label{diffusion_bessel}
\end{eqnarray}
where $\Gamma_{\rm inel}$ and $\Gamma_{\rm rad}$ denote respectively the
spallation rate over the interstellar medium and the radioactive decay
rate.
The Bessel expansion $q_i^{\rm source}(z)$ for a unspecified
source term $q(r,z)$ has also been used.

\subsection{Full solutions}

We do not recall here the procedure used to solve
Eq.~(\ref{eq_a_resoudre1}) and we refer the suspicious reader
to~\cite{Maurin01}.
Suffice to say that very general solutions can be written for each species.
They consist in the sum of two terms, the first involving sources
located in the disc (primary or secondary), and the second involving
sources located in the whole halo (radioactive origin).
For the sake of clarity, we present separately these two contributions.

\subsubsection{Contribution from progenitors in the disc}
The galactic disc is a particular place for GCR, as
it contains the standard primary sources $q(r,z)=q^{\rm disc}(r)\delta(z)$
(see Sec.~\ref{subsubsec:sources}), and
it is the place where
spallations occur.
The contribution to the density is given by
\begin{eqnarray}
          \label{SOL PRIMAIRES}
          \lefteqn{N(r,z)=\exp\left(
\frac{V_{c}z}{2K}\right)}\nonumber \\
          &&\times \sum_{i=0}^{\infty}\,  {\frac{{\cal Q}_i}{A_i}} \,
          \frac{\sinh\left[ S_i (L-z)/2 \right]}
          {\sinh\left[ S_i L/2 \right]}~ J_0
          \left(\zeta_i\frac{r}{R} \right)
\end{eqnarray}
with
\begin{eqnarray}
\nonumber
          \left\{
          \begin{tabular}{cl}
	$\displaystyle
	A_i= $& $ \displaystyle 2h \Gamma_{\rm inel} + V_c +
          KS_i \, \coth \left(\frac{S_iL}{2} \right)$~;\\
	$\displaystyle
	 S_i^2=$& $\displaystyle\frac{4 \zeta_i^2}{R^2} + \frac{V_c^2}{K^2}
          + 4\frac{\Gamma_{\rm rad}}{K}$~.
          \end{tabular}
          \right.
\end{eqnarray}

For a pure primary ${\cal Q}_i= q_0 Q(E)\times q_i^{\rm disc}$
 -- where $q_i^{\rm disc}$ denotes the Bessel expansion of $q(r)$ --
  and for a pure secondary
${\cal Q}_i=\sum_k^{m_k>m_j} \Gamma^{kj}N_i^{k}(0)$ --
where the superscript
$j$ denotes the nucleus evaluated, omitted in above expressions --.

When  energy losses and diffusive reacceleration are taken into
account, the differential equation is obtained, following the procedure
described e.g. in~\cite{Maurin01}, as
\begin{eqnarray}
          \lefteqn{A_iN_i(0)={\cal Q}_i} \\
          &&-2h\frac{\partial}{\partial E}
          \left\{b_{\rm tot}(E)N_i(0)- \beta^2K_{pp}\frac{\partial}{\partial E}
          N_i(0)\right\} \nonumber
          \label{eq_a_resoudre2}
\end{eqnarray}
where the energy loss and reacceleration terms were discussed in
Sec.~\ref{subsubsec:losses} and \ref{subsubsec:reac}.
Subscript $i$ refers to Fourier-Bessel coefficients for the above equation
expended on cylindrical Bessel basis $J_0(\zeta_i\rho)$.

\subsubsection{Contribution from the halo: radioactive \boldmath{$\beta$}
decay}
\label{subsub:rad_sol_halo}
All the nuclei treated here have at most one unstable progenitor.
The contribution
of these radioactive nuclei may be unimportant in some cases, but we should
take it into account as it is the dominant process for some others. In the
simple example of $^{10}$Be$\rightarrow^{10}$B, neglecting this channel
would give
an error of about 10\% on the B flux, whereas considering
that this term is only located in the disc would give an error
of about 3\% compared to the rigorous
treatment given above.
Due to lack of space, we do not reproduce the expression in this case,
but the reader is referred to~\cite{Maurin01}.

\subsubsection{Contribution from the halo: primary sources}
\label{subsubsec:contrib_halo_prim}
We will consider the hypothesis that primary sources of cosmic rays
may be present in the diffusive halo (see Sec.~\ref{subsubsec:exotic_sources}).
The solution of the diffusion
equation is somewhat different than in the standard case of sources
located in the disc. Denoting $q^{\rm halo}$ the corresponding source term,
the Bessel terms of the density are given by (see~\cite{barrau01})
\begin{eqnarray*}
          \lefteqn{N_i(z) = - \frac{y_i(z)}{K S_i}
          + \frac{y_i(L) e^{V_c (|z|-L)/2K} }{A_i \sinh (S_iL/2)} }\\
          && \times \left[ \cosh(S_iz/2) +
          \frac{V_c + 2h \Gamma_{\rm inel}}{K S_i}
          \sinh (S_iz/2) \right]
\end{eqnarray*}
with
\begin{eqnarray*}
          \lefteqn{y_i(z) = 2 \int_0^z \exp \left(\frac{V_c (z-z')}{2K}
\right)}\\
          && \times \sinh \left(\frac{S_i(z-z')}{2}\right) q^{\rm halo}_i(z')
          dz'\;.
\end{eqnarray*}
In particular, the density in the disc ($z=0$) is given by
\begin{displaymath}
          N_i(0) = \exp \left(-\frac{V_c L}{2K} \right)
          \frac{y_i(L)}{A_i \sinh (S_iL/2)} \; .
\end{displaymath}
It can be checked that with a source term localized in the disc, i.e.
$q^{\rm halo}(r,z) = 2h \delta(z) q^{\rm disc}(r)$, the expression 
(\ref{SOL PRIMAIRES})
is recovered.

\subsubsection{Modelling of the local bubble}

\begin{figure}[hbt!]
\centerline{\includegraphics[width=\columnwidth]{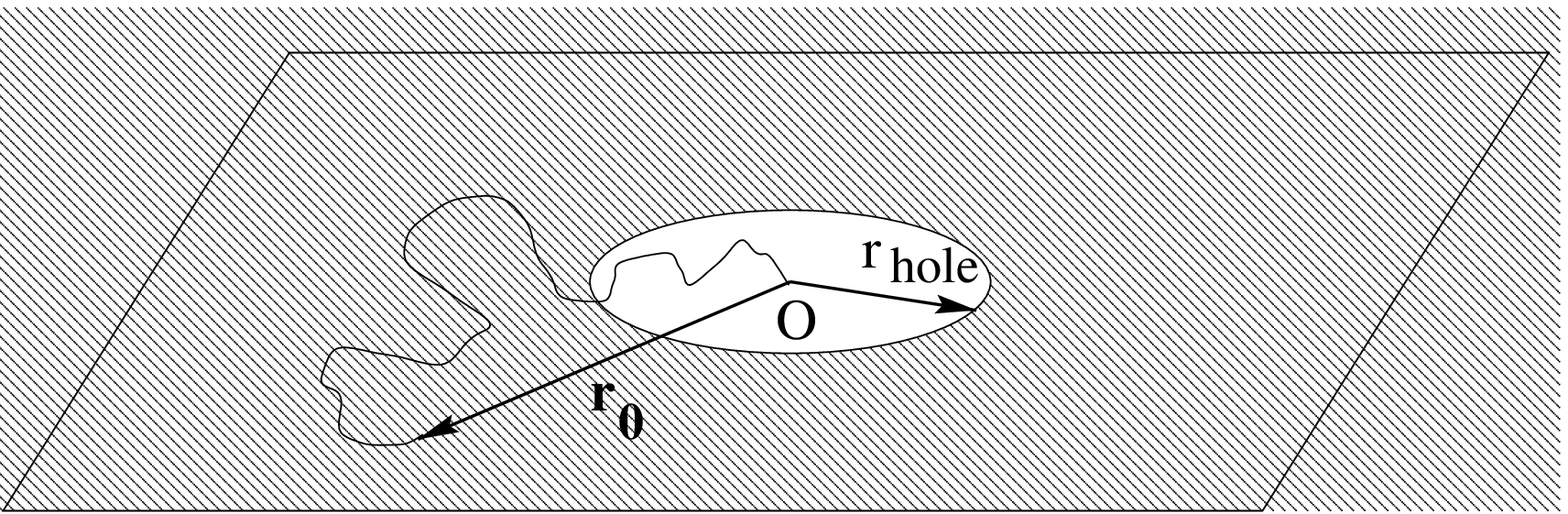}}
\caption{Schematic representation of the hole
model. The disc has zero
thickness with a hole of size $r_{\rm hole}$ in gas density.}
\label{fig:schema1}
\end{figure}

As mentioned above, it is very likely that the Sun lies in an
underdense bubble. The main effect on the cosmic rays is that
spallations, and hence the secondary production and destruction, are locally
strongly reduced. In the thin disc model, this can be taken into
account assuming position dependent spallation terms.
Actually, as will be discussed below (see Sec.~\ref{subsec:rad}),
this effect is not so
important for stable species, but it is crucial for the low energy radioactive
species, which are only sensitive to the local propagation conditions.
Once created at a given point, they diffuse in a small
region surrounding
this point before they decay. The density of their progenitors does
not vary much in this region, and it can be considered as homogeneous.
Moreover, the propagation of the radioactive is almost not altered by the
boundaries of the diffusive volume, so that we
can then put the center of the diffusive volume at the position of
the Sun. The bubble can then be modelled as a hole in the disc.
The rates
$\Gamma_{\rm spal}(r)=n_{\rm LISM}(r).v.\sigma_{\rm spal}$ and
$\Gamma_{\rm inel}(r)=n_{\rm LISM}(r).v.\sigma_{\rm inel}$ now depend
explicitly on $r$ {\em via}\/ the local interstellar density which reads
\begin{equation}
n_{\rm LISM}(r)=\Theta (r-r_{\rm hole}) \; n_{\rm ISM}
\end{equation}
where $\Theta$ is the Heaviside distribution (see Fig.~\ref{fig:schema1}).
As for the no-hole model, a solution is found by
expanding the density over the Bessel functions $J_0(\zeta_i r/R)$.
The expression for the solution may be found in~\cite{Donato01}.

\subsection{Physical interpretation of the formul\ae}

For the sake of clarity, we wish to further discuss two particular
cases.
To begin with, the case of primary stable species is of particular
interest. Their density in the disc is given by
\begin{displaymath}
          N(r,z=0) = \sum_{i=1}^\infty \frac{q_i^{\rm disc}}{A_i}
J_0(\zeta_i \rho)
\end{displaymath}
which, when compared to the Bessel expansion of the source
\begin{displaymath}
          q(r) = \sum_{i=1}^\infty q_i^{\rm disc} J_0(\zeta_i \rho)\;,
\end{displaymath}
contains basically all the physics of diffusion.
Diffusion in Bessel space may be seen as a filtering process, with a
filter $1/A_i$.
The small scale features, which correspond to large values of the index $i$,
are spread out and erased by diffusion as $A_i \rightarrow 2K \zeta_i \sim
\pi K i$ for large $i$.
The large scale structure is determined by the small $i$ behavior.
The effect of small $L$, spallations and galactic wind is to limit the
decrease
of $A_i$ for small values of $i$, i.e. they filter the
large scale features out.
In other words, they make diffusion a {\em local}\/ process.
More precisely for small $L$ and small $i$ we have
$A_i \approx 2h \Gamma_{\rm inel} + V_c + 2K/L $.
It appears that three characteristic lengths, although of different
origin, play a very similar role in erasing the large scale features of
the source distribution, namely $L$, $r_{\rm wind}\equiv 2K/V_c$ and
$r_{\rm spal} \equiv 2K/2h \Gamma_{\rm inel}$.
Structures larger than these are destroyed by diffusion.
This point is further discussed in
Secs.~\ref{subsec:origine_la_base} and~\ref{subsec:origin_realistic_case}.
		%---------------------------------%
		%---------------------------------%

\section{Origin of cosmic rays}
\label{sec:origin}

\subsection{The question of the spatial origin}

The cosmic rays emitted from distant sources are more likely to escape
from the diffusive halo or to be destroyed by spallation before
reaching Earth than those coming from nearby sources.
Conversely, a CR eventually detected on Earth has a greater
probability to come from a nearby source than a distant source.
In this section, we investigate the question of the spatial origin in
more details, and we find that under quite reasonable conditions,
most of the cosmic rays detected in Earth were emitted by sources located in
a radius less than $\sim3L$.

\subsection{Method}

The question we wish to address is the following:
a cosmic ray being detected at the position $\vec{r}_o$ of an observer
(in practice, this will be the position of the Sun), what is the
probability
\begin{eqnarray*}
             \lefteqn{d{\cal P} \left\{{\rm
emitted:~}\vec{r}_s,\vec{r}_s+d\vec{r}_s
             ~|~ {\rm observed:~} \vec{r}_o \right\}} \\
             && \equiv {\cal P} \left\{ \vec{r}_s |  \vec{r}_o  \right\}\;
d\vec{r}_s
\end{eqnarray*}
that it 
was emitted from a source located at the position
$\vec{r}_s$?
Such a question falls among classical problems of statistics.
A rigorous theoretical frame is provided by the Bayes approach that summarizes
proper use of conditional probabilities. A cruder but sufficient (and
equivalent) treatment is given through the frequency interpretation.
The probability is simply given by the number of particular realizations,
i.e. paths ${\cal N} \left[ r_s \rightarrow r_o \right]$
(the cosmic rays travels from the source at position $\vec{r}_s$ to
the observer
at position $\vec{r}_o$)
divided by the number of all realizations of ${\cal N} \left[ \rightarrow
\vec{r}_o \right]$
(the cosmic rays reach the observer position). Notice
that the latter number is just
a normalization so that the first quantity determines completely the
problem.
In the frequency interpretation, we thus have
\begin{equation}
             {\cal P} \left\{ \vec{r}_s |  \vec{r}_o  \right\}
             = \frac{{\cal N} \left[ r_s \rightarrow r_o \right] }
             {{\cal N} \left[ \rightarrow r_o \right] }\;\;.
\end{equation}
We finally notice that the number of paths
${\cal N} \left[ r_s \rightarrow r_o \right]$ determines the
number of cosmic rays that reach position $\vec{r}_o$, when a
source is placed at position $\vec{r}_s$.
We can thus write
\begin{equation}
             {\cal P} \left\{ \vec{r}_s |  \vec{r}_o  \right\}
            \propto N_{\rm r_s}(\vec{r}_o)\;\;,
\label{proba_elementaire}
\end{equation}
where the density $N_{\rm r_s}(\vec{r}_o)$ is the solution of the
propagation equation for a point source located at $\vec{r}_s$.
The normalization factor of this  relation is obtained by imposing that
${\cal P}$ actually is a probability, i.e. is normalized to unity.
If the sources are distributed according to $q(\vec{r}_s)$,
the probability that a cosmic rays detected at $\vec{r}_o$ was emitted
from a volume ${\cal V}'$ (or a disc surface) is given by
\begin{equation}
             {\cal P} \left\{ {\cal V}' | \vec{r}_o \right\}=
            \frac{\int_{\cal V'} q(\vec{r}_s) N_{\rm r_s}(\vec{r}_o)
d^3\vec{r}_s}
            {\int_{\cal V_{\rm tot}} q(\vec{r}_s) N_{\rm r_s}(\vec{r}_o)
d^3\vec{r}_s}\;\;.
\label{proba_integree}
\end{equation}
This probability contains all the physical information about the
spatial origin of cosmic rays and may be computed for different
situations (see~\cite{Taillet01} for further details).

\subsection{Pure diffusion case and key parameters for spallations and
wind}
\label{subsec:origine_la_base}

\begin{figure}[hbt!]
        \includegraphics[width=\columnwidth]{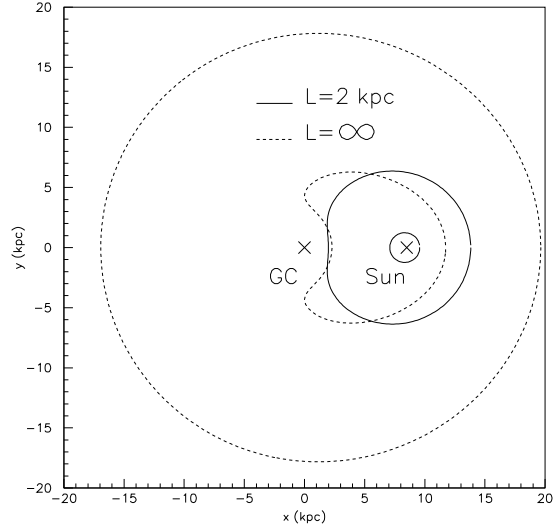}
        \caption{This figure shows the regions of the disc
	from which a given fraction
        (50 \% for the inner surface and 99 \% for the outer surface) of
        the cosmic rays reaching the Earth were emitted from. The dashed lines
        represent the case of a halo with no upper and lower boundaries ($L
        \rightarrow \infty$) whereas the solid line are for $L=2$ kpc.
        The spallations and galactic wind are not taken into account
        here, so that these regions overestimate the origin of the sources
	(they represent the geometrical limit).}
        \label{fig:carte_origine}
\end{figure}

The extent of the zone from which the cosmic rays detected on Earth
actually come from is limited by all the phenomena that keep them
from propagating over large distances: escape through the boundaries,
spallations and galactic wind.
It comes as no surprise, then, that we can define three typical lengths,
beyond which the cosmic rays do not reach us
\begin{displaymath}
          r_L = L\;;
\end{displaymath}
\vspace{-1mm}
\begin{eqnarray}
       \label{eq:r_spal}
          \lefteqn{r_{\rm spal}\equiv K/h \Gamma_{\rm inel} }\\
          && \approx 3.17 \unit{kpc}
          \times \frac{K/\beta}{0.03
          \unit{kpc}^2\unit{Myr}^{-1}}
          \, \frac{ 100 \unit{mb}}{\sigma_{inel}} \;;\nonumber
\end{eqnarray}
\begin{eqnarray}
	\label{eq:r_wind}
          \lefteqn{r_{\rm wind}\equiv 2K/V_c}\\
          &&\approx 5.87 \unit{kpc}  \times
          \frac{K}{0.03 \unit{kpc}^2\unit{Myr}^{-1}}\,
          \frac{10 \unit{km} \unit{s}^{-1}}{V_c}\;.\nonumber
\end{eqnarray}
The role of the length $r_{\rm wind}$ in the context of spatial origin was
recognized in~\cite{Jones_vent}.
For a given set of diffusion parameters, these lengths can be
computed, as well as the map of the probability that a particle reaching
Earth was emitted from each source located in the disc.
The lengths $r_{\rm wind}$ and $r_{\rm spal}$ depend on $K$, and  are larger at
higher energy.
This is of course not the case for $r_L=L$, so that at sufficiently high
energy, $r_{\rm wind}\gg L$ and $r_{\rm spal}\gg L$, the spatial origin being
solely dictated by the halo size $L$ (side boundary $r=R$
is unimportant).
The purely diffusive case is illustrated in
Fig.~\ref{fig:carte_origine}, which shows the regions from which a
given fraction of the cosmic rays reaching the Earth were emitted
from.
We refer the reader to~\cite{Taillet01} for the details.

Conversely, it turns out that when realistic diffusion parameters
are considered, these lengths may be rather small, in particular
for heavy nuclei
which are more sensitive to spallations, so that the observation of
cosmic rays  in the Solar neighborhood only gives information about
the {\em local}\/ conditions under which diffusion occurs.
This will be discussed in more details in
Sec.~\ref{subsec:origin_realistic_case}.

%%%%%%%%%%%%%%%%%%%%%%%%%%%%%%%%%%%%%%%%%%%%%%%%%%%%%%%%%%%%%%%%%%%%%%%%%%%%%%%%%%%%%
%%%%%%%%%%%%%%%%%%%%%%%%%%%%%%%%%%%%%%%%%%%%%%%%%%%%%%%%%%%%%%%%%%%%%%%%%%%%%%%%%%%%%

\section{Experimental determination of the diffusion parameters}
\label{sec:diff.param.determination}
We wish to study the constraints on the diffusion parameters
(the spectral index of sources $\alpha$, the normalization $K_0$ and
the spectral index $\delta$ of the diffusion
coefficient, the height of the diffusive halo $L$, the galactic convective
wind speed $V_c$ and the Alfv\'enic speed $V_a$) that come from the
measured fluxes of cosmic rays.
For the aim of our analysis, we can consider different classes
of flux ratios:  primary-to-primary (e.g. C/O),
secondary-to-primary (e.g. B/C or sub-Fe/Fe),
secondary-to-secondary (e.g. Li/B or Be/B),
ratios of  either stable (e.g. $^{10}$B/$^{11}$B)
or unstable (e.g. $^{10}$Be/$^9$Be) isotopes.
Each of these may be an indicator of some dominant
physical phenomenon and be particularly sensitive to the corresponding
diffusion parameters (see also Sec.~\ref{subsec:CR_overview}). The
ratio of two primaries
is  practically insensitive to changes in all the parameters,
since they have the same origin and undergo the same
physical processes (but keep in mind that extreme nuclei such as p and
Fe present drastically different destruction rates, see also
Sec.~\ref{subsec:iap01}).
Similar conclusions, even if less strong,  may be drawn
for the ratio of two isotopes of the same species, such as
$^{10}$B/$^{11}$B. Indeed, at very low energy values this
quantity is slightly affected
by changes in the injection spectra, but the effect is too weak
to constrain the diffusion parameters.

One of the most sensitive quantity is B/C,
as B is purely secondary and its main progenitors C and O are primaries.
The shape of this ratio is seriously modified by changes in the propagation
coefficients.
Moreover, it is also the quantity
measured with the best accuracy, so that it is ideal to test models.
Indeed, as a ratio of two nuclei with similar Z, it is less
sensitive to systematic errors and to Solar modulation
than single fluxes or other ratios of nuclei with more distant charges.
For the same reasons, the sub-Fe/Fe may also be useful.
Unfortunately, since existing data are still affected by sizeable
experimental errors, we can only use them to cross-check the
validity of B/C but not to further constrain the
parameters under scrutiny.

			%******************
			%******************

\subsection{The method}
For a given set of parameters, the source abundances of all nuclei
(i.e. primaries and mixed nuclei) are adjusted so that the propagated
top of atmosphere fluxes agree with  the data at 10.6 GeV/nuc
(see~\cite{Maurin01}).
The nuclear cascade is started from Sulfur, as we checked that the heavier
nuclei do not contribute significantly to the B/C ratio.
\begin{figure}[hbt!]
\includegraphics*[width=\columnwidth]{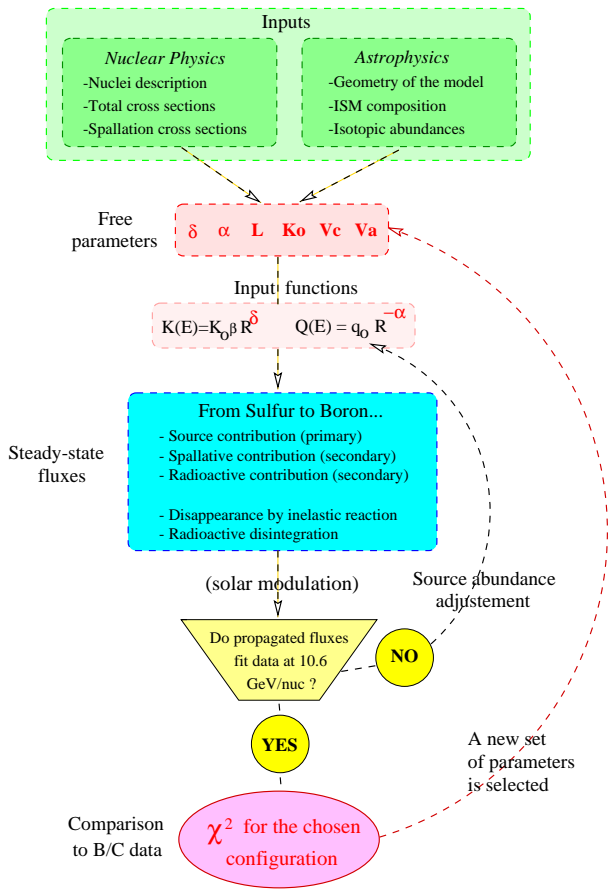}
\caption{Diagrammatic representation of the various steps of the
propagation code}
\label{cascade}
\end{figure}
The top of atmosphere fluxes are deduced from interstellar fluxes
using the force field
modulation scheme (see~\cite{Maurin01} and references therein).
The resulting B/C spectrum is then compared to the data (see below)
and the $\chi^2$ is computed for the chosen set of parameters, in our
case
\begin{equation}
           \chi^2 = \sum_i \frac{\left( ({\rm B/C})_{i, {\rm exp}} - ({\rm B/C})_{i,
{\rm model}}
           \right)^2}{\sigma_{i, {\rm exp}}^2}\;,
\end{equation}
where the sum runs over 26 experimental values from {\sc
heao}-3~\cite{Engelmann}.
In general, if the experimental set-up is such that the measured
(experimental)
values differ from the ``real" values by a non biased quantity
with a given probability distribution, then the
$\chi^2$ value gives a quantitative estimate of the probability that
the model is appropriate to describe the data.
However, this condition is probably not fulfilled for
{\sc heao}-3, as for some measured quantity,
the quoted errors $\sigma_{i, {\rm exp}}^2$ are much
smaller (e.g. oxygen fluxes) or much larger (e.g.
sub-Fe/Fe ratio) than the dispersion of data themselves.
For this reason, it is meaningless to associate a likelihood to given
$\chi^2$ values. As a consequence, in an ideal situation in which
very good and consistent data on B/C and
sub-Fe/Fe ratios were available, the best attitude would be to make a
statistical analysis of the combined set of data. Unfortunately,
this is not currently the case.
One can follow two ways to extract information from the sub-Fe/Fe data.
First, as a check, one can compare the sub-Fe/Fe
ratio predicted by our model -- using the parameters derived
from our above B/C analysis -- with data from the same experiment.
Second, one can search directly the minimum $\chi^2_{\rm Fe}$ of
the sub-Fe/Fe ratio, with no prior coming from B/C.
This procedure is more hazardous since the
statistical significance of the sub-Fe/Fe data is far from clear.

		%---------------------------------%
		%---------------------------------%

\subsection{The available data sets}
\label{subsec:data_sets}
\subsubsection{B/C and sub-Fe/Fe}

As emphasized above, the results presented here are mostly
based on the data taken by {\sc heao}-3~\cite{Engelmann}.
They have been taken in 1979-80,
for elements with charges from 4 to 28 and for
energies ranging from 0.6 to 35 GeV/nuc around a minimal Solar activity.
In the case of B/C,  the
quoted 1-$\sigma$ {\sc heao}-3 relative errors are 2-3\%.
We also considered  data from balloons~\cite{Dwyer}
and from the {\sc isee}-3 experiment~\cite{Krombel}, even if the relevant
error bars are wider.
The first one  collected data in 1973-75 for energies spanning from
around 1.7 to 7  GeV/nuc.
The second experiment was operating during 1979-81 on board a spacecraft,
in the energy range 100-200 MeV/nuc.
In some of the figures  presented below, we also
plot -- for purely illustrative goals -- the data point from
{\sc imp}-8~\cite{Modif_PLD}
and the {\sc voyager} experiments~\cite{Luk99} (we did
not include {\sc ulysses} data point~\cite{DuVernois96},
since it corresponds to a period
of maximal Solar activity). Nevertheless, we have
checked that the best $\chi^2$ values were not significantly
modified when these points were added.

As regards the sub-Fe(Sc+V+Ti)/Fe ratio, we used
data from {\sc heao}-3~\cite{Engelmann} and from
balloons~\cite{Dwyer}.
In both cases the error bars,  around 10\%, are
significantly larger than for B/C.

\subsubsection{Radioactive species}
\label{subsub:Radioactive_species}
Several experiments in the last twenty-thirty years have measured
radioactive isotopes in cosmic rays with increasing precision,
at energies of a few hundreds of MeV/nuc.
The first data -- usually presented as the ratio of some radioactive
isotope to its stable companion(s) -- were affected by errors of around
25-30\%.
The latest published data have error bars reduced
by a factor of two or three.
In the following, we implicitly refer to
three satellite experiments, namely {\sc voyager}, {\sc ulysses} and {\sc ace}.
Other experiments will sometimes be shown on figures
but they will be purely illustrative since their accuracy is far smaller.

The  best measured ratio is probably $^{10}$Be/$^9$Be
which corresponds to the lowest Z
$\beta$-radioactive nucleus.
Data from {\sc ulysses}~\cite{Connell} and from {\sc ace}~\cite{Binns99}
are
consistent, the quoted error bars being smaller for {\sc ace}.
They are also consistent with the {\sc voyager} data point~\cite{Luk99}
for which the quoted error is larger.
We do not use the {\sc smili} data, as the possibility that they are
plagued by statistical fluctuations is not ruled out~\cite{Ahlen}.

As regards the radioactive chlorine isotope $^{36}$Cl,
results are usually
provided as $^{36}$Cl to total Cl ratio. The only available data, to
our knowledge, are those from {\sc ulysses}~\cite{Connell2} with a
1-$\sigma$
error of about 35\%, and {\sc ace}~\cite{Binns99}
whose errors (even taken at 3-$\sigma$) are completely included in the
{\sc ulysses} 1-$\sigma$ upper error band.

Finally, the measurement of the $^{26}$Al/$^{27}$Al ratio
is more problematic.
Indeed, the data from {\sc ulysses}~\cite{Simpson_Connell}
and {\sc ace}~\cite{Binns99}
do not seem to be compatible
(the {\sc ace} central point is much lower than {\sc ulysses}' one).
Even enlarging {\sc ace} error bars (which are smaller than {\sc ulysses}) to
3-$\sigma$ does not improve significantly the compatibility.
On the other side, the {\sc ulysses} data are
fully compatible with 1-$\sigma$ {\sc voyager}~\cite{Luk94} ones,
whose uncertainty is still much greater than for the other two experiments.
The possible discrepancy between some of these data is
addressed in~\cite{Donato01}.

\subsubsection{p and He fluxes}
\label{subsubsec:p_He}

We do not study here the compatibility of p and He fluxes with the
prediction from diffusion models. However, we need the observed values
of these fluxes to compute the  secondary antiproton fluxes.
The contribution of heavier nuclei to the antiproton production is negligible.
Until recently, the spectra of p and He were known with a modest accuracy and
the data from different experiments were often
incompatible at high energy. This induced an uncertainty of
some tens of percents in the predicted antiproton spectrum.
Recent measurements made by the balloon-borne spectrometer
{\sc bess}~\cite{proton_helium_bess} and by the {\sc ams} detector
during the space shuttle flight~\cite{proton_ams}
dramatically reduced the
uncertainties both on  proton and helium spectra.
We fitted the high energy (T $>$ 20 GeV/nuc) part of these measured spectra
with the power law:
\begin{equation}
\Phi (\rm T) = N\,(\rm T/\unit{GeV/nuc})^{-\gamma}\;\;,
\label{eq:cosmic}
\end{equation}
where the kinetic energy per nucleon T is given in units of GeV/nuc and the
normalization factor N in units of $\unit{m}^{-2} \unit{s}^{-1}
\unit{sr}^{-1}\unit{(GeV/nuc)}^{-1}$.
This provides a good description down to
the threshold energy for the  antiproton production.

We fitted the {\sc bess} and {\sc ams} data both separately and combined,
obtaining very similar results.
This is obvious since the data from the two experiments are now totally
compatible. The best fit corresponds to
$N = 13249\unit{m}^{-2} \unit{s}^{-1}
\unit{sr}^{-1}\unit{(GeV/nuc)}^{-1}$ and
$\gamma = 2.72$; spectra obtained from the best fits
on the single {\sc bess} and {\sc ams} data completely overlap.
We did the same for helium
and the corresponding numbers are $N = 721\unit{m}^{-2} \unit{s}^{-1}
\unit{sr}^{-1}\unit{(GeV/nuc)}^{-1}$ and $\gamma = 2.74$.
The 1-$\sigma$ deviation from the best fit spectrum does not exceed
1\% for both species.
The uncertainty induced on the antiproton
spectrum is smaller than the ones induced by the uncertainties on
nuclear physics and
diffusion parameters.
The situation has significantly improved since 1998,
when an error of $\pm$ 25\% was quoted~\cite{Bottino_Salati}.

		%---------------------------------%
		%---------------------------------%

\subsection{The diffusion parameters estimated from {\sc heao}-3 B/C
measurements}
\label{subsec:res.from_B/C}

\shortversion{%% version longue
\subsubsection{Partial exploration of the parameter space}
In a first step~\cite{Maurin01}, we did not consider $\alpha_j$, the
slope of the source spectrum of each
species $j$ (see Sec.~\ref{subsubsec:sources},
p.~\pageref{page:source_spectrum}),
as a free parameter but instead we set it
to the value obtained by subtracting $\delta$ from the  spectra measured
at high energy  (see~\cite{Maurin01}).
We display models that have $\chi^2<40$.

\paragraph{Parameters are strongly degenerated}
Many configurations for $L$, $K_0$, $\delta$, $V_c$ and $V_a$
lead to the same B/C ratio. This can be understood as meaning that they
all lead to the same effective description in terms of
the associated $\lambda_{\rm esc}^{\rm LB}$ and grammage (see discussion in
Sec.~\ref{subsubsec:weighted_slab}).
In Fig.~\ref{f1_stables}, each value of $\delta$
gives a different contour plot in the $K_0/L-L$ plane. It appears that
they all can be superimposed to a single curve by a rescaling
$K_0/L \rightarrow K_0/L \times f(\delta)$, where $f$ is a function of
$\delta$ only. For the contours displayed in Fig.~\ref{f1_stables},
it takes the values $f(0.46)=0.51$, $f(0.5)=0.62$, $f(0.6)\equiv1$,
$f(0.7)= 1.54$ and
$f(0.85)=2.78$.
In the $V_a/\sqrt{K_0}-V_c$ plane (Fig.~\ref{f2_stables}), the values
of $V_c$ are shifted
downward as $\delta$ is decreased but the allowed range of
$V_a/\sqrt{K_0}$ does not significantly move.
In particular, with the above criterion, we do not find
any model having a good $\chi^2$ without convection ($V_c=0$) or
without reacceleration ($V_a=0$). We emphasize the fact that
when $\alpha$ and $\delta$ are fixed, the
four remaining parameters are strongly correlated so that, when one of them is
determined, in principle the allowed ranges for the others will be
narrower than what could naively appear from the figures.
\begin{figure}[hbt!]
          \includegraphics*[width=\columnwidth]{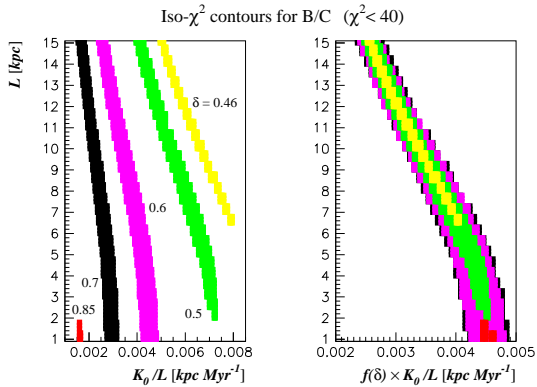}
          \caption{Models with different values of $\delta$ are shown.
          As in the previous figures, for each value of $L$ and $K_0/L$,
          only the best $\chi^2$ value is retained when the other
parameters $V_c$ and
          $V_a/\sqrt{K_0}$ are varied.
          The figure in the left panel displays the contour levels for
          $\chi^2 < 40$ for the indicated values of $\delta$. It is
possible to scale the
          $K_0/L$ values by a function $f(\delta)$ to superimpose the
contours corresponding
          to different values of $\delta$ (see text). This is displayed in
          the right panel.}
          \label{f1_stables}
\end{figure}
\begin{figure}[hbt!]
          \includegraphics*[width=\columnwidth]{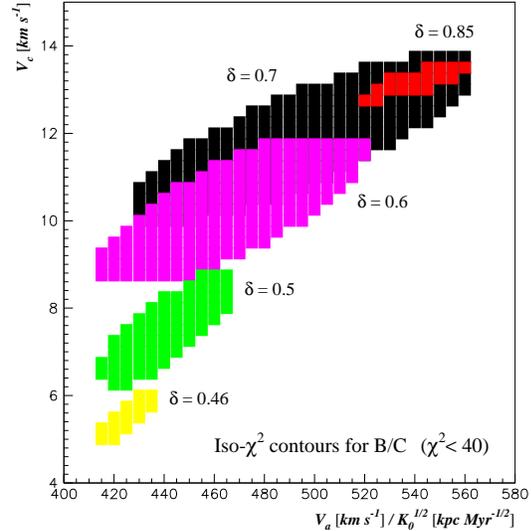}
          \caption{Models with different values of $\delta$,
          the diffusion coefficient spectral
          index, are shown. For each value of $V_c$ and $V_a/\sqrt{K_0}$ ,
          only the best $\chi^2$ value is retained when the other
parameters $L$ and
          $K_0/L$ are varied.  The figure displays the contour levels for
          $\chi^2 < 40$ for the indicated values of $\delta$.}
          \label{f2_stables}
\end{figure}

\paragraph{Partial conclusion}
Several comments about these results can be made.
First, the diffusion slope
is constrained and the particular value $\delta=1/3$ is excluded by our
analysis (even if a less stringent criterion on $\chi^2$ is adopted).
We find that the power law index for the diffusion coefficient is
restricted to the interval $0.45-0.85$, the best $\chi^2$
being $25.5$ for $\delta=0.70$, leading to $\alpha\sim2.1$.
For any $\delta$ in this interval, the good parameters in the
$K_0/L-L$ and $V_a/\sqrt{K_0}-V_c$ planes can be straightforwardly
deduced from the corresponding values for $\delta=0.6$ by a simple scaling
law. We also exclude any model without a convective velocity or without
reacceleration for any combination of the three other diffusion
parameters.
These conclusions could get more stringent by new measurements in
the whole energy spectrum for all nuclei, as all our results were obtained
using the best data, which are rather scarce and more than 20 year-old.
% These results
% are somewhat in contradiction with quite well accepted
% theoretical and other B/C induced values (see e.g. Strong et al, 1998).
}
{%% version courte
A first analysis was performed in~\cite{Maurin01}. The parameter space
was partially explored, the spectral index of the sources being kept
constant. We only present here the full analysis.
}

\shortversion{ %% version longue
\subsubsection{Full exploration of the parameter space}
This study was then refined~\cite{Maurin02} by
adding four updates: i) a more conventional reacceleration
term is used (i.e. similar to terms used by other authors);
ii) effect of deviation from a
pure power law of source spectra at low energy is inspected
-- see Sec.~\ref{subsubsec:sources}, p.~\pageref{page:source_spectrum} --;
iii)
different diffusion schemes (slab Alfv\'en wave turbulence, isotropic fast
magnetosonic turbulence and a mixture of the two case) are tested and,
iv) above all, the source slope $\alpha_j\equiv \alpha$ for all species
is set as a free parameter.
The net result, is that all the above trends are confirmed. As a bonus,
one recovers that for small diffusion slope, convection is unfavored,
as found in~\cite{Strong_Moskalenko}.
}
{%% version courte
}

\begin{figure*}[hbt!]
          \includegraphics*[width=\textwidth]{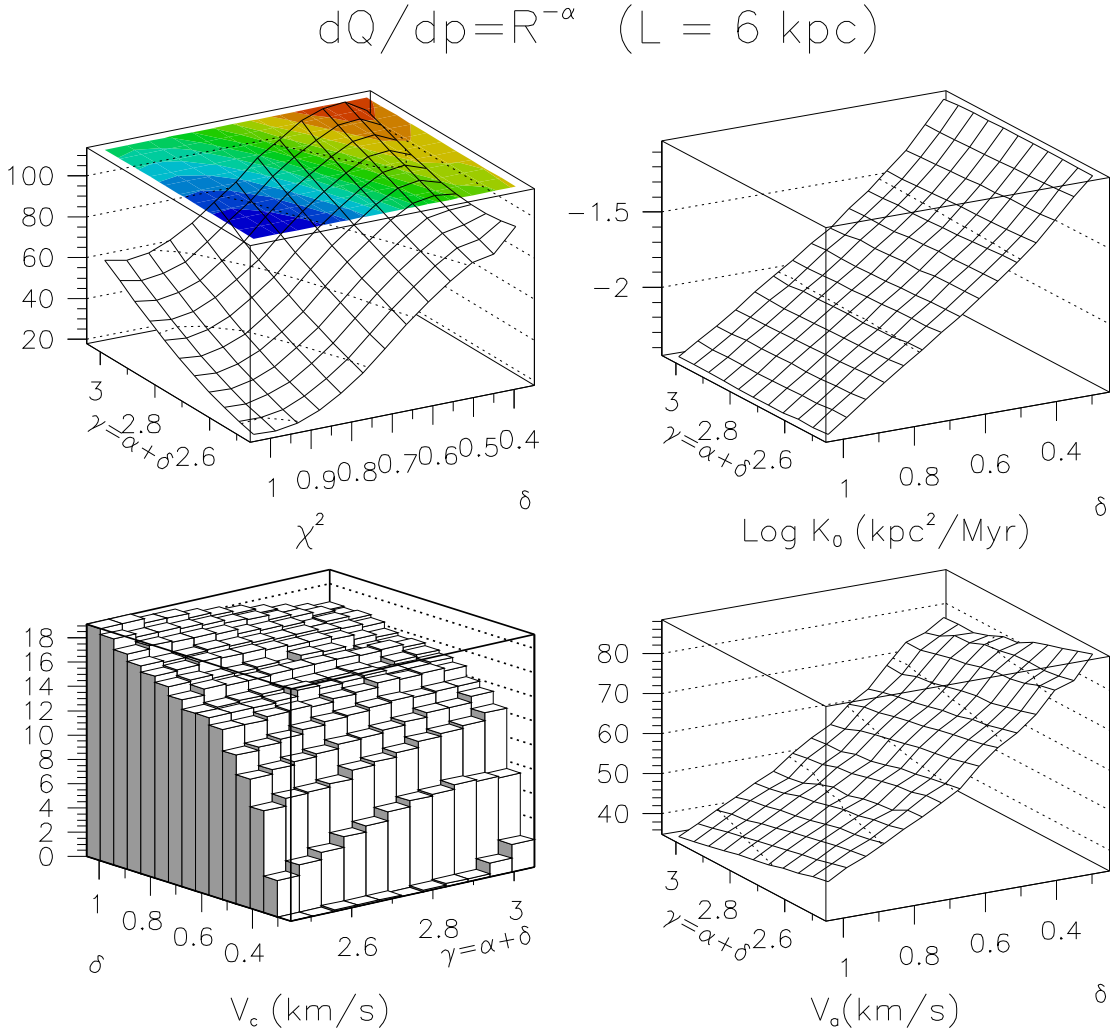}
          \caption{From top to bottom: for each best $\chi^2$ in the plane
          $\delta-\gamma$
          ($L=6$ kpc), the corresponding values of $\log(K_0)$, $V_c$ and $V_a$
          are plotted for both source spectrum types.}
          \label{f3_stables}
\end{figure*}

\begin{figure*}[hbt!]
\includegraphics*[width=\textwidth]{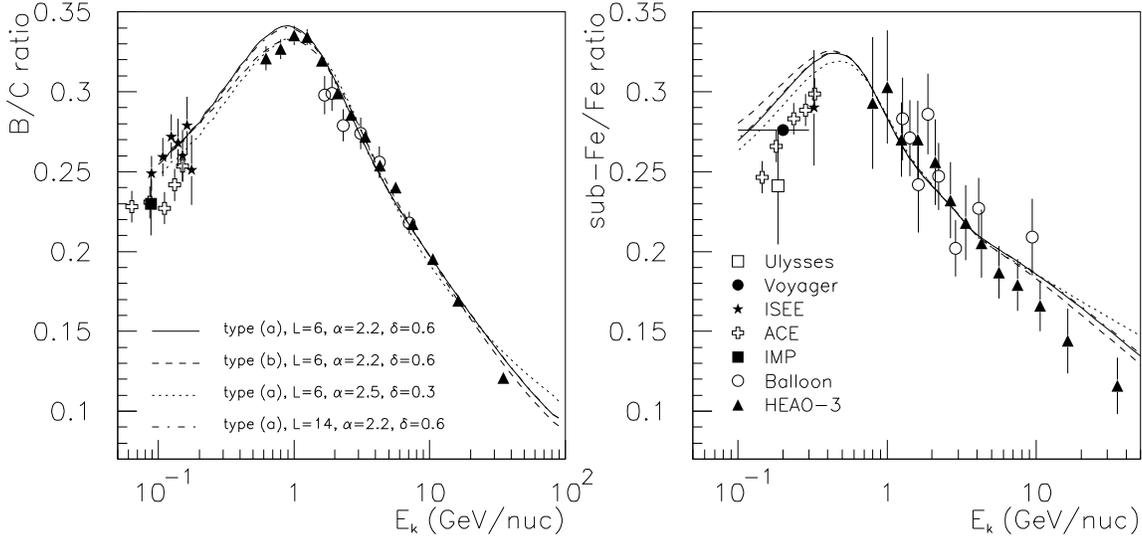}
\caption{The B/C and sub-Fe/Fe spectra (modulated at $\phi=500$~MV)
for several sets of parameters (giving
the best fit to B/C for these values) are displayed, along with
experimental data
from {\sc heao}-3~\cite{Engelmann}, balloon flights~\cite{Dwyer},
{\sc heat} on {\sc ulysses}~\cite{DuVernois_Thayer},
{\sc hkh} on {\sc isee}-3 (Leske~\cite{reac_traceur})
and {\sc voyager}~\cite{Webber_flan}. Notice
that {\sc ace} data~\cite{Davis} correspond to a modulation parameter
$\phi\approx750$~MV.}
\label{BC_et_subFe}
\end{figure*}

\paragraph{Summary of the trends}
In Fig.~\ref{f3_stables} we show the preferred values of the three
diffusion parameters $K_0$, $V_c$ and $V_a$, for each best $\chi^2$ in the
$\delta-\gamma$ plane, when $L$ has been fixed to 6 kpc (the behavior
does not particularly depend on $L$).
The two upper panels show that the evolution of $\alpha$ does not affect $K_0$.
On the other hand, we clearly see the (anti)correlation
between the two parameters $K_0$ and $\delta$ entering the diffusion
coefficient formula, because they need to give about the same
normalization
at high energy ($K_0\times E_{\rm thresh}^\delta\approx cte$).
Almost the same numbers are obtained for the pure power law and for the
modified source spectra (these were defined in Sec.~\ref{subsubsec:sources}).
$K_0$ spans the range 0.003 to 0.1 kpc$^2$~Myr$^{-1}$.
The middle panels show the values for the convective velocity. Only very
few configurations include $V_c=0$~km~s$^{-1}$, always when $\delta=0.3$, for
both types of source spectra. Increasing $\gamma$ and $\delta$ at
the same time makes $V_c$ change its trend. Finally,
the Alfv\'en  velocity $V_a$ doubles from $\delta = 1.0$ to 0.3, whereas it
is almost unchanged by a variation in the parameter $\gamma$ (or
equivalently $\alpha$). B/C and sub-Fe ratios are displayed
for several configurations in Fig.\ref{BC_et_subFe}.

The three parameters $K_0$, $V_c$ and $V_a$
behave very similarly with respect to a change in the source spectrum
from ``pure power law" to ``modified".
It can be explained as the influence
on the primary and secondary fluxes which can be factored out if
energy changes are discarded (their effect is
actually small on the derived parameters).
The currently available data on B/C do not allow to discriminate
clearly between these
two shapes for the acceleration spectrum.
This could be achieved by means
of better data not only for B/C but also for primary nuclei (Donato
{\em et al.}\/, in preparation). We finally notice
that pursuing the analysis with the sub-Fe/Fe ratio does not allow to go
farther for reasons invoked above.

\paragraph{Conclusions}
Thus, forgetting for a while some of our theoretical {\em a priori}\/ about
the diffusion power spectrum, a new picture of cosmic ray propagation
seems to emerge, in which high values for the diffusion coefficient spectral
index ($\delta\gtrsim 0.6-0.7$)~\footnote{It is noticeable
that Strong {\em et al.}\/ that used and preferred $\delta=1/3$ for
the five past
years took $\delta=0.42-0.52$ in their recent study ({\tt astroph/0210480}).}
and source spectral indices $\alpha \sim 2.0$ are favored.
The conclusions of the full analysis can be
summarized as follows:
(i)  the values $\delta\sim 0.7-0.9$ and $\alpha\sim2.0$ are preferred;
(ii) this preference holds whatever the specific form of the spectrum at low
energy;
the numerical values of the other parameters are also only slightly
modified by this low energy dependence even though deviation from a power-law
at low energy is preferred.
The study of fluxes should give a more definite answer;
(iii) $K_0$ scales logarithmically with $\delta$ and models with small
halos tend to one-dimensional models with a simple relation between 
the surface mass density $\mu$, $K_0$, $L$ and $V_c$;
(iv) several existing models are compared and the qualitative and
quantitative differences
between them are studied and partially explained
(the reader is referred to~\cite{Maurin02} for a deeper discussion).

%%%%%%%%%%%%%%%%%%%%%%%%%%%%%%%%%%%%%%%%%%%%%%%%%%%%%%%%%%%%%%%%%%%%%%%%%%%%%%%
%%%%%%%%%%%%%%%%%%%%%%%%%%%%%%%%%%%%%%%%%%%%%%%%%%%%%%%%%%%%%%%%%%%%%%%%%%%%%%%

\section{Diffusion parameters applied to ``standard" CR physics}
\label{sec:standard_CR_physics}

Armed with our set of parameters derived from the B/C ratio,
it seems reasonable to suppose that other charged nuclei follow
the same propagation history; this is already the case for sub-Fe/Fe nuclei
as shown in Fig.~\ref{BC_et_subFe}.
We therefore use the same diffusion model, with these
sets of parameters, to study the propagation of all species.
In this section we focus on the subjects pertaining to the
astrophysics field.
We start with secondary $\bar{p}$~\cite{Donato02}
(Sec.~\ref{sub:sec_anti_p_d}) that is a further cross-check of the
validity of the propagation parameter derived. This nucleus is also
particularly important to possibly constrain exotic primary sources in
the diffusive halo~\cite{barrau01,Bottino_Salati}
-- this will be presented in Sec.~\ref{sec:astroparticle_physics} --.
We devote Sec.~\ref{subsec:origin_realistic_case}
to the question of the spatial origin of all these stable 
nuclei~\cite{Taillet01}.
Then we focus on the study of radioactive species~\cite{Donato01} in
Sec.~\ref{subsec:rad}, where some clues about  the local
properties of the interstellar medium will be given.
Finally, in Sec.~\ref{subsec:iap01} we study some
consequences of our model on the composition at higher 
energies~\cite{Maurin03}.

It should be reminded that a strong degeneracy between the
propagation parameters was pointed out in Sec.~\ref{subsec:res.from_B/C}.
Consequently, much of the following work has to deal with inspecting
carefully the consequences of this degeneracy, and if possible, to
break it.

\subsection{Secondary antiprotons and antideuterons}
\label{sub:sec_anti_p_d}

The secondary antiprotons and antideuterons are yielded by the
spallation of cosmic ray nuclei
over the interstellar medium, mostly protons and helium whose spectra
$\Phi_{\rm p}$ and $\Phi_{\rm He}$ are by now well determined \cite{Donato02}
as discussed in Sec.~\ref{subsubsec:p_He}.
The production cross sections have been discussed in
Sec.~\ref{subsubsec:sec_sources}.

	\subsubsection{Secondary antiprotons}
	\label{subsub:sec_anti_p}
Unlike the secondary nuclei that are produced at fixed energy per nucleon,
the source term for secondary antiprotons is obtained from the convolution of
the energy spectra of the incident cosmic rays with the relevant differential
production cross sections. In the case of impinging protons, this leads to
\begin{eqnarray*}
\lefteqn{q_{\bar{p}}^{\rm sec}(r,E_{\bar{p}}) = 4 \pi
{\int_{E_{\rm th}}^{+ \infty}}\, dE_{\rm p} \,
\Phi_{\rm p}(r,E_{\rm p})}\\
&& \times \left\{
n_{\rm H} \,
{ \frac{d \sigma_{\rm p H \to \bar{p}}}{d E_{\bar{p}}}}
       +  n_{\rm He} \,
{\displaystyle \frac{d \sigma_{\rm p He \to \bar{p}}}{d E_{\bar{p}}}}
\right\}
\left( E_{\rm p} \to E_{\bar{p}} \right)\;.
\label{sec_pbar_source}
\end{eqnarray*}
A similar term arises from the cosmic ray helium radiation.

\paragraph{Tertiary production}
Once they have been created, antiprotons may interact with the
interstellar material in three different ways.
First, they may undergo elastic scatterings on galactic hydrogen.
The cross section for that reaction has been shown to peak in the
forward direction~\cite{eisenhandler} so that the corresponding
antiproton energy loss is negligible. Antiprotons are not perturbed
by these elastic scatterings as they survive them while their energy
does not change.
Then, they may also annihilate on interstellar protons. This process
dominates at low energy, and its cross section $\sigma^{\bar{p}p}_{\rm ann}$
is given in \cite{TanNg83}.
Last but not least, antiprotons may survive inelastic scatterings
where the target proton is excited to a resonance. Antiprotons
do not annihilate in that case but lose a significant amount of their
kinetic energy. The cross section for these inelastic yet non-annihilating
interactions is
\begin{equation}
\sigma^{\bar{p}p}_{\rm non-ann} \; = \;
\sigma^{\bar{p}p}_{\rm inel} \, - \, \sigma^{\bar{p}p}_{\rm ann} \;\; .
\end{equation}
The energy distribution of antiprotons that have undergone such
reactions has not been measured. It may be assumed to be similar
to the proton energy distribution after p-p inelastic scattering. An
impinging antiproton with kinetic energy $T'_{\bar{\rm p}}$
has then a differential probability
\begin{equation}
{\displaystyle \frac{dN_{\bar{p}}}{dE_{\bar{p}}}} \; = \;
{\displaystyle \frac{1}{T'_{\bar{\rm p}}}}
\end{equation}
to end up with the final energy $E_{\bar{p}}$, hence the differential
cross section
\begin{equation}
{\displaystyle
\frac{d \sigma_{\bar{p} \, H \to \bar{p} \, X}}{dE_{\bar{p}}}}
\; = \; {\displaystyle
\frac{\sigma^{\bar{p}p}_{\rm non-ann}}{T'_{\bar{p}}}} \;\; .
\end{equation}
The corresponding source term for these so-called tertiary antiprotons
may be expressed as
\begin{eqnarray}
          \label{Belle_et_Sebastien}
          \lefteqn{q_{\bar{p}}^{\rm ter}(r , E) =-4\pi n_{\rm H}
          \sigma^{\bar{p}p}_{\rm non-ann}(E) \, \Phi_{\bar{p}}(r , E)}\\
          && +4 \pi  n_{\rm H}  \int_{E}^\infty
          \frac{\sigma^{\bar{p}p}_{\rm non-ann}(E')}{T'} \,
          \Phi_{\bar{p}}(r , E') \, dE'\;. \nonumber
\end{eqnarray}
The integral over the antiproton energy $E$ of $q_{\bar{p}}^{\rm ter}(E)$
vanishes. This mechanism does not actually create new antiprotons.
It merely redistributes them towards lower energies and tends therefore
to flatten their spectrum. Notice in that respect that the secondary
antiproton spectrum that results from the interaction of cosmic ray
protons impinging on interstellar helium is already fairly flat below
a few GeV. Since it contributes a small fraction to the final result,
the effect under scrutiny here may not be as large as previously thought
\cite{bergstrom99}.
Helium is taken into account by replacing
the hydrogen density in relation (\ref{Belle_et_Sebastien}) by a
geometrically-inspired factor of $n_{\rm H}+4^{2/3}\:n_{\rm He}$.

\begin{figure}[hbt!]
         \includegraphics*[width=\columnwidth]{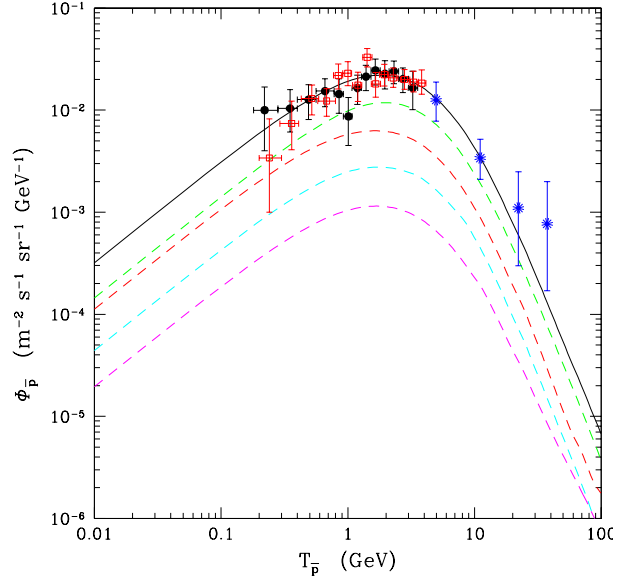}
          \caption{Solid line shows the total top-of-atmosphere (TOA)
          secondary antiproton spectrum
          for the reference set of diffusion parameters (see text for details).
          Dashed lines are the contributions to this total flux from various
          nuclear reactions (from top to bottom: p-p, p-He, He-p and He-He).
          In data points, circles correspond to the combined
	  {\sc bess}
          1995 and 1997 data~\cite{bess_pbar 1999}, squares to
          the 1998 ones~\cite{bess_pbar 2000} and stars to
	  {\sc caprice} data~\cite{caprice}.}
          \label{pbar_sec_f1}
\end{figure}

\paragraph{Conclusions}
First, the values of all the inputs being either extracted from the
analysis of nuclei (diffusion parameters $\delta$, $L$, $K_0$, $V_c$
and $V_a$) or measured
(proton and helium fluxes), all the cosmic antiproton fluxes naturally
coming out of the calculation are completely contained within the experimental
error bars of {\sc bess} data (see Fig.~\ref{pbar_sec_f1}).

\begin{figure}[hbt!]
\includegraphics*[width=\columnwidth]{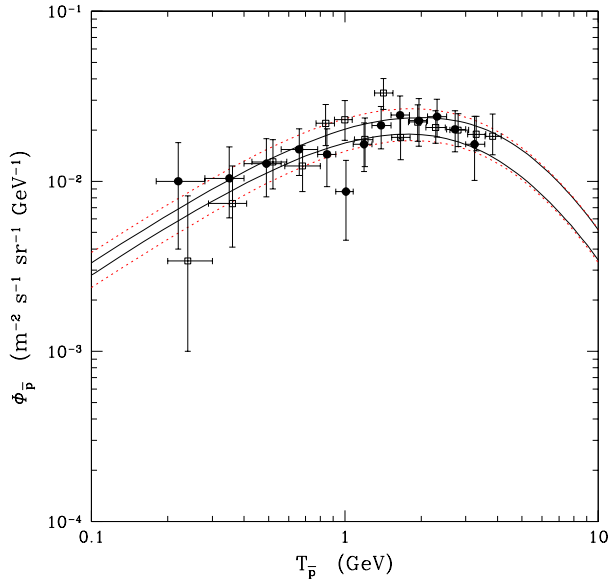}
\caption{Solid lines show the envelope of the TOA antiproton
spectra generated
with the sets of diffusion parameters consistent with B/C. The dotted lines
give an estimation of the uncertainty due to the indeterminacy of the nuclear
parameters.}
\label{fig:all_unc}
\end{figure}

The major uncertainties come from nuclear physics and are already
comparable to observational error bars. In Fig.~\ref{fig:all_unc},
the two dotted lines feature the uncertainties related to nuclear physics.
The upper curve is obtained with the set of maximal p-He, He-p, He-He
cross sections while increasing the p-p cross section -- as given by
the Tan \& Ng parameterization \cite{TanNg82,TanNg83} -- by a generous
10\%. Similarly, the lower curve is obtained with the minimal values
for those cross sections while decreasing the p-p cross section by 10\%.
Indeed, such a variation for p-p has been included for the sake of
completeness even if it modifies the antiproton spectrum only by a few
percents.  The uncertainties which the upper and lower dotted curves
exhibit are of the order of 22-25~\% over the energy range 0.1-100 GeV.
Another source of uncertainty comes from the fact
that the propagation parameters are degenerate, even if they
are severely constrained by the analysis of B/C experimental results.
This induces an error which is also displayed in Fig.~\ref{fig:all_unc} 
(solid lines).

As antiproton spectrum measurements
should better in the near future, antiproton studies could be limited
by nuclear indeterminacies. Further work and especially new measurements
of antiproton production in the p-He channel would be of great interest.

			%-----------------------

\subsubsection{Secondary antideuterons}

The production of secondary antideuterons also proceeds through the collision
of cosmic ray high-energy particles -- mostly protons and helium nuclei --
on the atoms of the interstellar medium.
The source term $q_{\bar{D}}^{\rm sec}(r,E_{\bar{D}})$
for secondary antideuterons can then be readily obtained, using the
coalescence scheme described in Sec.~\ref{subsub:dbar}.

The result is that the spectrum of secondary antideuterons
sharply drops below a few GeV/nuc (see Fig.~\ref{fig:dbar},
p.~\pageref{fig:dbar}), mostly because in the galactic frame, the
production threshold is 17 $m_{\rm p}$.
Energy losses tend to shift slightly the antideuteron spectrum towards
lower energies with the effect of replenishing the low-energy tail with the
more abundant species which, initially, had higher energies. That effect is
nevertheless mild. In particular, there is no tertiary contribution
$q_{\bar{D}}^{\rm ter}(r,E_{\bar{D}})$. An antideuteron that
undergoes an inelastic
scattering is likely to be destroyed unlike antiprotons for which inelastic
yet non-annihilating interactions are possible (see Sec.~\ref{subsub:sec_anti_p}).

As a consequence, the flux of secondary antideuterons is quite suppressed
with respect to antiprotons. However, a preliminary study, 
which does not take advantage of the full set of propagation parameters
allowed by B/C shows that a dozen events may be
collected by the {\sc ams} collaboration during its space station stage,
in the energy range extending up to 100 GeV/nuc.
These antideuterons correspond to IS energies in excess of $\sim 3$~GeV/nuc,
a region free from the effects of Solar modulation. This result takes into
account geomagnetic suppression as discussed in \cite{dbar_99}.

		%%%%%%%%%%%%%%%%%%%%%%%%%%%%%%%%

\subsection{Spatial origin}
\label{subsec:origin_realistic_case}

The question of the spatial origin of cosmic rays in our stationary diffusion
model was addressed in Sec.~\ref{sec:origin}.
It was shown that the origin is fully described by three
parameters $L$, $r_{\rm wind}$ and $r_{\rm spal}$
(see also Sec.~\ref{subsec:origine_la_base}), which depend on the
values of the propagation parameters.
Having determined a definite range of realistic parameters in
Sec.~\ref{sec:diff.param.determination},
we can go one step further in characterizing the origin of cosmic rays.

\begin{figure}[hbt!]
\centerline{
\includegraphics*[width=\columnwidth]{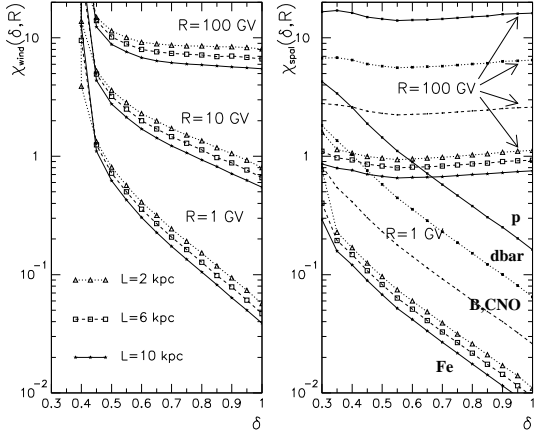}}
\caption{Left panel: $\chi_{\rm w}(\delta,{\cal R})$ as a function of the
diffusion spectral index $\delta$ for different rigidities ${\cal R}$
(it does not depend on species);
from top to bottom, ${\cal R}=100$~GV, ${\cal R}=10$~GV and ${\cal R}=1$~GV.
The parameter $\chi_{\rm w}$, as well as $\chi_{\rm spal}$, is not very
sensitive to the halo size $L$. Right panel: $\chi_{\rm spal}(\delta,{\cal R})$
as a function of $\delta$ for  ${\cal R}=100$~GV (upper curves) and
${\cal R}=1$~GV (lower curves) for four species: p ($\sigma_{\rm
inel}\sim 40$~mb),
$\bar{D}$ ($\sigma_{\rm inel}\sim 100$~mb), B-CNO ($\sigma_{\rm
inel}\sim 250$~mb) and
Fe ($\sigma_{\rm inel}\sim 700$~mb). For the latter species we
plotted the same three
$L$ values as in left panel.}
\label{fig:rw_rs_delta}
\end{figure}

\paragraph{Realistic values for $r_{\rm spal}$ and $r_{\rm wind}$}
In Fig.~\ref{fig:rw_rs_delta} are
plotted $\chi_{\rm w}(\delta,L) \equiv 2L/r_{\rm wind}$
and $\chi_{\rm spal}(\delta,L) \equiv L/r_{\rm spal}$ for several
energies and species. Actually, $\chi_{\rm w}(\delta,L)$ is a function
of rigidity; it is more clever to use the latter choice instead of kinetic
energy per nucleus, since $\chi_{\rm w}(\delta,L)$ then does not depend
on the species. The left panel of Fig.~\ref{fig:rw_rs_delta}
displays $\chi_{\rm w}(\delta,L)$ for
three rigidities: 1~GV, 10~GV and 100~GV.
The quantity $\chi_{\rm w}(\delta,L)$ is an indicator of the
competing role  of convection and diffusion to keep the remote cosmic rays
to reach Earth.
The figure shows that above several tens of GV,  diffusion has the
main role, whereas convection may dominate below this value,
at least for large values of $\delta$.
Different values of the halo size $L$ yield the same conclusions.
The right panel of Fig.~\ref{fig:rw_rs_delta} displays
$\chi_{\rm spal}(\delta,L)$ which is related to the spallation
efficiency. It appears that at low energy,
heavy nuclei such as Fe are preferentially destructed
rather than swept out by the convective wind or by the escape through
the boundaries.

\begin{figure}[hbt!]
\centerline{
\includegraphics*[width=\columnwidth]{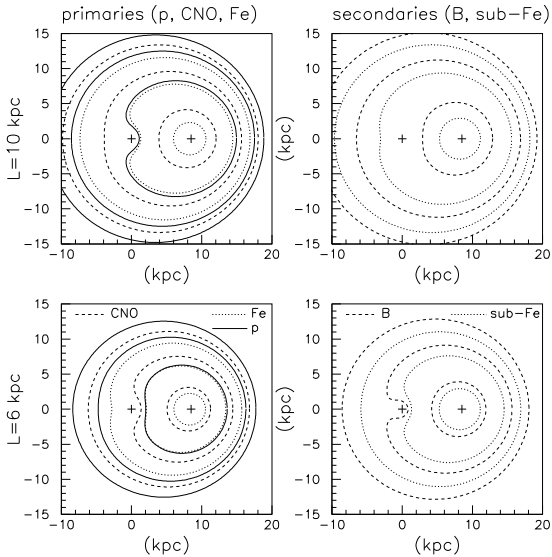}}
\caption{Contours from where 99\% of primaries (left panels) or
secondaries (right panels) come, for two halo sizes $L=10$~kpc
(upper panels) or $L=6$~kpc (lower panels). From external to internal
lines, and for each nucleus, these contours correspond to the
diffusion slope values $\delta=0.35-0.60-0.85$.}
\label{fig:origine_realiste}
\end{figure}

\paragraph{Origin of primaries and secondaries}
With the knowledge
of $r_{\rm wind}$, $r_{\rm spal}$ and $L$ at
1 GeV/nuc for all species one can now answer the question
of the origin of CR: results for p, CNO and Fe nuclei are shown
in Fig.~\ref{fig:origine_realiste}. The peculiar behavior of secondaries
is fully explained in~\cite{Taillet01}.

To summarize the situation, most cosmic rays detected on Earth were
emitted from a limited zone of the disc. This is even more true for
heavy species such as Fe, which are very sensitive to spallations and
thus are unlikely to travel long distances.
This implies that the information on the diffusive processes inferred
from the study of the ratio sub-Fe/Fe are only valid on a very
{\em local}\/ region, all the more that
the diffusion slope $\delta$
is large. Even if the propagation conditions were very
different outside of this region, the observations made in the Solar
neighborhood would almost not be affected, pointing
preferentially once more towards large $\delta$ values. Otherwise,
as several anomalies in some CR radiations indicate,  CR fluxes
could be variable with location, making all stationary models meaningless.

		%%%%%%%%%%%%%%%%%%%%%%%%%%%%%%%%

\subsection{The radioactive species}
\label{subsec:rad}

The radioactive cosmic ray nuclei deserve a specific treatment.
     From their creation, they diffuse on a typical distance $l_{\rm rad}\equiv
\sqrt{K\gamma \tau_0}$ before decaying.
In this expression, not only the diffusion coefficient $K$, but also
the lifetime $\gamma \tau_0$, depend on energy, due to the relativistic
time stretch. The following table gives some values of this distance
for three species (typical values $K_0 = 0.033
\unit{kpc}^2\unit{Myr}^{-1}$ and $\delta = 0.6$ have been assumed
in Tab.~\ref{tab:rad}).
\begin{table}[hbt!]
          \begin{center}
	\begin{tabular}{|c|c|cc|}   \hline
	    & $\tau_0$ (Myr) & $1 \unit{GeV/nuc}$ & $10
\unit{GeV/nuc}$ \\ \hline
	    $^{10}$Be & 2.17 & $220\unit{pc}$ &  $950 \unit{pc}$\\
	    $^{26}$Al & 1.31  & $110\unit{pc}$ & $470\unit{pc}$\\
	    $^{36}$Cl & 0.443 & $56\unit{pc}$ & $250\unit{pc}$ \\
	    \hline
	\end{tabular}
	\caption{Rest frame lifetimes and corresponding values of $l_{rad}$ for
	several nuclei at selected energies.}
          \label{tab:rad}
          \end{center}
\end{table}
These nuclei are therefore very sensitive to the presence of the
local bubble which has an extent of about $50-200$~pc
(see Sec.~\ref{subsub:ILM_LISM}).
As regards the diffusion process itself, as described by the
coefficient $K(E)$,
radio and $\gamma$-ray observations, which can test {\em in situ}\/
the spectrum and density of cosmic rays~\cite{LISMcoll.2},
indicate that it is not affected by the presence of the bubble, i.e.
diffusion is homogeneous.

The bubble has nevertheless an effect on the propagation.
First it leads to a decrease in the spallation source
term of the radioactive species.
Second, it also leads to a local decrease of destructive spallations.
Third, as there is less interstellar matter to interact with, the
energy losses are also lowered.
Because the typical propagation time of radioactive nuclei 
is short,
the energy redistributions are negligible, as well as destruction,
and the first effect is dominant.
As mentioned above, the bubble is modelled as a hole, in the thin
disc approximation (see Fig.~\ref{fig:schema1}, p.~\pageref{fig:schema1}).
The radius of this hole is considered as an unknown parameter in the analysis.
\begin{figure}[hbt!]
\includegraphics*[width=\columnwidth]{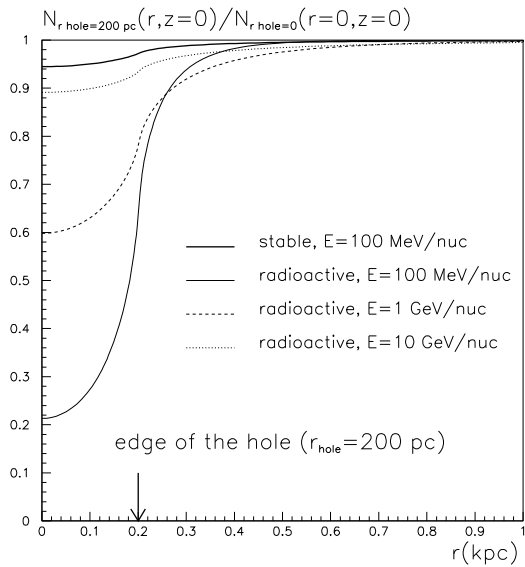}
\caption{Radial distribution of a radioactive species in the disc, across
the hole, for $r_{\rm hole}=200 \unit{pc}$. Numerical values were taken for
$^{26}$Al (radioactive) and $^{28}$Si (stable).
The distribution of radioactive is very sensitive to the presence of
the hole. However this effect is local and vanishes at several
$r_{\rm hole}$.
On the other side, the distribution of stable species is not much affected.}
\label{fig:hole}
\end{figure}
Fig.~\ref{fig:hole} shows, for $r_{\rm hole}=200$~pc,
how the fluxes both for stable and radioactive species are affected
in the neighborhood of the hole. The stables remain grossly unaffected
by its presence, whereas radioactive are strongly suppressed
for the reason just mentioned.

\begin{figure*}[hbt!]
          \centerline{\includegraphics*[width=\textwidth]{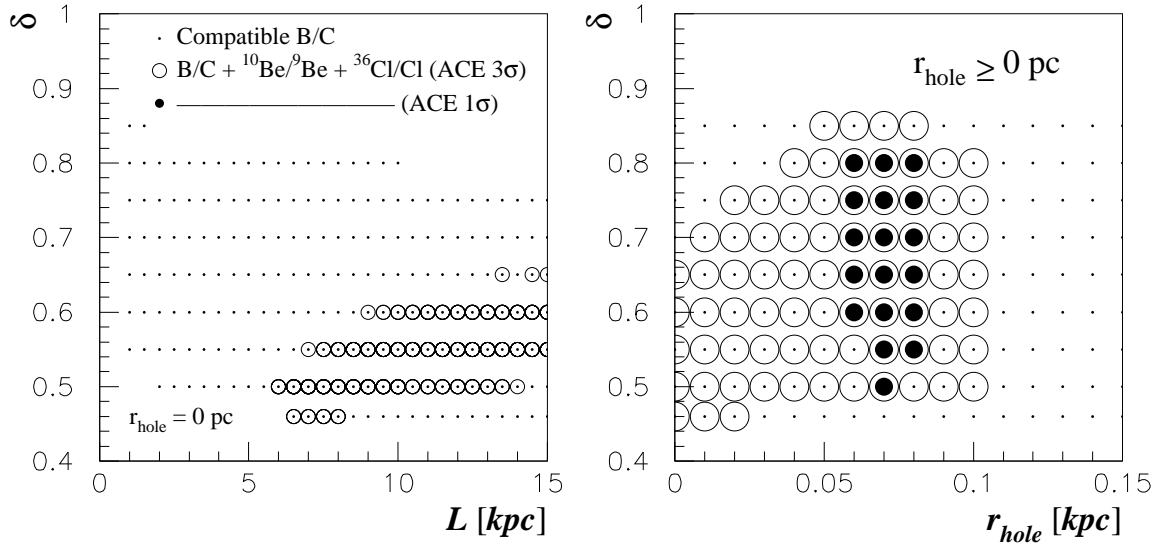}}
          \caption{Representation of the models compatible with B/C plus both
          $^{10}$Be/$^9$Be and $^{36}$Cl/Cl {\sc ace} 3-$\sigma$ (open
circles) and
          1-$\sigma$ (filled circles).
          Left panel displays homogeneous models ($r_{\rm hole}=0$) in the plane
          $L-\delta$.
          Right panel displays inhomogeneous models ($r_{\rm hole}\geq 
0$) in the
          plane $r_{\rm hole}-\delta$.}
          \label{fig:be_et_cl}
\end{figure*}

\paragraph{Effect and size of the hole}
The major result we found~\cite{Donato01} is that at the center of
the bubble, the radioactive fluxes are decreased by a factor which
can be approximated as
\begin{displaymath}
       \frac{N^{r_{\rm hole}}}{N^{r_{\rm hole}=0}}\propto \exp(-r_{\rm
hole}/l_{\rm rad})\;.
\end{displaymath}
When all the sets of diffusion parameters allowed by
the B/C data are used to
compute the $^{10}$Be/$^9$Be, $^{36}$Cl/Cl ratios,
we find that each of the radioactive nuclei independently points towards
a bubble of radius $\lesssim 100 \unit{pc}$, in relatively good agreement
with direct observations.
If these nuclei are considered simultaneously, only models with a bubble
radius $r_{\rm hole} \sim 60 - 100 \unit{pc}$ are consistent with the data.
In particular, the standard case $r_{\rm hole}=0 \unit{pc}$ is disfavored.
This is shown in Fig.~\ref{fig:be_et_cl}, which is a projection of
the parameter subspace allowed by B/C, $^{10}$Be/$^9$Be and $^{36}$Cl/Cl
on the $L-\delta$ plane (left panel, no hole) or $r_{\rm hole}-\delta$
plane (right panel, hole $r_{\rm hole}$).
When the ratio $^{27}$Al/$^{27}$Al is
considered along with the two others mentioned above, the results
become less clear~\cite{Donato01}, and it is suspected that the data 
(nuclear or astrophysical) on which they rely should not be trusted.

\begin{figure*}[hbt!]
\includegraphics*[width=\textwidth]{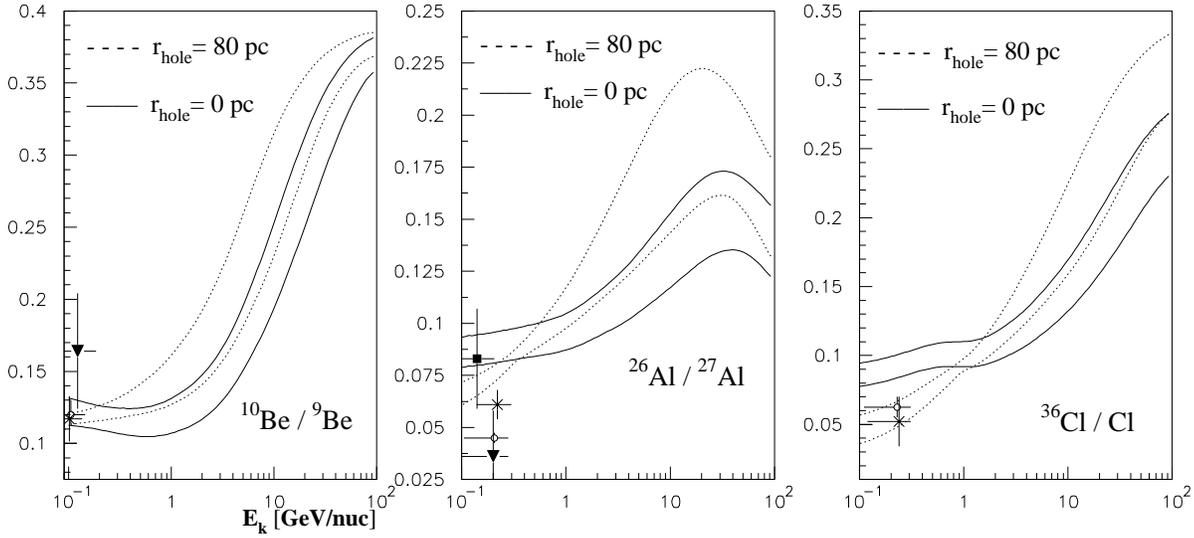}
\caption{Envelopes of the spectra obtained with all the models
compatible with $^{10}$Be/$^9$Be {\sc ace} 1-$\sigma$ for the three
radioactive species.
Solid lines are for homogeneous models ($r_{\rm hole}=0$ pc) and dashed lines
are for inhomogeneous models ($r_{\rm hole} = 80$ pc).
Data are from {\sc ace} (circles), {\sc ulysses}
(crosses), {\sc voyager} (filled squares) and {\sc isee}
(see Sec.~\ref{subsub:Radioactive_species}).}
\label{fig:red_spectre}
\end{figure*}

\paragraph{Whole energy spectrum}
As $^{10}$Be/$^9$Be is probably the best measured
radioactive ratio (see Sec.~\ref{subsub:Radioactive_species}),
it is interesting to see how ratios depend on energy, given
the constraint that parameters must fit both B/C and
$^{10}$Be/$^9$Be data. This is shown in Fig.~\ref{fig:red_spectre}
for the homogeneous model and a hole $r_{\rm hole}=80$~pc.
These curves show that it would be very important to have accurate
measurements of  the energy spectra of radioactive species such as
$^{10}$Be, $^{36}$Cl  and $^{27}$Al.

\paragraph{Conclusions}
Most studies use the radioactive nuclei to constrain the
halo size $L$ (see discussion in Sec.~\ref{subsubsec:halo_et_la}).
The meaning of the results are very much dependent on the
the treatment of the LISM. In particular, the presence of the local
bubble leads to an
exponential attenuation of the radioactive fluxes.
This is of special importance for short-lived nuclei such as
$^{14}$C, which is attenuated by a factor $\ll 1$, unless a local
source is present. This may explain the fact that {\sc ace} did not
detect any such nucleus.
Finally, if one takes $r_{\rm hole}$ as an input fixed by observations,
the radioactive progenitor/parent may give further constraints on the
diffusion parameters.

		%%%%%%%%%%%%%%%%%%%%%%%%%%%%%%%%%%%%

\subsection{PeV fluxes}
\label{subsec:iap01}

Recent measurements of the cosmic ray average logarithmic mass and
all-particle spectrum  around $10^{15}$~eV~\cite{Apanasenko,Shirasaki}
gave new clues to understand the origin of the cosmic rays
       and in particular the puzzle of the knee in the energy spectrum.
       The highest energy particles
are almost certainly extragalactic. A similar origin is not excluded
near the knee, but it is difficult to account for the observed softening
of the spectrum in this region. As a
consequence, the intermediate region between $10^{15}$ and
       $10^{19}$~eV should be analyzed in terms of the same physical
       mechanisms than lower energy particles.

At these energies, fluxes are too low to be directly measured
($\sim$ m$^{-2}$ sr$^{-1}$ yr$^{-1}$). With present techniques, only
two quantities can be extracted in large ground array detectors
(e.g.~\cite{Antoni1}); namely all-particle spectrum and
$\langle \ln A\rangle$. They are given by linear combination of
the individual fluxes with different weights;
\begin{equation}
\Phi^{\rm all} \equiv \sum_j \Phi_j
\;\;\; \mbox{and} \;\;\;
\langle \ln A \rangle \equiv \frac{\sum_j \ln A_j \Phi_j}{\sum_j \Phi_j}\;.
\label{maurin_et_al:lnA}
\end{equation}
As the experimental data are given in total energy
rather than in kinetic
energy per nucleus, we adopt the same presentation for the results
above 100~GeV.

			%###############%

\subsubsection{Separation of key ingredients}
We first estimate the evolution of $\langle\ln
A\rangle$  in the Leaky Box frame.
This allows to apprehend more easily some basic ingredients that drive
the mass evolution, but not all of them.
In particular, the geometrical effects cannot be considered.
Trends are also more easily understood with only one light and one
heavy nucleus (e.g. p and Fe).
If energy gains and losses are discarded, the Leaky Box equation
for primary species, such as p and Fe, reads
\begin{equation}
          -\frac{N^j}{\tau_{\rm esc}}+ \bar{q}^j_0Q^j(E)
          -\Gamma^j  N^j =0\;.
          \label{maurin_et_al:leaky-box}
\end{equation}
The proportion of light and heavy species evolves with
energy because of two effects.
The source spectra $\bar{q}^j_0Q^j(E)$ may differ, and the spallative
destruction rate $\Gamma^j$ are definitely not the same.
The escape time $\tau_{\rm esc}=\tau_0 {\cal R}^{-\delta}$ is supposed to be
independent of species.

\paragraph{Source spectrum effect}
The first effect is studied by neglecting the spallation term in
Eq.~(\ref{maurin_et_al:leaky-box}).
We suppose that the two species p and Fe are injected with different slopes
related by $\alpha_{\rm Fe}-\alpha_{\rm p}=0.1$~\cite{Wiebel-SoothI}.
If we start from the experimental value $\left({\rm Fe}/{\rm
p}\right)_{100~\rm GeV}=1/20$
(e.g.~\cite{Wiebel-SoothII}),
Eq.~(\ref{maurin_et_al:lnA}) gives the following evolution,
independently of $\delta$,
\begin{eqnarray*}
          \lefteqn{\langle\ln A \rangle (E={\rm 100\:GeV /\: 10\:TeV
/\: 1\:PeV/\:
          \infty)}} \\
          && \sim 0.19\: / \: 0.30 \: / \:0.50\: / \:4.03\;.
          \label{maurin_et_al:result1}
\end{eqnarray*}

\paragraph{Escape plus selective destruction}
If the spallations are taken into account,
Eq.~(\ref{maurin_et_al:lnA}) can be written as
\begin{eqnarray}
              \lefteqn{\langle\ln A \rangle (E)=\ln (A_{\rm Fe})\times\left[
              1+
              (q^0_{\rm p}/q^0_{\rm Fe}) \times
               \left(\frac{Z_{\rm p}}{Z_{\rm Fe}}\right)^{\alpha}
               \right. }
               \nonumber \\
              && \left.  \times
              \frac{(\sigma_0(E/Z_{\rm Fe})^{\delta}+
          \sigma^{\rm inel}_{\rm Fe})}{(\sigma_0(E/Z_{\rm p})^{\delta}+
          \sigma^{\rm inel}_{\rm p})}\right]^{-1}\;.
\label{maurin_et_al:Pure_propagation}
\end{eqnarray}
where the cross sections are $\sigma^{\rm inel}_{\rm p}\sim 30$
mb and $\sigma^{\rm inel}_{\rm Fe}\sim 710$ mb).
The quantity $\tau_{\rm esc}$ has been converted in mb, with a
typical value $\sigma_0=47$~mb and $\delta=0.61$~\cite{Webber_spal98}.
We are not interested here in the effect of injection spectra and the slopes
have now been assumed to  be the same for all species ($\alpha_{\rm p}=
\alpha_{\rm Fe}\equiv \alpha$=2.2).
With  $\langle\ln A \rangle (100~\rm GeV)= 0.19$, this gives
\begin{eqnarray*}
          \lefteqn{\langle\ln A \rangle (E={\rm 100\:GeV /\: 10\:TeV /\: 1\:PeV
          \:/\: \infty)}}\\
          && \sim 0.19\: / \:0.83 \: / \:1.05 \: / \:1.08\;.
          \label{maurin_et_al:result2}
\end{eqnarray*}

			%--------------%

\paragraph{Geometrical source effect}
As emphasized above, the study of the influence of geometry is not
straightforward in Leaky Box models.
      From the previous discussion in the framework of diffusive
propagation (see Sec.~\ref{sec:origin}), we know that
cosmic rays detected on Earth come from an average
distance $\langle r\rangle(E)$ which is larger at higher energy.
As a metallicity gradient is present, at the level of
$\sim -0.05$~dex~kpc$^{-1}$, i.e.
$[{\rm Fe}/{\rm H}]=-0.05\: (r-R_{\odot}) \equiv
\log \left(q_{\rm Fe}/q_{\rm p}\right)^{\rm eff}
-\log ({\rm Fe}/{\rm H})_{\odot}$
(see Sec.~\ref{subsubsec:sources}), the nuclei with different energies
were emitted from sources having different properties.
This leads to (the two previous effects are ignored)
\begin{displaymath}
          \langle\ln A \rangle (E)= \frac{\ln (A_{\rm Fe})}{1+
          \left(q_{\rm p}/q_{\rm Fe}\right)^{\rm eff}}\;~;
\end{displaymath}
where
\begin{displaymath}
          \left(\frac{q_{\rm p}}{q_{\rm Fe}}\right)^{\rm eff}=
          \left(\frac{q_{\rm p}}{q_{\rm Fe}}\right)_{\odot}\times
          10^{0.05\:(r-R_\odot)}\:.
\end{displaymath}
An upper bound is obtained for this effect by assuming that
$\langle r\rangle(100~{\rm GeV})=R_\odot$ and
$\langle r\rangle(\infty)=0$ (galactic center), which yields
\begin{equation}
          \langle\ln A \rangle (E={\rm 100\:GeV /\:\infty)}
          \sim 0.19\: / \: 0.54\;.
          \label{maurin_et_al:result3}
\end{equation}
Even in this optimal case, this effect is seen to be negligible compared
to the two others (see below).
			%###############%

\subsubsection{Results from our propagation model}

\shortversion{%% version longue
The quantity $\langle\ln A \rangle$ can also be computed
with our propagation model, considering the different sets of
propagation parameters derived from the B/C analysis
(see Sec.~\ref{subsec:res.from_B/C}).
}

\begin{figure}[hbt!]
          \includegraphics*[width=\columnwidth]{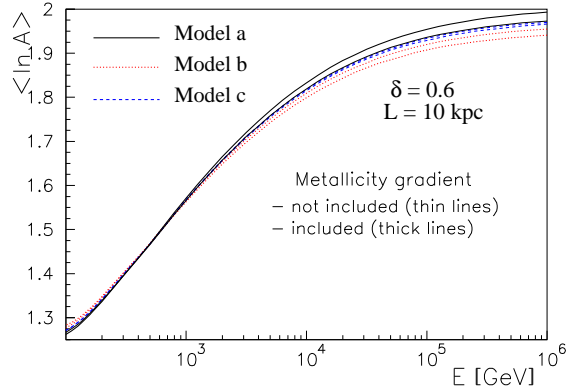}
          \caption{Average logarithmic mass $\langle \ln A\rangle$
          for the radial distributions depicted in
          Sec.~\ref{subsubsec:sources}, p.~\pageref{subsubsec:sources}
          ($\delta=0.6$ and $L=10$ kpc for all models).}
          \label{fig:michel_elisa}
\end{figure}

\paragraph{Geometrical effects are subdominant}
Fig.~\ref{fig:michel_elisa} shows
spallations effects
and geometrical effects.
The first strong conclusion is that spallation effects affect
dramatically the composition of cosmic rays.
Furthermore, at sufficiently high energy, as expected with the Leaky Box toy
model, the asymptotic
regime is reached and propagation ceases to affect $\langle \ln A\rangle$.
The geometrical effects (source distribution plus metallicity
gradient) are less important than the others.
They induce a change of at most 5\% in the results.
This would be even more true for lower $L$.
For $L=3$~kpc, these geometrical effects are completely negligible.

\begin{figure}[hbt!]
          \includegraphics[width=\columnwidth]{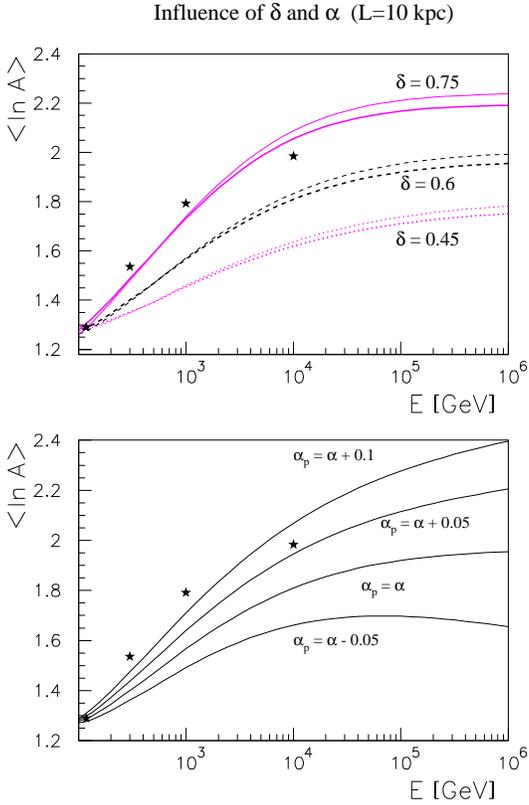}
          \caption{For different sets of parameters compatible with the
          B/C ratio, the average mass $\langle\ln A\rangle(E)$ normalized  to
          100 GeV observations is displayed as a function of energy.
          Upper panel: the diffusion slope $\delta$ is varied keeping
          $\alpha_j\equiv\alpha$ for all species (thin lines, no geometrical
          effects; thick lines, geometrical effects are included). Lower panel:
          a different source slope is considered for p ($\delta$ is set to 0.6).
          Stars show experimental data~\cite{Apanasenko}
          (see~\cite{Maurin03} for details).}
          \label{fig:pev2}
\end{figure}

\paragraph{Evolution above PeV energies}
The evolution of $\langle \ln A\rangle$ obtained with different sets
of parameters compatible with B/C is presented in Fig.~\ref{fig:pev2}.
The right panel focuses on the source effect and shows that
$\alpha_{\rm p}>\alpha$ is preferred. As suggested in our toy model, only
source effects enable evolution of $\langle\ln A\rangle$ above 1~PeV.
It can also be seen that large diffusion slopes $\delta$ are
preferred, as was already hinted in Sec.~\ref{subsec:res.from_B/C}.

\begin{figure}[hbt!]
          \includegraphics[width=\columnwidth]{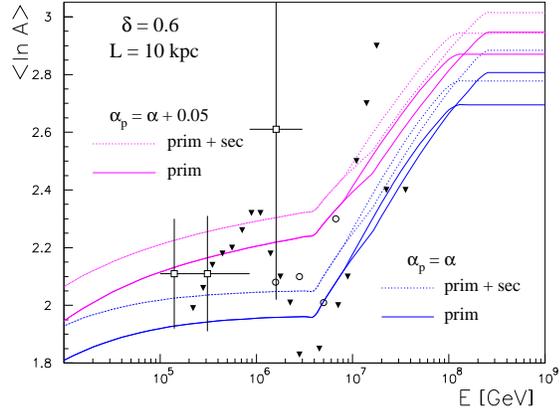}
          \caption{Average logarithmic mass for models including a break either
          in rigidity ($R=4$~PeV) or in total energy per nucleus
(E/nuc~$=4$~PeV).
          Solid lines correspond to primaries only, whereas dotted lines
          correspond to primaries plus secondaries.
          For illustrative purpose, some data from ground
          arrays have been displayed: triangles are {\sc casablanca}'s data with
          Monte-Carlo {\sc hdpm}~\cite{Fowler}, empty circles are from
{\sc kascade}
          collaboration with {\sc qgsjet} simulation~\cite{Antoni1} (empty
          squares are from {\sc jacee} direct experiment); {\sc runjob}
data plus some
          {\sc jacee} data are lower than 1.8 and do not appear on the graph,
          see~\cite{Apanasenko}).
          }
          \label{fig:blurpfinal}
\end{figure}

\paragraph{Inclusion of the knee}
An important experimental fact, referred to as the {\em knee}\/,
has not be taken into account in the analysis presented above.
It is a drop of the slope of $\Delta\gamma\sim0.4$
in the observed spectrum of cosmic rays
at an energy of a few PeV.
This may be due to a change in the source spectra or in the diffusion
coefficient.
We now want to study the effect of the knee on the evolution
of $\langle\ln A\rangle$.
To this aim, we propose to model the knee  in a very naive way, by a
break in  the source spectral indices at an energy $E_{\rm knee}$
that depends on the species (notice that a change
in diffusive regime, e.g. as recently re-inspected in~\cite{Candia},
is also possible).
We consider two cases, a break at a given rigidity $E_{\rm knee}=Z\times4$~PeV
and a break at given energy per nucleon $E_{\rm knee}=A\times4$~PeV.
The resulting average mass $\langle \ln A\rangle$ is shown in
Fig.~\ref{fig:blurpfinal} (the all particle flux
$\Phi^{\rm all}$ can be found in~\cite{Maurin03}).
Several features are visible.
They are first generated by protons changing their spectral index,
followed by others: the helium component, the CNO group, and Fe.
At high energy, all species have the same post-knee spectral index.
The transition is even more gradual when secondary species (upper curve)
are taken into account (they also change the normalization).
The choice of the break (in rigidity or in energy per nucleon)
affects mostly the  all-particle spectrum.
The conclusion at this stage is that the scatter in the data is too
large to provide a clear test of our theoretical predictions
(see Fig.~\ref{fig:blurpfinal}).
		%---------------------------------%
		%---------------------------------%

%%%%%%%%%%%%%%%%%%%%%%%%%%%%%%%%%%%%%%%%%%%%%%%%%%%%%%%%%%%%%%%%%%%%%%%%%%%%%%%%%%%%%
%%%%%%%%%%%%%%%%%%%%%%%%%%%%%%%%%%%%%%%%%%%%%%%%%%%%%%%%%%%%%%%%%%%%%%%%%%%%%%%%%%%%%

\section{Propagation models applied to astroparticle physics}
\label{sec:astroparticle_physics}
Connections between high-energy cosmic ray physics and cosmology are
explored in these three sub-sections. We start by a general
introduction and illustrate the above-mentioned ties by a discussion
(i) on the limits on the cold invisible gas~\cite{salati96}, (ii)
on the baryon asymmetry in the universe (see e.g.~\cite{chardonnet99})
and finally (iii) on the presence of putative
neutralinos that could make up the
astronomical dark matter~\cite{chardonnet96}
or primordial black holes that could have been formed in the early Universe.
Then, in Sec.~\ref{subsec:origine_primaires_halo} we explore the question of
the spatial origin of these exotic nuclei.
Finally, Sec.~\ref{subsec:exotic_susy} and Sec.~\ref{subsec:pbh}
respectively deal with SUSY~~\cite{Bottino_Salati,dbar_99}
and PBH~\cite{barrau02,barrau01} signatures in
antiprotons and antideuterons.
		%---------------------------------%
		%---------------------------------%

\subsection{Introduction}
\label{subsection_7_1}

Ties are strong between high-energy cosmic ray physics and cosmology.
In this section, we explore a few examples that illustrate these connections
and show that a deep understanding on how cosmic rays propagate turns out
to be crucial.

%Dark Matter
A long standing problem in astronomy lies in the existence of large amounts
of unseen material whose gravitational effects are nevertheless large.
To commence, the Milky Way is surrounded by an extended halo that induces
a flat rotation curve in its plane \cite{ALBADA_SANCISI}. This trend has
also been observed in many spiral systems. On larger scales, the presence
of dark matter inside galactic clusters has been noticed since many
decades \cite{OORT}. Finally, the observations of the Cosmic Microwave
Background (CMB) anisotropies \cite{boomerang}, combined either with the
determination of the relation between the distance of luminosity and the
redshift of type Ia supernovae \cite{supernovae_omegaL}, or with the large
scale structure (LSS) information from galaxy and cluster surveys \cite{2dF},
give independent evidence for a cosmological dark matter density in the range
$\Omega_{\rm \, cdm} \, h^{2} = 0.13 \pm 0.05$ \cite{boomerang},
to be compared to a baryon density of
$\Omega_{\rm b} \, h^{2} = 0.019 \pm 0.002$ as
indicated by nucleosynthesis
\cite{nucleosynthesis} and the relative heights of the first acoustic peaks
in the CMB data, $\Omega_{\rm b} \, h^{2} = 0.022 \pm 0.003$~\cite{archeops}.
The nature of the astronomical dark matter is still
unresolved and most of it could be made of non-baryonic species
or primordial black holes. But observations also point towards
the presence of dark baryons insofar as the amount of luminous material
is significantly smaller than $\Omega_{\rm b}$.

%Dark Gas inside the Milky Way and cosmic ray propagation.
\paragraph{Limits on the Dark Gas inside the Milky Way}
As a dark matter of fact, De Paolis {\em et al.}\/~\cite{DE_PAOLIS}
have outlined a scenario in which dark clusters of compact objects pervade
the halo of the Milky Way together with clouds of molecular hydrogen, at
distances larger than 10 to 20 kpc. Pfenniger {\em et al.}\/~\cite{COMBES1}
have also suggested that a flat rotation curve beyond the Solar circle
could be explained by a thin disc of cold molecular hydrogen,
widening at large distances from the center. If unseen gas was present
in the galactic halo or in the ridge of the Milky Way, it would be impacted
by cosmic rays originating from the disc. This would lead to a strong
$\gamma$-ray signal showing up as a new component in the galactic
diffuse radiation.
By using the {\sc cosB} results at high galactic latitude, Gilmore
\cite{GILMORE}
has already inferred a conservative limit $\sim$ 15 \% on the amount
of hidden gas in the halo.
That analysis can be significantly refined because a key ingredient --
the propagation of cosmic rays -- is now better understood. A reliable
calculation of their density as a function of galactocentric radius
$r$ and height $z$ above the plane is definitely possible, as
discussed in the previous sections.
It allows
for the determination of the gamma ray hydrogen emissivity $I_{H}$ per
hydrogen atom as a function of location
\begin{displaymath}
          I_{H}(E_{\gamma} , r , z) \; = \;
          \left\{ \frac{N(r , z)}{N(r_{\odot} , 0)} \right\} \,
          I_{H}(E_{\gamma} , \odot) \,\, ,
\end{displaymath}
where the spectral index of cosmic ray protons -- the dominant
contribution -- is found to remain roughly constant throughout
the Galaxy. In the local range, the emissivity is
\begin{eqnarray*}
          \lefteqn{
          I_{H}(E_{\gamma} > 100 \unit{MeV}, \odot)  =}\\
          && (1.84 \pm 0.10) \times 10^{-26}
          \; {\rm photons \, H^{-1} \, s^{-1} \, sr^{-1}}
\end{eqnarray*}
The spallation of cosmic rays with the gas potentially concealed in the
Milky Way produces an extra $\gamma$-ray diffuse emission whose flux
obtained from the convolution along the line of sight of the density
$n_{\rm H}^{\rm dark}$ of dark gas with the local $\gamma$-ray emissivity
$I_{H}$ is
\begin{equation}
\Phi_{\gamma}^{\rm extra}(E_{\gamma}) \; = \;
{\displaystyle \int} \,
I_{H}(E_{\gamma} , s) \; n_{\rm H}^{\rm dark}(s) \; ds\,\, .
\end{equation}
%
%
%%%%%%%%%%%%%%%%%%%%%%%%%%%%%%%%%%%%%%%%%%%%%%%%%%%%%%%%%%%%%%%%%%%%%%%%%%%%%%%%
%%
\begin{figure}[hbt!]
\includegraphics[bb=33 84 560 760,clip,width=0.85\columnwidth]{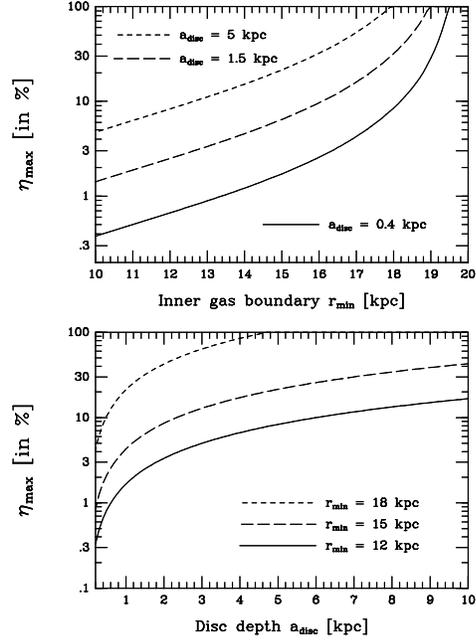}
          \caption{A Mestel's disc generates the flat galactic rotation
          curve. The central part of that disc, up to the radius $r_{\rm min}$,
          does not contain any cold gas so far undetected in CO. In the
left panel,
          the inner gas boundary $r_{\rm min}$ is varied for various
values of the
          disc thickness $a_{\rm disc}$. Conversely, in the right
panel, the bound
          $\eta_{\rm max}$ is presented as a function of
          $a_{\rm disc}$~\cite{salati96}.}
          \label{section_7_1a}
\end{figure}
%%%%%%%%%%%%%%%%%%%%%%%%%%%%%%%%%%%%%%%%%%%%%%%%%%%%%%%%%%%%%%%%%%%%%%%%%%%%%%%%
%%
%
Observations collected by the {\sc egret} instrument on board
{\sc cgro} have been compared to the $\gamma$-ray diffuse radiation
that may be modeled with both HI and H$_{2}$ contributions.
An isotropic component is found in the residuals of the signal.
Its integral flux between 500 MeV and 1 GeV is found to be
$2.5 \times 10^{-6}$ photons cm$^{-2}$ s$^{-1}$ sr$^{-1}$
in the direction of the Perseus arm
(b~=~0$^{\circ}$ , l~=~115$^{\circ}$). By requiring that
$\Phi_{\gamma}^{\rm extra}$ does not exceed that value,
a constraint may be derived on the amount of cold molecular clouds
-- so far untraced by CO -- which the Milky Way ridge could
conceal. A Mestel's disc with local surface mass density
$\Sigma(\odot) \sim$ 200 ${\rm M}_{\odot}$ pc$^{-2}$ is assumed.
The spatial density of that distribution is given by
\begin{equation}
\rho(r) \; = \; \left( \frac{\Sigma(\odot)}{a_{\rm disc}} \right)
\left( \frac{r_{\odot}}{r} \right) \,\, ,
\end{equation}
where $a_{\rm disc}$ denotes the disc thickness.
Fig.~\ref{section_7_1a} features the upper limit $\eta_{\rm max}$
on the fraction of cold gas which the outer parts of the disc
($r > r_{\rm min}$) may contain. In particular, for $a_{\rm disc} = 400$ pc,
the limit relaxes to 1 only if $r_{\rm min}$ exceeds 19.5 kpc.
We therefore conclude that molecular hydrogen, sufficiently cold
to have escaped detection and so far untraced in CO, should actually be
sparse up to $\sim$ 20 kpc from the galactic center and that
it cannot account for the rotation curve of the Milky Way.

\paragraph{Neutralino as a Dark Matter Candidate, PBH as a cosmological
remnant}
One of the favorite candidates to the astronomical missing mass
is a neutral weakly interacting particle. Such a species is predicted
in particular by supersymmetry, a theory that is actively tested at
accelerators. There are indirect clues for supersymmetry, noticeably
the existence of a single high-energy unification scale for
supersymmetric grand-unified models alone. It is conceivable
therefore that most of the dark matter in the halo of the Milky Way
is made of such neutral particles. The relic abundance of these so-called
neutralinos is relevant to cosmology. If present in the galactic halo,
they should still annihilate mutually to yield, among a few other
indirect signatures, a flux of antiprotons whose spectrum is discussed
in Sec.~\ref{subsec:exotic_susy}.

%Primordial Black Holes as Dark Matter Candidates
The same line of argument applies if black holes make up a fraction of
the galactic dark matter. Primordial black holes could have formed in the
early universe and
could have concealed baryons that are not accounted for by primordial
nucleosynthesis since they would not have participated in the nuclear fusion
of helium. The actual value for $\Omega_{\rm b}$ 
could therefore be much larger
than $\sim$ 5\% and even reach $\Omega_{\rm \, cdm}$, sparing us the need
for non-baryonic particles. If the black holes are light, they could
evaporate today and yield also antiprotons as is discussed in
Sec.~\ref{subsec:pbh}.

%The secondary flux is important.
The primary antiprotons produced by neutralino annihilations or
black hole evaporations will get mixed with a conventional
population of secondary antiprotons that are produced by the spallation of
cosmic ray nuclei on the interstellar gas of the Milky Way ridge. It is
crucial to investigate in some detail this later mechanism in order
to ascertain which of the primary or secondary species is dominant.
This has been done in Sec.~\ref{sub:sec_anti_p_d}.

\begin{figure}[hbt!]
       \includegraphics[width=\columnwidth]{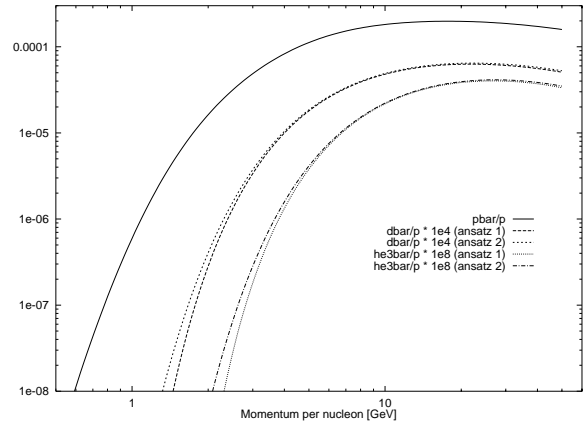}
       \caption{The fluxes of cosmic ray antiprotons and of anti-deuterium and
       anti-tritium nuclei, relative to the proton flux, are presented
       as a function
       of the momentum per nucleon~\cite{Orloff}. To fit on the same
       diagram, the curves
       have been scaled by a factor of $10^{4}$ for anti-deuterium and
       of $10^{8}$
       for anti-helium $\bar{\rm He}$. The doubling of curves
       corresponds to different
       factorization schemes.}
       \label{section_7_1b}
\end{figure}

%Antinuclei
\paragraph{Antinuclei in the Galaxy}
Another crucial problem of cosmology is the existence of a possible
asymmetry between matter and antimatter. The current wisdom is that
we live in a matter dominated universe and that antimatter islands
seem to be excluded. The ongoing annihilations that would take place
at the frontiers would create an intense gamma-ray emission that is
not seen. The mechanism which would be required to account
for a separation of matter
and antimatter domains is not known. However, one should not subscribe
to any dogmatic point of view. The amount of antimatter in cosmic rays is
about to be measured with unequalled accuracy by the space station borne
spectrometer of the {\sc ams} collaboration \cite{AMS}. One of the
most exciting
goals of the experiment is the possible detection of antinuclei in the
cosmic radiation. It is generally believed that the observation of a single
anti-helium or anti-carbon would undoubtedly signal the presence of
antistars because the conventional production of antinuclei through
spallation is negligible. This point has been examined in
Sec.~\ref{sub:sec_anti_p_d} where we mostly concentrated on
secondary antideuterons. Suffice it to say that the creation of
any additional antinucleon during a single spallation event is
suppressed by a factor of $\sim 10^{-5} - 10^{-4}$ \cite{Orloff}.
As featured by Fig.~\ref{section_7_1b}, the $\bar{D}/\bar{p}$ ratio  is
$\sim 3 \times 10^{-5}$ while $\bar{\rm He}/\bar{D} \sim 7 \times 10^{-5}$
for a momentum per nucleon in excess of 10 GeV. Secondary antinuclei
with atomic number $A \geq 3$ are so much suppressed that they are out
of reach of the near future instruments.
Another possibility would be that antiglobular clusters lies
in the halo of the Galaxy and could make a substantial contribution to
the anti-helium flux~\cite{anti-He}.
There is more hope for antideuterons. Their primary production will be
discussed in Sec.~\ref{subsec:exotic_susy} for supersymmetric neutralinos
and briefly mentioned in Sec.~\ref{subsec:pbh} in the case of
evaporating black holes.

%%%%%%%%%%%%%%%%%%%%%%%%%%%%%%%%%%%%%%%%%%%%%%%%%%%%%%%%%%%%%%%%%%%%%%%%%%%%%%%%
%%%%%%%%%%%%%%%%%%%%%%%%%%%%%%%%%%%%%%%%%%%%%%%%%%%%%%%%%%%%%%%%%%%%%%%%%%%%%%%%

\subsection{Spatial origin of the PBH or SUSY exotic primaries}
\label{subsec:origine_primaires_halo}
As for species created in the disc (see
Sec.~\ref{subsec:origin_realistic_case}),
one can infer the closed surfaces from which most of the cosmic rays 
detected on Earth~\cite{Maurin04} were emitted from. 
We are now confronted with a volume
distribution of sources (see Sec.~\ref{subsubsec:exotic_sources}). As
for standard sources in the disc, geometrical
effects are very important (see Sec.~\ref{subsec:origine_la_base}).
Apart from these geometrical restrictions, spallations and convective wind
greatly affect the shape of these isodensity surfaces
and, e.g., location of the particularly interesting contour defined by
${\cal P}({\cal V}(x)|O)=99$\% (contour from where 99\% of exotic cosmic ray
originate). 
As a consequence, not only the characteristic size
$L$ is important to estimate the $x=99$\% surface, but also
spallations and convection typical extension, i.e. parameters $r_{\rm spal}$
and $r_{\rm wind}$ such as defined in Eqs.~(\ref{eq:r_spal})
and~(\ref{eq:r_wind}).
These parameters strongly depend on $\delta$ (the diffusion
coefficient slope) when realistic configurations are retained.

\paragraph{Realistic values of $r_{\rm spal}$ and $r_{\rm wind}$ for
antideuterons}
Antideuteron signal seems to be the most promising
species to consider. We choose the interstellar energy
1~GeV/nuc; nuclei that reach the detector are Solar modulated
so that the final energy corresponds roughly to $200-800$~MeV/nuc
depending of the modulation parameter. This is the window where the
exotic signal becomes interesting. Tab.~\ref{tab:tab3} summarizes the values
of $r_{\rm wind}$ and $r_{\rm spal}$ at this energy for antideuterons
for three halo sizes
and three values of $\delta$.
\begin{table}[hbt!]
    \begin{center}
	\begin{tabular}{|c||c|ccc|}   \hline
	    &(kpc) &{\scriptsize $\delta=0.35$} & {\scriptsize $\delta=0.6$} & 
	    {\scriptsize $\delta=0.85$}\\\hline\hline
	    $L=10$~kpc & $r_{\rm wind}=$ & $\infty$ & 8.95 & 3.29\\
	    & $r_{\rm spal}=$ & 24.19  & 8.69 & 4.02\\\hline
	    $L=2$~kpc    & $r_{\rm wind}=$ & $\infty$  & 2.42  & 0.95\\
	    & $r_{\rm spal}=$ & 6.97   & 2.49  & 1.18 \\\hline
	\end{tabular}
	\caption{$r_{\rm wind}$ and $r_{\rm spal}$ for two halo 
	sizes $L$ and
	three diffusion slopes $\delta$: these numbers are for 1~GeV/nuc
	(interstellar energy) antideuterons.}
	\label{tab:tab3}
    \end{center}
\end{table}
We notice  that the situation is very different for small or large $\delta$.
For Kolmogorov power spectrum, only spallations act on the propagation
and only weakly for this light nucleus: for large $\delta$  -- the value
$\delta=0.85$ is the one preferred in our B/C analysis (see
Sec.~\ref{subsec:res.from_B/C})
-- models are convection/spallation dominated with spallation and
convection acting at about the same footing.

\begin{figure}[hbt!]
          \includegraphics[width=\columnwidth]{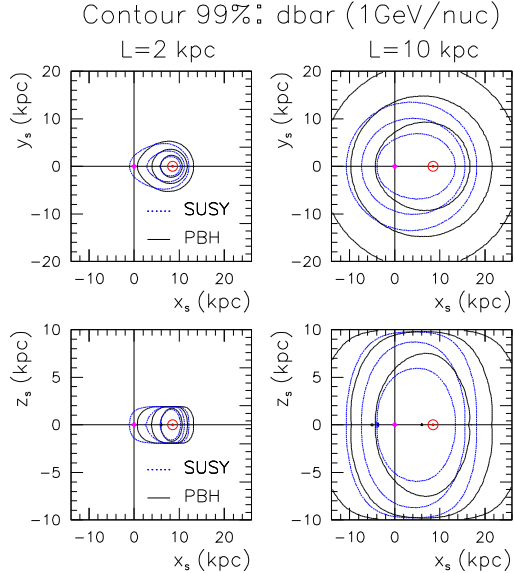}
          \caption{Contours defining surfaces from where 99\% of 
exotic primaries
          come (no side boundaries).
          Upper panels: cut in the $z_s=0$~kpc plane; lower panels: cut in the
          $y_s=0$~kpc plane. Left panels correspond to $L=2$~kpc and
          right panels to $L=10$~kpc. In each panel, we plot
          either the PBH case (solid lines) or the SUSY case (dotted lines).
          From external lines to internal lines correspond the values of
the diffusion
          coefficient slope $\delta=0.35$, $\delta=0.60$, $\delta=0.85$.}
          \label{fig:final1}
\end{figure}

\paragraph{Results}
The production of particles by PBH is related to the dark matter
profile whose weight  is
$w_{\rm PBH}(r,z)= f(r,z)$, whereas for supersymmetric
annihilating particles, the production is related to the weight
$w_{\rm SUSY}(r,z)=f(r,z)^2$ (see Sec.~\ref{subsubsec:exotic_sources}). 
Weighting the elementary probability
by this production rate leads to the contours displayed in
Fig.~\ref{fig:final1} for ${\cal P}_{\rm cyl}({\cal V}(x)|O)=99$\%
(parameters are taken from Tab.~\ref{tab:tab3}). For $\delta=0.35$ (external
contours), we recover basically  contours that one would obtain in the high
energy limit, i.e. contours driven by geometrical effects, pure diffusion
(see also~\cite{Taillet01}). However, for larger $\delta$ (internal
contours), these contours shrink naturally and
 all surfaces are distorted
towards the galactic center, where the maximum of production occurs.
One has also to keep in mind that whatever
these surfaces, most cosmic rays emitted from inside these volumes
actually do not reach us. Escape preferentially occurs and it is obvious
that the closer the source, a smaller part of the emitted 
flux escapes.
The fraction that reaches us can be estimated to be about
0.01\%-0.1\% of the total emission in volume defined by $x=99$\%, 
whatever $\delta$ (it slightly depends on the halo size $L$ for PBHs 
as well as for SUSY particles).
Another meaningful number is the fraction of particles created in the volume
$x=99$\% with respect
to the total number of particles
created in the entire dark halo. This quantity is
strongly dependent on the diffusive halo size and, for $L=10$~kpc compared
to $L=2$~kpc, numbers are 2.\% vs 0.2\% (PBH) and 80\% vs 20\% (SUSY).
This is a mere consequence of the different production terms.

%%%%%%%%%%%%%%%%%%%%%%%%%%%%%%%%%%%%%%%%%%%%%%%%%%%%%%%%%%%%%%%%%%%%%%%%%%%%%%%%
%%%%%%%%%%%%%%%%%%%%%%%%%%%%%%%%%%%%%%%%%%%%%%%%%%%%%%%%%%%%%%%%%%%%%%%%%%%%%%%%

\subsection{Primary antiprotons and antideuterons from supersymmetric sources}
\label{subsec:exotic_susy}

%The production takes place everywhere
The neutralinos that could be concealed in the halo of the
Milky Way -- and be responsible for the flatness of its rotation curve --
should be steadily annihilating and produce antiprotons together with
antideuterons. A key difference with respect to the production mechanisms that
have already been discussed lies in the fact that this new yet putative source
of antinuclei is not confined to the galactic disc. It spreads all over the
halo far above and beneath the ridge. Its contribution to the local flux
may nevertheless be easily derived with the formalism presented in
Sec.~\ref{subsubsec:contrib_halo_prim}.
The propagation of primary species from the remote regions of the Milky Way
neighborhood to the Earth has actually been treated in~\cite{Bottino_Salati}.

%SUPERSYMMETRIC ANTIPROTONS
%The MSSM
\paragraph{SUSY model}
The neutralino naturally appears in the framework of the Minimal Supersymmetric
extension of the Standard Model (MSSM) \cite{susy} as the lowest-mass linear
superposition of the photino $\tilde \gamma$, the zino $\tilde Z$ and the two
higgsino $\tilde H_1^{\circ}$ and $\tilde H_2^{\circ}$ states
\begin{equation}
          \chi \equiv a_1 \tilde \gamma + a_2 \tilde Z + a_3 \tilde H_1^{\circ}
          + a_4 \tilde H_2^{\circ} \;\; .
          \label{eq:neu}
\end{equation}
This particle is neutral and interacts weakly. It generally turns out to be
the lightest supersymmetric state and is therefore stable if R-parity is
conserved.
%
%The rates of production

\paragraph{Antiproton multiplicity}
The differential multiplicity for antiproton production
in a neutralino pair annihilation may be expressed as
\begin{displaymath}
          {\displaystyle \frac{dN_{\bar{p}}}{d E_{\bar{p}}}} \; = \;
          {\displaystyle \sum_{\rm F , h}} \; B_{\rm \chi h}^{\rm (F)} \;
          {\displaystyle \frac{dN_{\bar{p}}^{\rm h}}{d E_{\bar{p}}}} \;\; .
\end{displaymath}

The annihilation proceeds -- through the various final states F -- towards
the quark or the gluon h with the branching ratio $B_{\rm \chi h}^{\rm (F)}$.
Quarks or gluons may be directly produced. They may alternatively result from
the intermediate production of a Higgs or a gauge boson as well as of a top
quark.
Each quark or gluon h generates in turn a jet whose subsequent fragmentation
and hadronization yields the antiproton energy spectrum
${dN_{\bar{p}}^{\rm h}} / {d E_{\bar{p}}}$.
%
%The source term
The source term for supersymmetric antiprotons
\begin{displaymath}
          q_{\bar{p}}^{\rm susy}(r,z,E_{\bar{p}}) =
          \langle \sigma_{\rm ann} v \rangle \,
          {\displaystyle \frac {dN_{\bar{p}}}{d E_{\bar{p}}}} \,
          \left\{ {\displaystyle
          \frac{\displaystyle \rho_{\chi}(r,z)}{m_{\chi}}} \right\}^{2}
\end{displaymath}
supplements the spallation contributions $q_{\bar{p}}^{\rm sec}$ and
$q_{\bar{p}}^{\rm ter}$.

%%% BABABA
%
%The spectra and the results
Because the annihilation of a neutralino pair $\chi-\chi$ occurs
at rest with respect to the Galaxy, most of the antiprotons -- and
antideuterons for that matter -- are produced at low energies. The
resulting spectrum is fairly flat below a few GeV. This has important
observational consequences insofar as the secondary antiproton spectrum is
already fairly flat as shown in Fig.~\ref{fig:all_unc},
p.~\pageref{fig:all_unc}.
It may therefore prove difficult  to disentangle a primary component from
the secondary antiproton radiation.
The primary flux nevertheless
modifies the magnitude -- if not the spectrum -- of the cosmic ray
antiproton radiation and could contribute significantly as shown in
Tab.~\ref{table:susy}.
This effect can be used to constrain the supersymmetric
parameter space.
%
%%%%%%%%%%%%%%%%%%%%%%%%%%%%%%%%%%%%%%%%%%%%%%%%%%%%%%%%%%%%%%%%%%%%%%%%%%%%%%%%
%%
\begin{table*}[hbt!]
\[
\begin{array}{|c|c|c|c|c|c|c|c|} \hline
{\rm case} &
m_{\chi} &
P_{g} (\%) &
\Omega_\chi h^{2} &
\Phi_{\bar{p}}^{\rm min} \left( 0.24 \; {\rm GeV} \right) &
\Phi_{\bar{D}}^{\rm min} \left( 0.24 \; {\rm GeV/nuc} \right) &
\Phi_{\bar{D}}^{\rm max} \left( 0.24 \; {\rm GeV/nuc} \right) &
N_{\bar{D}}^{\rm max}
%\left( 0.1-3 \; {\rm GeV/n} \right)
\\
& & & & & & &
\\
\hline \hline
a & 36.5 & 96.9 & 0.20  & 1.2 \times 10^{-3} & 1.0 \times 10^{-7} &
2.9 \times 10^{-8} & 0.6\\
b & 61.2 & 95.3 & 0.13 & 3.9 \times 10^{-3} & 3.5 \times 10^{-7}&
1.1 \times 10^{-7} & 2.9 \\
c & 90.4 & 53.7 & 0.03 & 1.1 \times 10^{-3} & 1.8 \times 10^{-7} &
6.1 \times 10^{-8} & 2.0 \\
d & 120 & 98.9 & 0.53  & 2.9 \times 10^{-4} & 2.5 \times 10^{-8}&
8.6 \times 10^{-9} & 0.3  \\
\hline \hline
\end{array} \]
\caption{
These four cases illustrate the richness of the supersymmetric
parameter space. There is no obvious correlation between the
antiproton and antideuteron Earth fluxes with the neutralino mass
$ m_{\chi}$. Case (c) is a gaugino-higgsino mixture and still
yields signals comparable to those of case (a), yet a pure gaugino.
Antideuteron fluxes are estimated at both Solar minimum and
maximum, for a modulated energy of 0.24 GeV/nuc. The last column
features the corresponding number of $\bar{D}$'s which {\sc ams} on board
ISS can collect below an IS energy of 3 GeV/nuc.
}
\label{table:susy}
\end{table*}

%SUPERSYMMETRIC ANTIDEUTERONS
%
%The rates of production
\paragraph{Supersymmetric $\bar{D}$ signal}
As regards the antideuteron production, the factorization-coalescence scheme
discussed above leads to the antideuteron differential multiplicity
\begin{eqnarray*}
          \lefteqn{\frac{dN_{\bar{D}}}{dE_{\bar{D}}} =
          \frac{4 \, P_{\rm coal}^{3}}{3 \, k_{\bar{D}}}
          \,
          \frac{m_{\bar{D}}}{m_{\bar{p}} \, m_{\bar{n}}}}\\
          && \times \sum_{\rm F , h} \, B_{\rm \chi h}^{\rm (F)} \,
          \left\{
          \frac{dN_{\bar{p}}^{\rm h}}{d E_{\bar{p}}}
          \left( E_{\bar{p}} = \frac{E_{\bar{D}}}{2} \right)
          \right\}^{2} \;\; .
          \label{dNdbar_on_dEdbar_susy}
\end{eqnarray*}
It may be expressed as a sum -- extending over the various quarks and
gluons h as well as over the different annihilation channels F -- of the
square of the antiproton differential multiplicity. That sum is weighted
by the relevant branching ratios. The antineutron and antiproton differential
distributions have been assumed to be identical.
%The source term
This readily leads to the source term for supersymmetric antideuterons
\begin{displaymath}
          q_{\bar{D}}^{\rm susy}(r,z,E_{\bar{D}}) \; = \;
          \langle \sigma_{\rm ann} v \rangle \,
          {\displaystyle \frac {dN_{\bar{D}}}{d E_{\bar{D}}}} \,
          \left\{ {\displaystyle
          \frac{\displaystyle \rho_{\chi}(r,z)}{m_{\chi}}} \right\}^{2} \;\; .
\end{displaymath}
\begin{figure}[hbt!]
\resizebox{\columnwidth}{!}
{\includegraphics*[1.5cm,6.5cm][18.5cm,22.2cm]{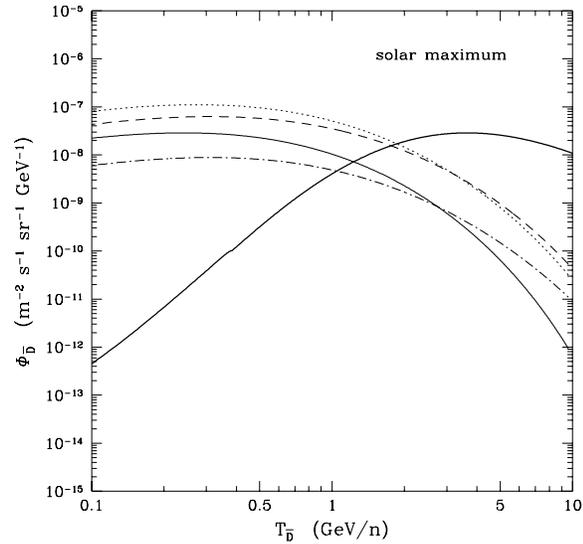}}
          \caption{
          The IS flux of secondary antideuterons (heavier solid curve)
          decreases at low energy whereas the energy spectrum of the
          antideuterons from supersymmetric origin tends to flatten
          (see Sec.~\ref{subsec:exotic_susy}).
          The four cases of Tab.~\ref{table:susy} are respectively featured
          by the solid (a), dotted (b), dashed (c) and dot-dashed (d) curves.
          Solar modulation has been taken at maximum when the {\sc ams}
observatory
          operates on board ISS.
          }
          \label{fig:dbar}
\end{figure}

%
%The spectra and the results

The four supersymmetric configurations of Tab.~\ref{table:susy} are
presented in Fig.~\ref{fig:dbar}
together with the calculated secondary spectrum 
(see Sect.\ref{sub:sec_anti_p_d}). The antideuteron spectra are
fairly flat at low energy. They yield a few events for the {\sc ams}
observatory on board ISS below an IS energy of 3 GeV/nuc -- a region
from which secondary antideuterons are absent --.

\begin{figure}[hbt!]
\centerline{
\resizebox{\columnwidth}{!}
{\includegraphics*[1.5cm,6.5cm][18.5cm,22.2cm]{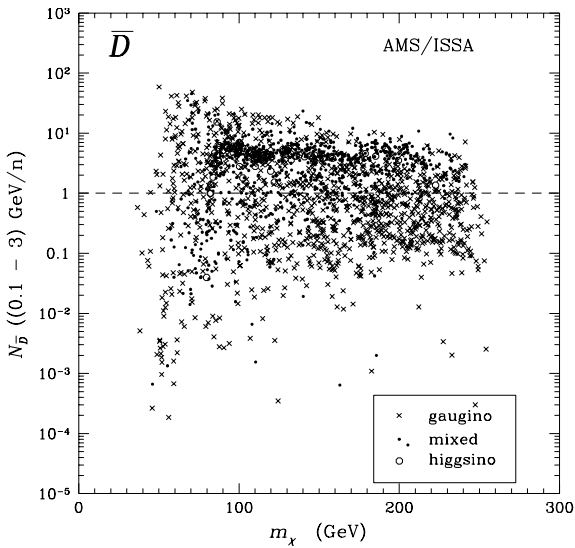}}
}
          \caption{
          The supersymmetric $\bar{D}$ flux has been integrated over
          the range of IS energies extending from 0.1 up to 3 GeV/nuc.
          The resulting yield $N_{\bar{D}}$ of antideuterons which {\sc ams}
          on board ISS can collect is plotted as a function of the neutralino
          mass $m_{\chi}$. Modulation has been considered at Solar maximum.
          }
          \label{fig:ndbar_versus_mchi}
\end{figure}

%
%Supersymmetric yields below an IS energy of 3 GeV/n
\paragraph{Consequence for SUSY parameter space}
For each configuration  of the whole supersymmetric parameter 
space, the $\bar{D}$ flux has been integrated
over that low-energy range. The resulting yield $N_{\bar{D}}$ which
{\sc ams} may
collect is presented as a function of the neutralino mass $m_{\chi}$
in the scatter
plot of Fig.~\ref{fig:ndbar_versus_mchi}. During the {\sc ams}
shuttle mission, the Solar cycle was close to maximum.
 Most of the configurations are gaugino like (crosses)
or mixed combinations of gaugino and higgsino states (dots). A significant
portion
of the parameter space is associated to a signal exceeding one antideuteron
-- horizontal dashed line --. In a few cases, {\sc ams} may even
collect more than a
dozen
of low-energy $\bar{D}$ nuclei.
%
%Le coup de Ben Moore
Notice finally that numerical simulations have shown that neutralinos
-- and more generally cold dark matter -- should cluster in very dense
and numerous clumps \cite{moore}. Because neutralino annihilations proceed
through a two-body reaction, their antiproton and antideuteron signatures
would be enhanced by a large factor that could even reach up a few hundred
in the case of density profiles {\em \`a la} Moore.
%superbe jeu de mot non ?
If so, the entire supersymmetric constellation in
Fig.~\ref{fig:ndbar_versus_mchi} would be shifted
upwards by at least two orders of magnitude and detection would become
crystal clear for the most optimistic configurations.
Another important question is related to the clumpiness of the dark halo.
If the dark matter is concentrated into clumps, then the cosmic ray signals on
Earth may be sensitive to the spatial distribution of the clumps.
This aspect can be studied with the formalism used to determine the
spatial origin of cosmic rays,  as described above (Bottino et al., in preparation).

	%%%%%%%%%%%%%%%%%%%%%%%%%%%%%%%%%%%%%%%
	%%%%%%%%%%%%%%%%%%%%%%%%%%%%%%%%%%%%%%%

\subsection{Primary antiprotons and antideuterons from Primordial Black Holes}
\label{subsec:pbh}

Very small black holes should have formed in the early Universe from
initial density inhomogeneities (Hawking \cite{Hawking2}). They should now
evaporate intensely through the Hawking mechanism (Hawking
\cite{Hawking5}) if their
initial masses were around $M_* \approx 5\times 10^{14}$ g. Detecting
such objects
is a great challenge of modern physics and cosmology as it would both allow to
give experimental grounds to the Hawking radiation which is one of the only
tentative achievement of semi-classical quantum gravity, and to probe the very
small scales of the early Universe that remains totally inaccessible to other
observations. Although  the standard cosmological model of structure formation,
assuming a pure scale-invariant Harisson-Zeldovitch
power spectrum normalized to CMB amplitudes, would lead to a very
small amount of PBH dark-matter in the present Universe, several
realistic inflationary scenarios and phase-transition phenomena can produce a
significant amount of PBHs.

\subsubsection{Hawking evaporation and antiproton source term}

The Hawking black hole evaporation process can be intuitively understood
as a quantum creation of particles from the vacuum by an external field.
The basic characteristics can be easily seen through a simplified model,
and the interested reader is referred to~\cite{Frolov} for
more details.

\paragraph{Elementary energy spectrum}
The accurate emission process was derived by Hawking, using the usual quantum
mechanical wave equation for a collapsing object with a post-collapse
classical curved metric.
He found that the emission spectrum for particles of energy $Q$ per unit
of time $t$ actually mimics a Planck law and is given by, for each
degree of freedom:
\begin{equation}
          \frac{{\rm d}^2N}{{\rm d}Q{\rm
d}t}=\frac{\Gamma_s}{h\left(\exp\left(\frac{Q}{h\kappa/4\pi^2c}
\right)-(-1)^{2s}\right)}\;,
          \label{tatata}
\end{equation}
where the effect of angular velocity and electric potential have
been neglected since the black hole discharges and finishes its rotation
much faster than it evaporates~\cite{Gibbons_Page1}.
In this expression, $\kappa$ denotes the surface gravity, $s$ is the spin
of the emitted species and $\Gamma_s$ is the absorption probability, given by
\begin{equation}
          \Gamma_s=\frac{4\pi \sigma_s(Q,M,\mu)}{h^2c^2}(Q^2-\mu^2)\;,
\end{equation}
where $\sigma_s$ is the absorption cross section computed
numerically~\cite{Page2} and $\mu$ the rest mass of the emitted particle.
It is also convenient to introduce the Hawking temperature, defined by
\begin{equation}
          T=\frac{hc^3}{16\pi k G M}\approx\frac{10^{13}{\rm 
~g}}{M}{\rm ~GeV}\;.
\end{equation}
The expression  (\ref{tatata}) may then be written simply as a
function of $Q/kT$.

\paragraph{Antiproton multiplicity}
As it was shown in~\cite{MacGibbon1}, when the
black hole temperature is
greater than the quantum chromodynamics confinement scale
$\Lambda_{\rm QCD}$, quarks and gluons jets are
emitted instead of composite hadrons. To evaluate the number of
emitted antiprotons $\bar{p}$ , one therefore
needs to perform the following convolution:
\begin{eqnarray*}
          \lefteqn{\frac{{\rm d}^2N_{\bar{p}}}{{\rm d}E{\rm d}t}=
          \sum_j\int_{Q=E}^{\infty}\frac{\alpha_j}{e^{Q/kT}-(-1)^{2s_j}}
          \frac{\Gamma_{s_j}(Q,T)}{h}}\\
          &&\times\frac{{\rm d}g_{j\bar{p}}(Q,E)}{{\rm d}E}{\rm d}Q\;,
\end{eqnarray*}
where $\alpha_j$ is the number of degrees of freedom, $E$ is the
antiproton energy and
${\rm d}g_{j\bar{p}}(Q,E)/{\rm d}E$ is the normalized differential
fragmentation function, i.e.
the number of antiprotons  created with an energy between $E$ and $E+dE$ by a
parton jet of type $j$ and energy $Q$.
The fragmentation functions have been evaluated with the
high-energy physics frequently used event generator
{\sc pythia/jetset}~\cite{Tj}.

\paragraph{Whole spatial and spectral source distribution}
Once the spectrum of emitted antiprotons is known for a single PBH of
given mass, the
source term used for propagation is given by
\begin{equation}
          \frac{d^3 N_{\bar{p}}(E) }{dEdtdV}= \int_{0}^{\infty}
          \frac{d^2N_{\bar{p}}}{dEdt}(M,t_0)\frac{d^2n}{dMdV}dM\;,
\end{equation}
where $d^2n/dMdV$ is the PBH mass spectrum today. It can be deduced from the
initial mass spectrum
through the mass loss rate which reads as $dM/dt=-\alpha(M)/M^2$ (by a simple
integration of the Hawking spectrum multiplied by the energy of the
emitted quantum)
where $\alpha(M)$ accounts for the available degrees of freedom at a
given mass.
In the assumption $\alpha(M)\approx const$ it leads to:
\begin{equation}
          \frac{d^2n(M)}{dMdV_i}=\frac{M^2}{(3\alpha t +
          M^3)^{2/3}}\cdot
          \left.\frac{d^2n_i}{dM_idV_i}\right|_{M_i = M_i^0}
\end{equation}
where the derivative is evaluated at $M_i^0 = (3\alpha t + M^3)^{1/3}$.
The initial values $d^2n_i/dM_idV_i$ were shown~\cite{Carr1}
to scale as
$M_i^{-5/2}$ for PBHs formed, as expected, in a radiation dominated Universe
from a scale invariant power spectrum.
The nowadays spectrum is therefore mostly identical to the initial one above
$M_*\equiv 3\alpha t_0\approx 5\times 10^{14}$g and proportional to $M^2$
below.

The last problem is that the spatial distribution of these PBHs is
basically unknown.
However, these objects should have formed in the very early stages of
the history of the Universe:
their initial mass is very close to
the horizon mass at the formation epoch ($M_H\approx M_{\rm Pl}
\times t_{\rm form} /
t_{\rm Pl}$
which means that the formation time $t_{\rm form}$ is of the order of
$10^{-23}$ s
for $M=M_*$).
Once formed, they only interact through gravitation, so that they
behave as cold dark matter, and their spatial distribution should be
similar.
As a consequence, we use the same profile for the PBHs distribution as
for the neutralinos, as given in Eq.~(\ref{distribution_isotherme}).

\subsubsection{Resulting upper limit and cosmological consequences}

As shown in~\cite{barrau02} the main uncertainties
on the estimation of the antiproton flux from PBHs are associated
with astrophysical parameters. Several possible effects due to the quite large
size of the horizon mass at the end of the inflation period (mostly imposed by
gravitinos and moduli fields constraints) and to the possible photosphere that
could form around hot PBHs were studied in~\cite{barrau02}
and show that the present estimates could be substantially revised in
the future.
The emitted antiproton spectrum is mostly resulting from PBHs with masses
between $10^{12}$ g and $10^{14}$ g: the lightest ones are not numerous enough
because of the $M^2$ shape of the nowadays mass spectrum and the heaviest ones
do not emit much because of their low temperature.

\begin{figure}[hbt!]
          \includegraphics[width=\columnwidth]{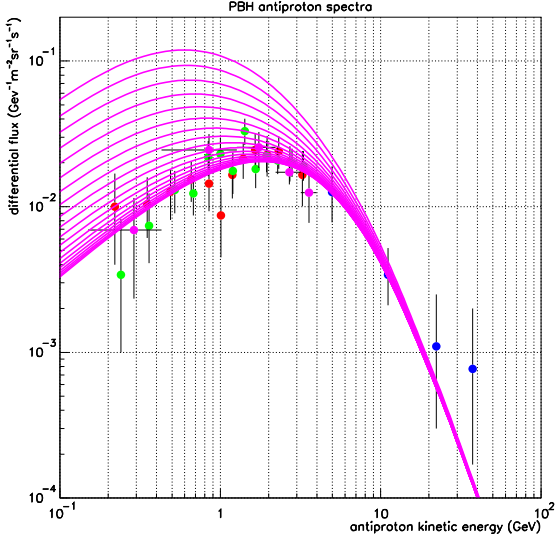}
          \caption{Experimental data from {\sc bess} 95 (filled circles),
{\sc bess} 98 (circles),
          {\sc caprice} (triangles) and {\sc ams} (squares) superimposed
          with mean theoretical PBH spectra for $\rho_{\odot}^{\rm PBH}$ between
          $5\cdot10^{-35}$~g~cm$^{-3}$ (lower curve) and $10^{-32}$~g~cm$^{-3}$
          (upper curves).
          }\label{pbar_tot}
\end{figure}

\paragraph{Density exclusion criterion}
Superimposed with {\sc bess}, {\sc caprice} and {\sc ams} data, the
full antiproton flux,
including the secondary component and the primary component, is shown
in Fig.~\ref{pbar_tot} for 20 values of $\rho_{\odot}^{\rm PBH}$
logarithmically spaced between $5 \times 10^{-35}$
and $10^{-32}~{\rm g} \,{\rm cm^{-3}}$ with fixed astrophysical parameters. The
lowest curves are clearly in agreement with data whereas the upper ones
contradict experimental results. To derive a reliable upper
limit, and to account for asymmetric
error bars in data, we define a generalized $\chi^2$ as
\begin{eqnarray}
    \lefteqn{
    \chi^2= \sum_i
    \frac{(\Phi^{\rm th}(Q_i)-\Phi_i^{\rm e})^2}
    {(\sigma^{\rm e+}_i+\sigma^{\rm th+}(Q_i))^2}\Theta(\Phi^{\rm
    th}(Q_i)-\Phi^{\rm e}_i)} \nonumber \\
    && +\sum_i \frac{(\Phi^{\rm th}(Q_i)-\Phi_i^{\rm e})^2}
    {(\sigma^{\rm e-}_i+\sigma^{\rm th-}(Q_i))^2}\Theta(\Phi^{\rm
    e}_i-\Phi^{\rm
    th}(Q_i)).\nonumber 
\end{eqnarray}
% \begin{displaymath}
%             \begin{array}{ll}
%               \chi^2= &\sum_i
%               \frac{(\Phi^{\rm th}(Q_i)-\Phi_i^{\rm exp})^2}
% {(\sigma^{\rm exp+}_i+\sigma^{\rm th+}(Q_i))^2}\Theta(\Phi^{\rm
% th}(Q_i)-\Phi^{\rm exp}_i)\\
%               & +\sum_i \frac{(\Phi^{\rm th}(Q_i)-\Phi_i^{\rm exp})^2}
% {(\sigma^{\rm exp-}_i+\sigma^{\rm th-}(Q_i))^2}\Theta(\Phi^{\rm
% exp}_i-\Phi^{\rm
% th}(Q_i)).\\
%             \end{array}
% \end{displaymath}
where $\sigma^{\rm th+}$ and $\sigma^{\rm e+}$ ($\sigma^{\rm th-}$ and
$\sigma^{\rm e-}$)
are the theoretical and experimental
positive (negative) uncertainties.
An upper limit on
the primary flux for each value of the magnetic halo thickness is computed.
The theoretical errors included in the $\chi^2$ function
come from nuclear physics ($p+He \rightarrow \bar{p}+X$
         and $He+He \rightarrow \bar{p}+X$~\cite{Donato02}), and from the
astrophysical parameters analysis
which were added linearly, in order to remain conservative. The resulting
$\chi^2$ leads to very safe results as it assumes that limits on the
parameters correspond to 1 sigma.

\begin{figure}[hbt!]
          \includegraphics[width=\columnwidth]{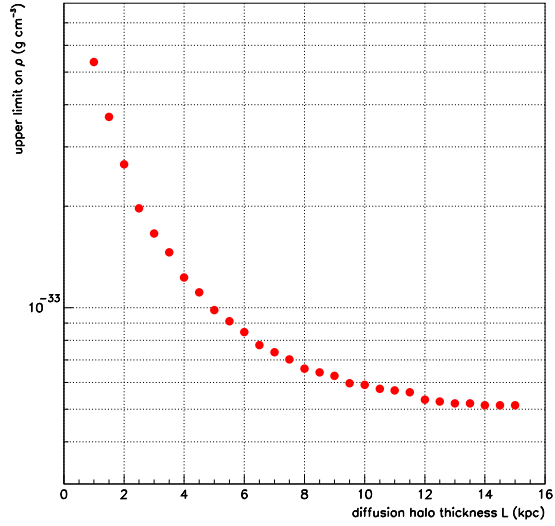}
          \caption{Upper limits on the local density of PBHs as a function of
          the magnetic halo thickness $L$.
          }\label{chi2}
\end{figure}

\paragraph{Resulting constraints}
Fig.~\ref{chi2} gives the upper limits on the local density of PBHs as
a function of $L$. It is a
decreasing
function of the halo size because a larger diffusion region means a higher
number of PBHs inside the magnetic zone for a given local density.
Between $L=1$ and $L=15$ kpc (extreme astrophysical values), the 99\%
confidence
level upper limit goes from $5.3\cdot 10^{-33}$~g~cm$^{-3}$ to
$5.1\cdot 10^{-34}$~g~cm$^{-3}$.

Such limits are of great cosmological relevance. They are a unique probe of the
early Universe on very small scales that could be equaled only, in a far
future, by the detection of primordial gravitational waves. Based on
complementary, but nearly equivalent, limits coming from the 100 MeV gamma-ray
background, numerous constraints were derived on the spectral index of scalar
perturbations (see, e.g.~\cite{Kim}): $n<1.27$. A too ``blue" spectrum
would lead to an
overproduction of PBHs in conflict with observations.

This upper limit on the local PBH density obtained with antiprotons
can also be used to
severely constrain models with Broken Scale Invariance (BSI). A jump in the
first derivative of the inflaton potential~\cite{Starobinsky} should
lead to a huge increase of the PBHs formation probability when the
corresponding
scale re-enters the horizon~\cite{Blais}. The
antiproton limit can, therefore, be
directly translated into an exclusion area in the parameter space of BSI
inflationary models~\cite{Barrau_soumis}. Finally, it allows
interesting prospects for PBH dark matter~\cite{Blais2} or
Planck relics investigations~\cite{Alexeyev}.

\subsubsection{A new window for detection: antideuterons}

\begin{figure}[hbt!]
          \includegraphics[width=\columnwidth]{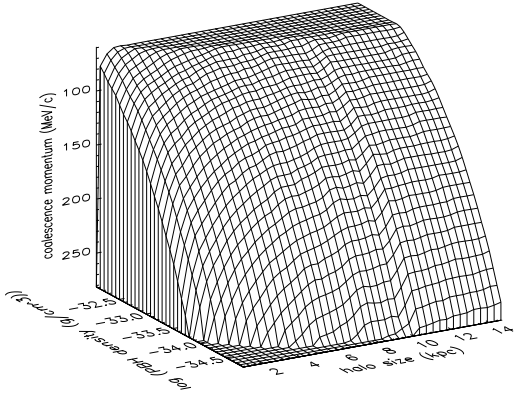}
          \caption{Parameter space (halo thickness $L$: 1-15 kpc;
coalescence momentum
          $p_0$: 60-285 MeV/c; PBH density $\rho_{\odot}$:
          $10^{-35}-10^{-31}$~g~cm$^{-3}$)
          within the {\sc ams} sensitivity (3 years of data taking). The allowed
region lies
          below the surface.\label{fig:3d_ams}}
\end{figure}

To go beyond an upper limit and try to detect PBHs it seems very
interesting to look for
antideuterons. Below a few GeV, there is nearly no background for
kinematical reasons~\cite{dbar_99} and the possible signal due to
PBHs evaporation could be easy to detect.
The emission scheme is nearly the same but the probability that an antiproton
and an antineutron merge into an antideuteron is taken into account.
The possible detection range for the {\sc ams} experiment~\cite{Barrau5} can be
evaluated. It is shown
on Fig. \ref{fig:3d_ams} as a function of the three unknown parameters: $L$,
the height of the magnetic halo, $p_0$, the coalescence momentum to form an
antideuteron, and
$\rho^{\rm PBH}_{\odot}$, the local density of primordial black holes.
The sensitivity of the experiment should allow, for averaged
parameters, an improvement in the current best upper limit by a
factor of six, if not a
positive detection. The situation is very different than for antiproton: the
main limitation is due to the experimental sensitivity and not to the
unavoidable physical background. Great improvements can, therefore,
be expected in the
future and this investigation means seems very promising.

%%%%%%%%%%%%%%%%%%%%%%%%%%%%%%%%%%%%%%%%%%%%%%%%%%%%%%%%%%%%%%%%%%%%%%%%%%%%%%%%%%%%%
%%%%%%%%%%%%%%%%%%%%%%%%%%%%%%%%%%%%%%%%%%%%%%%%%%%%%%%%%%%%%%%%%%%%%%%%%%%%%%%%%%%%%

\section{Summary, conclusions and perspectives}
\label{sec:concl}

A consistent framework to understand the propagation of CR nuclei 
in the energy range 100 MeV--100 GeV was presented in this paper. 
The observed fluxes
of most species can be explained by assuming that once emitted
from some sources located in the galactic disc, these nuclei undergo a
diffusive propagation altered by escape through the boundaries, spallations,
reacceleration, energy losses and galactic wind.
The magnitude of these effects has been constrained using the B/C data,
and the consistency of the model has been tested against the observed
antiproton flux and by the study of radioactive species.

This well-tested model has then been used to study the propagation of
cosmic rays of a more hypothetical origin, such as light antinuclei
produced by SUSY galactic Dark Matter or Primordial Black Holes.
In particular, limits on the abundance of primordial
black holes in the galactic halo could be set.

There are several directions in which this work may be extended.
First, the constraints on the propagation parameters could be refined
by considering other species, stable or secondary.
However, this approach is currently limited by the accuracy of the
available data on cosmic ray fluxes and on the nuclear cross sections.
Second, a specific study of the EC unstable species could provide
valuable information about the  processes responsible for the
acceleration of cosmic rays.
Third, the existing limits on the SUSY induced antiproton signal can
be bettered by using the constraints on the propagation parameters in
a fully consistent way.
Finally, the propagation code we use should ultimately be able to
yield the flux of all cosmic ray species (including gamma rays,
electrons and positrons) at every position in the Galaxy.

\begin{figure}[hbt!]
          \includegraphics[width=\columnwidth]{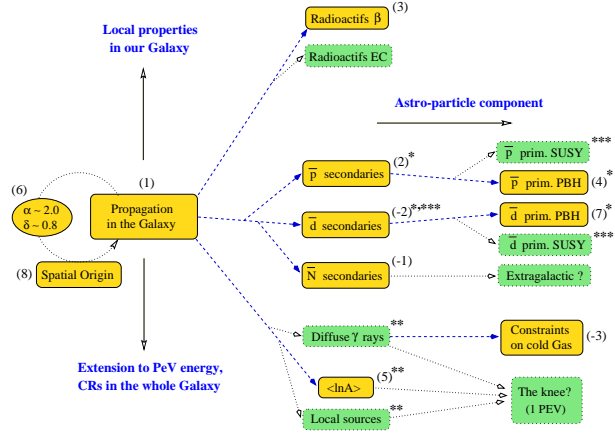}
          \caption{Schematic view of the subjects discussed in this paper.
          The stars indicate collaborations of {\sc lapth} (Annecy-le-Vieux,
          France) members with
          {\sc isn} ($\star$, Grenoble, France), {\sc iap} ($\star \star$, Paris,
          France) or {\sc infn} ($\star \star \star$, Turin, Italy).
          Dashed boxes represent future projects.
          The numbers in parenthesis represent the publications. The
          starting point (1) is the first use of the elaborate propagation
          model presented here \cite{Maurin01}, (2) is \cite{Donato02}
          (3) is \cite{Donato01}, (4) is \cite{barrau01}, (5) is 
\cite{Maurin03},
          (6) is \cite{Maurin02}, (7) is \cite{barrau02} and (8) is
\cite{Taillet01,Maurin04}.
          The pre-\cite{Maurin01} works are labelled as (-1) for
\cite{Bottino_Salati}
          (-2) for \cite{Orloff} and (-3) for \cite{chardonnet96}.}
          \label{fig:organigram}
\end{figure}

%%%%%%%%%%%%%%%%%%%%%%%%%%%%%%%%%%%%%%%%%%%%%%%%%%%%%%%%%%%%%%%%%%%%%%%%%%%%%%%%%%%%
%%%%%%%%%%%%%%%%%%%%%%%%%%%%%%%%%%%%%%%%%%%%%%%%%%%%%%%%%%%%%%%%%%%%%%%%%%%%%%%%%%%%
%%%%%%%%%%%%%%%%%%%%%%%%%%%%%%%%%%%%%%%%%%%%%%%%%%%%%%%%%%%%%%%%%%%%%%%%%%%%%%%%%%%%

\section*{Acknowledgments}

This work has benefited from the support of the Programme National
de Cosmologie and PICS 1076, CNRS. The authors are grateful to E. Vangioni-Flam \&
M. Cass\'e for enduring support and fruitful collaboration
on several aspects of CR sources and propagation. We acknowledge
G. Pelletier for clarifying discussion on diffusion coefficients.
One of the authors, D.M., is deeply indebted to A. Soutoul, E.
Vangioni-Flam \& M. Cass\'e for illuminating discussions and for
their constant support.

%%%%%%%%%%%%%%%%%%%%%%%%%%%%%%%%%%%%%%%%%%%%%%%%%%%%%%%%%%%%%%%%%%%%%%%%%%%%%%%%%%%%
%%%%%%%%%%%%%%%%%%%%%%%%%%%%%%%%%%%%%%%%%%%%%%%%%%%%%%%%%%%%%%%%%%%%%%%%%%%%%%%%%%%%

\end{document}